\documentclass[reprint,nofootinbib,amsmath,amssymb,prd]{revtex4-1}

\usepackage{graphicx}
\graphicspath{{figures/}}

\let\savedcor\corresponds
\let\corresponds\relax
\usepackage{mathabx}
\let\corresponds\savedcor

\usepackage{url}
\usepackage{mathtools}
\usepackage[version=4]{mhchem}
\usepackage{makecell}
\usepackage[hyperindex]{hyperref}
\usepackage{dcolumn}
\usepackage{bm}
\usepackage{times}
\usepackage{multirow}
\usepackage{epstopdf}
\usepackage{color}
\usepackage[normalem]{ulem}
\usepackage{xspace}
\usepackage{capt-of}
\usepackage{multirow}

\definecolor{cmntblue}{rgb}{0.0, 0.58, 0.71}
\definecolor{cmntgreen}{rgb}{0.0, 0.42, 0.24}

\usepackage[draft, inline, nomargin]{fixme}
\fxsetup{theme=color, mode=multiuser}
\FXRegisterAuthor{bl}{1}{\color{blue}BL}  
\FXRegisterAuthor{pts}{anb}{\color{red}PTS}

\begin{document}


\title{Benefits of MeV-scale reconstruction capabilities in large liquid argon time projection chambers} 

\author{W. Castiglioni}
\affiliation{Physics Department, Illinois Institute of Technology, Chicago, IL 60616, USA}

\author{W. Foreman}
\email{wforeman@iit.edu}
\affiliation{Physics Department, Illinois Institute of Technology, Chicago, IL 60616, USA}

\author{I. Lepetic}
\email{ivan.lepetic@rutgers.edu}
\affiliation{Department of Physics and Astronomy, Rutgers University, Piscataway, NJ 08854, USA}

\author{B. R. Littlejohn}
\email{blittlej@iit.edu}
\affiliation{Physics Department, Illinois Institute of Technology, Chicago, IL 60616, USA}

\author{M. Malaker}
\affiliation{Physics Department, Illinois Institute of Technology, Chicago, IL 60616, USA}

\author{A. Mastbaum}
\affiliation{Department of Physics and Astronomy, Rutgers University, Piscataway, NJ 08854, USA}

\date{\today}

\begin{abstract}
Using truth-level Monte Carlo simulations of particle interactions in a large volume of liquid argon, we demonstrate physics capabilities enabled by reconstruction of topologically compact and isolated low-energy features, or `blips,' in large liquid argon time projection chamber (LArTPC) events.
These features are mostly produced by electron products of photon interactions depositing ionization energy.  The blip identification capability of the LArTPC is enabled by its unique combination of size, position resolution precision, and low energy thresholds. We show that consideration of reconstructed blips in LArTPC physics analyses can result in substantial improvements in calorimetry for neutrino and new physics interactions and for final-state particles ranging in energy from the MeV to the GeV scale.  Blip activity analysis is also shown to enable discrimination between interaction channels and final-state particle types.  In addition to demonstrating these gains in calorimetry and discrimination, some limitations of blip reconstruction capabilities and physics outcomes are also discussed.  
\end{abstract}

\maketitle


\section{Introduction}
\label{sec:introduction}

Large liquid argon time projection chambers (LArTPCs) have become one of the primary detector technologies used for performing neutrino physics experiments.  
At present, a suite of sub-kiloton scale LArTPCs, SBND~\cite{sbnd_web}, MicroBooNE~\cite{ub_det}, and ICARUS~\cite{icarus_web}, collectively called the Fermilab SBN Program, are operating or being constructed along Fermilab's Booster Neutrino Beamline at sub-km baselines for the primary purposes of probing short-baseline neutrino appearance/disappearance~\cite{sbn} and measuring neutrino-nucleus interaction cross sections on argon~\cite{sbnd_phys}.  
Within the next decade, the 40 kiloton Deep Underground Neutrino Experiment (DUNE) LArTPC~\cite{DUNE_tdr1} will be deployed underground in the Sanford Underground Research Facility in South Dakota along a new Fermilab-based neutrino beamline~\cite{pip_web} for the primary purposes of measuring long-baseline neutrino oscillations and leptonic CP-violation, probing a host of beyond-the-Standard-Model (BSM) physics models, and measuring neutrinos from astrophysical sources, such as core-collapse supernovae~\cite{dune_tdr2}.  

The primary technological advantage most exploited thus far in existing large LArTPC measurements and physics sensitivity studies is arguably its millimeter-scale spatial resolution.  
The LArTPC's uniform electric field, low electron diffusion, and sub-centimeter charge readout element spacing enable the conservation and recording of initial ionization electron topologies produced by particle interactions in the argon.
For GeV-scale neutrinos, the rich topologies of interaction final-state tracks and showers can be used to distinguish electron- and muon-type neutrino interactions, enabling sensitive electron-type neutrino searches in conventional $\nu_{\mu}$-dominated neutrino beamlines~\cite{argo_nue,argo_nue_xsec,uB_numi}.  
Simple but precise mm-scale topological analysis of interaction vertices and final-state tracks has also enabled LArTPCs to provide sensitive BSM searches~\cite{ub_hnl} and new insight into neutrino interaction models~\cite{argo_hammer,ub_mult}; major improvements on the latter front are expected as a larger set of exclusive cross-section measurements are developed and published by MicroBooNE and other experiments.  
Small-angle muon scattering clearly visible in high-resolution images~\cite{ub_mcs}, as well as track length, have been used in MicroBooNE as primary methods of energy reconstruction for BSM and GeV-scale neutrino interaction cross section measurements in LAr~\cite{ub_cc,ub_hnl}.  
Many software tools, based on a range of operational principles, have been developed that use LArTPC image topologies to identify track and shower objects and reconstruct their kinematics~\cite{ub_pandora, ub_dl1,ub_dl2}.  
Many of the stated centerpiece goals of short-baseline and long-baseline LArTPC experiments will be achieved by combining mm-scale resolution and calorimetric capabilities in analyzing charged particle interactions in LAr ranging from the tens to thousands of MeV~scales.  
This recipe provides the kinematic and particle identification details necessary to perform the long list of neutrino LArTPC physics goals given above.  

A technological advantage of the large neutrino LArTPC that has received comparatively less attention is its low-energy-threshold detection capability.  
This capability is enabled by the modest 24~eV mean ionization energy of liquid argon, the high ionization electron collection efficiency of the TPC, and the low levels of  noise achievable in modern readout electronics.  
Through studies of Michel electrons~\cite{ub_michel,lariat_light,icarus_michel}, nuclear de-excitation photons~\cite{argo_mev}, and $^{39}$Ar $\beta$-particles~\cite{uB_ar39}, it has been established that single-phase neutrino LArTPCs are capable of identifying physics signatures at and well below the MeV energy scale.  
In one of these studies, performed by the ArgoNeuT single-phase LArTPC experiment~\cite{argo_det}, a physics detection threshold of 200--300~keV was established by comparing simulated and measured de-excitation photons generated by final-state nuclei and neutrons from neutrino interactions~\cite{argo_mev}.  
This study specifically highlights the uniqueness of LArTPCs among all demonstrated massive neutrino detector technologies in achieving a \emph{combination} of mm-scale position resolution \emph{and} sub-MeV energy thresholds.  

The aim of this paper is to explore how these low-energy-threshold, high-position-resolution combined capabilities might be put to use in a broad variety of contexts relevant to large neutrino LArTPC physics goals.  
More specifically, we will describe how LArTPC signatures that are compact (sub-cm scale), low-energy (below the few~MeV scale) and/or topologically isolated (separated from larger topological objects by cm~or more) are produced in physics events of interest, and how these signatures can be used to enhance capabilities in calorimetry, energy calibration, and discrimination of particle type or interaction type.  
We will show that analysis of these low-energy, compact LArTPC signatures, which we refer to as `blips' throughout the paper, can be broadly beneficial in supernova neutrino, solar neutrino, long-baseline oscillation, and BSM studies in LArTPCs.   
All studies are performed using a common framework of truth-level Monte Carlo simulations in a generic liquid argon environment.  

This paper will begin in Section~\ref{sec:methods} by describing the Monte Carlo methods used to define blip activity in LArTPC events, and to describe blip physics metrics of interest used throughout the paper.  
The benefits of considering reconstructed blip activity in supernova and solar neutrino energy reconstruction and interaction channel identification (Section~\ref{sec:supernova}), neutron identification and calorimetry (Section~\ref{sec:neutrons}), electromagnetic shower calorimetry~(Section~\ref{sec:em}), particle discrimination~(Section~\ref{sec:particleid}), BSM physics (Section~\ref{sec:bsm}), and single $\gamma$-ray spectroscopy (\ref{sec:spec}) will then be presented and discussed.  
Some primary detector-related effects limiting the utility of blip activity for physics purposes will be studied in Section~\ref{sec:limitations}.  
Summarizing remarks will be given in Section~\ref{sec:conc}.

\section{Study Definitions and Procedures}
\label{sec:methods}

In this section, we will summarize the physics processes that define the LArTPC blip signals to be studied in this paper, and then describe the Monte Carlo simulation used to generate the truth-level reconstructed physics quantities we will use to demonstrate the physics potential of blip signals.

\subsection{Physics of Low-Energy Depositions in Liquid Argon}

The blip features in LArTPC images that will receive most of the focus of this paper are the product of ionization of argon by low-energy ($\sim$50~keV to $\sim$5~MeV) electrons.  
True ionization topologies from these electrons have extents on the mm~scale to cm~scale range, and are generally smaller than would be expected based on a straight line path following the electron's CSDA range in LAr~\cite{csda}, due to repeated hard scatters of the electron while thermalizing.  

In large single-phase neutrino LArTPCs using wire-based charge readout, such as the Fermilab SBN experiments and DUNE far detector modules, wires are spaced in 3-5~mm intervals, meaning that most low-energy electrons will produce a measurable signal, or hit, on a small number (usually 1 or 2) wires in each LArTPC readout plane.  
Reconstruction of the 2D or 3D position of the ionization feature in this case is enabled by matching collected signals between non-parallel readout planes at common readout times.  
Due to the small number of hit wires, reconstruction of the original direction or energy deposition density profile of the electron with any level of precision is likely not possible with current LArTPC technology.  
Energy deposited by the electron can be reconstructed by integrating the total digitized signal amplitude on the wires (usually the collection plane wires), and scaling the result to take into account the LArTPC's ADC-to-electron calibration and recombination and quenching factors; this process is described in Ref.~\cite{lepetic_phd}. 

Low-energy electron signals will of course also be produced in large LArTPCs using other detector charge signal readout technologies.  
Of particular interest are single-phase detectors using a 2D pixel readout, as is envisioned for the DUNE Near Detector LArTPC and a possibility in a future far detector module, and dual-phase LArTPCs with large electron multiplier (LEM) readout, as planned for at least one DUNE far detector module ~\cite{dune_tdr2}.  
Each of these applications of charge readout technologies will offer similar spatial resolution to wire-based readout systems, as well as similar energy thresholds within roughly a factor of two~\cite{Dwyer_2018,dune_idr3}.  
Thus, all demonstrations and procedures to be provided in this paper should be viewed as equally applicable to wire-based, pixel-based, and dual-phase large neutrino LArTPCs.  

Low-energy electrons of interest for this study are mainly produced by electromagnetic interactions of photons via the Compton scattering process, although the photoelectric effect and pair-production also provide non-negligible contributions.  
See Refs.~\cite{caratelli_phd,ub_michel,ub_pi0} for a more detailed description of photon interaction cross-sections on argon in this energy regime.  
As uncharged photons themselves do not generate ionization signatures in the LAr, the ionizing Compton electrons they produce during interactions appear as topologically isolated features in LArTPC images.  
In this study, we will focus primarily on MeV-scale $\gamma$-rays generated via de-excitation of nuclei following inelastic interactions with neutrinos, neutrons, pions, and muons, and via bremsstrahlung interactions of  electrons.  

Isolated blip-like features can also be produced in LArTPC events via proton-producing inelastic interactions of high-energy neutrons with argon.  
Given their higher deposition density profiles, proton-produced blips can have reconstructed energies higher than what can be produced by an electron.  
However, MeV-scale proton-produced blips are largely indistinguishable from electron-produced blips.  

\subsection{Blip Simulation and Truth-level Reconstruction}

For this study, primary particles are generated in and propagated through a large, essentially infinite uniform volume of liquid argon using the Geant4-based~\cite{g4} LArSoft simulation package~\cite{larsoft}.  
Primary electrons, neutrons, protons, pions, muons, and $\gamma$-rays are mainly generated using the standard Geant4 gun generator; for supernova-related studies, neutrino interaction final-state products are generated using the MARLEY neutrino interaction software package~\cite{marley}.  
In propagating particles through the argon with LArSoft, care is taken to implement the correct physics libraries and threshold settings required for properly simulating high-energy physics processes, such as neutron inelastic scattering and pion and muon capture.  
For high-energy physics processes, the \texttt{NeutronHP} and \texttt{QGSP\_BERT\_HP} Geant4 libraries are implemented.
During particle transport, a wide variety of particle history information is stored for further analysis, including the true starting and ending energies and locations for all particles, their identities and the physics processes resulting in their creation or destruction, and the properties of their parent and daughter particles. While Geant4 and LArSoft propagate particles of all energies, only particles with energy above 10~keV are stored for later analysis.

Rather than simulating the full readout, low-level processing, and higher-level object reconstruction pathways of a specific LArTPC experiment to generate reconstructed blip objects for analysis, we adopt a detector-agnostic truth-level approach.  
To begin, any topologically-isolated electron depositing more than 75~keV of energy in the liquid argon is considered as an identified, reconstructed blip.  
This threshold is lower than that achieved in previous blip analyses in ArgoNeuT~\cite{lepetic_phd} to reflect the reduction in noise levels that are achievable in large LArTPCs using cold electronics~\cite{ub_noise}.  
The reconstructed position and energy of each blip is taken to be the true start location of and true energy deposited by the electron interaction, respectively.  
The addition of a finite blip energy resolution or higher detection threshold only marginally impacts the physics results to be described; these detector effects will be explored in more detail in Section~\ref{sec:limitations}.  

The reconstructed blip quantities of interest for physics analysis in this paper are the total blip multiplicity and individual and summed blip energies.  
To consider the likely need to avoid inclusion of blip activity from radiogenic $^{39}$Ar $\beta$ decays, only blips within a set proximity to points of interest in an event, such as a neutrino interaction vertex, will be considered.  
In practice in this paper, this point of interest will be defined as the true generation vertex of the relevant primary particle.  
In most physics cases considered in this paper, reference points of interest will be easy to identify to mm-scale precision with conventional large-feature reconstruction algorithms.  
For this study, blip distances of 20, 30, and 60~cm are usually considered; for reference, the attenuation (interaction) length for a 1~MeV photon (10~MeV neutron) in LAr is approximately 15~cm (30~cm).  
The presence and impact of $^{39}$Ar contamination of blip samples will not be considered in most of the following sections, and is instead separately examined in Section~\ref{sec:limitations}.  
Beyond blip proximity, other topological features of blips are not considered in our analysis.  

\begin{figure}
\includegraphics[trim = 24.0cm 5.7cm 12.0cm 10.0cm, clip=true, 
width=0.49\textwidth]{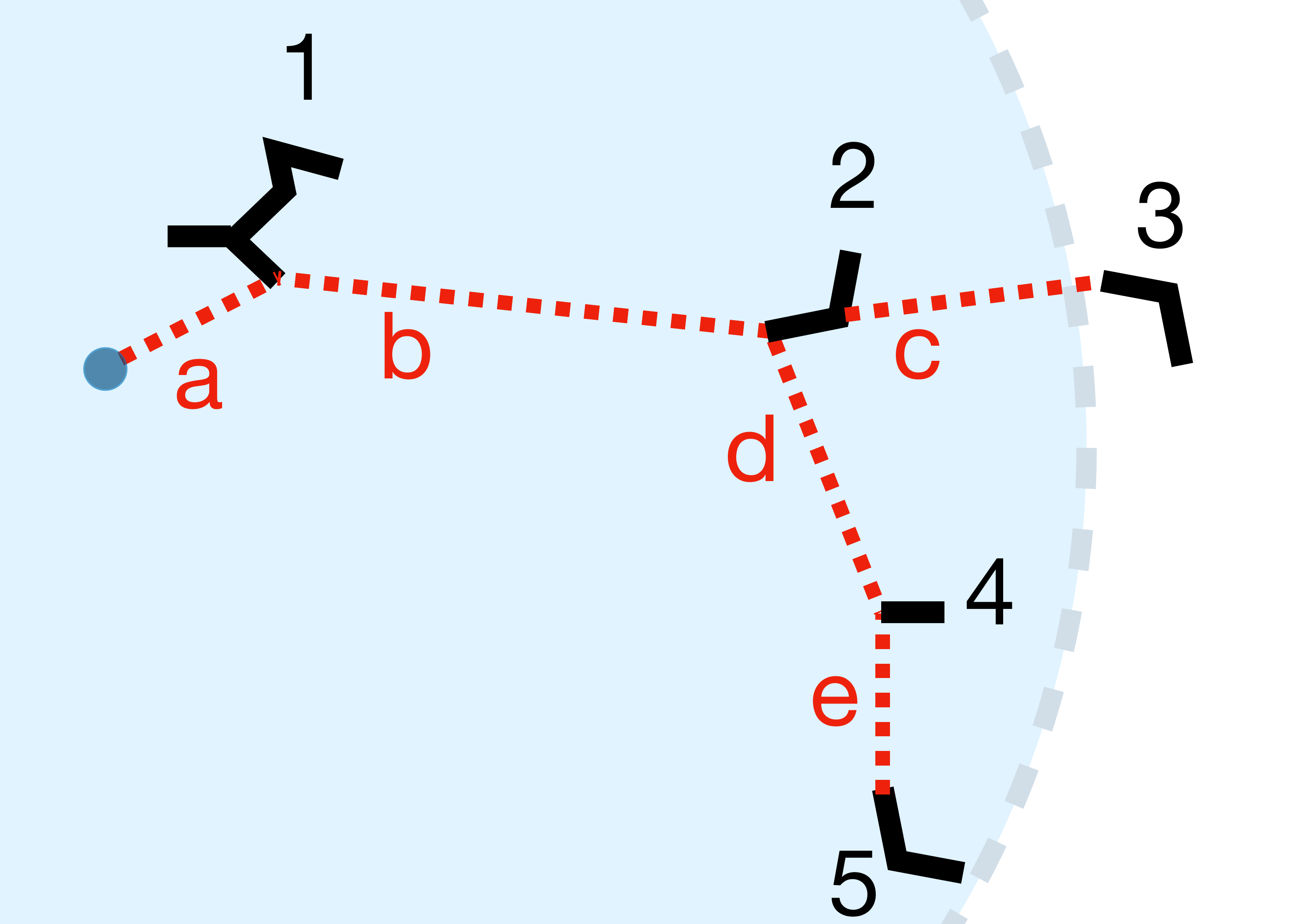}
\caption{Illustrated history of an example 3.0 MeV $\gamma$-ray interaction in liquid argon.  Red (black) dotted (solid) lines indicate $\gamma$-ray (electron) trajectories; energies of these particles are indicated in Table~\ref{tab:history}.  A blue circle illustrates a 30~cm proximity requirement with respect to the $\gamma$-ray generation point (dark blue dot).  Four of five black electron groups are identified as reconstructed blips, and only three of these, `1,' `2,' and `5,' meet the proximity requirement.}
\label{fig:history}

\vspace{2ex}

\begin{tabular}{|c|c|l|c|c|c|}
\hline 
Item & Type & Creator Process & \makecell{E$_\text{start}$\\(MeV)}&\makecell{E$_\text{end}$\\(MeV)} &\makecell{E$_\text{blip}$\\(MeV)} \\ \hline
a & $\gamma$ & Primary & 3.00 & 3.00 & - \\
1 & 2~e$^-$ & Compton scatter & 1.50 & 0 & 1.50 \\
b & $\gamma$ & Compton scatter & 1.50 & 1.50 & - \\
2 & e$^-$ & Compton scatter & 1.00 & 0 & 0.75 \\
c & $\gamma$ & Bremsstrahlung & 0.25 & 0.25 & - \\
3 & e$^-$ & Photoelectric effect & 0.25 & 0 & 0.25 \\
d & $\gamma$ & Compton scatter & 0.50 & 0.50 & - \\
4 & e$^-$ & Compton scatter & 0.05 & 0 & - \\
e &  $\gamma$ & Compton scatter & 0.45 & 0.45 & - \\
5 & e$^-$ & Photoelectric effect & 0.45 & 0 & 0.45 \\ \hline 
\end{tabular}
\captionof{table}{Particles followed in the recorded history of the example 3~MeV $\gamma$-ray event shown in Figure~\ref{fig:history}, which is used to illustrate the details of the blip reconstruction procedure.  This event would produce four identified blips, with reconstructed energies of 1.5, 0.75, 0.25, and 0.45~MeV.  Of the primary $\gamma$-ray's 3~MeV total energy, 2.7~MeV of summed  blip energy is considered after proximity requirements are applied.}
\label{tab:history}

\end{figure}

In this truth-level blip reconstruction scheme, care must be taken in  summing energies for contiguous electron interactions and in considering electrons undergoing bremsstrahlung interactions.  
These and other aspects of the blip reconstruction procedure are illustrated in Figure~\ref{fig:history} and Table~\ref{tab:history} using an example history of a simulated 3~MeV $\gamma$-ray.  
This $\gamma$-ray undergoes three Compton scatters and ends its life via the photoelectric effect.  
Reconstructed blip `1,' formed at the first Compton scatter vertex, must have an energy that includes the depositions of both the initial Compton electron as well as the hard scattered electron it produced.  
This is accomplished by taking the Compton electron's starting energy (E$_\text{start}$ in Table~\ref{tab:history}) as the reconstructed blip energy.  
For a test sample of simulated 1.5~MeV $\gamma$ rays, less than 1\% of all tracked electrons were produced in hard scatters of a parent electron; this fraction is likely to be higher for higher-energy simulated primary particles.  

For reconstructed blip `2,' energy lost via bremsstrahlung interaction of the Compton electron must accounted for by subtracting from E$_\text{start}$ the energies of any produced daughter $\gamma$-ray. Both the first electron (`2') and that produced by the bremsstrahlung photon interaction (`3') are considered as separate candidate reconstructed blips.  
For the 1.5~MeV $\gamma$-ray dataset mentioned above, bremsstrahlung photons will occasionally convert within a distance smaller than the position resolution of a LArTPC ($\sim$0.5~cm) and produce two overlapping, indistinguishable blips.  
As this tight blip spacing applies to less than 1\% of all tracked electrons in the test dataset, we do not treat it as a special case.  

The Compton electron produced at point `4,' at 50~keV, is below the default blip detection threshold, and is excluded from the analysis.  
Although it is above the detection threshold, blip `3' is beyond the pictured proximity requirement, and will not be included when determining summed blip multiplicity and energy for this event.  
Overall, this 3~MeV $\gamma$-ray produces a blip multiplicity of 3, with individual blip energies of 1.50, 0.75, and 0.45~MeV, and a summed blip energy of 2.7~MeV.  

As mentioned in the previous section, neutron inelastic interactions in argon can produce final-state protons, which will also appear as blips in LArTPC images.  
To realistically include these signatures in our truth-level reconstruction scheme, any proton blip below 3~MeV in total energy is treated as a reconstructed electron blip.  
Any proton above this energy would produce a single-hit blip too high in energy to reasonably be produced by an electron~\cite{lepetic_phd}; thus, these protons are excluded from the set of considered blips.

\section{Supernova and Solar Neutrinos}
\label{sec:supernova}

A DUNE-based analysis of solar neutrinos will improve the solar-based measurement of $\Delta m^2_{12}$, enabling precise tests of the Standard Model neutrino mixing picture~\cite{dune_solar}.
In the case of supernova burst neutrinos, the primary reconstructed physics metrics of interest are the independent energy and time profile of fluxes for the different neutrino flavors, in addition to the total number of detected neutrinos.
Using our truth-level blip reconstruction technique on MARLEY- and Geant4-generated low-energy neutrino interaction final states, we will demonstrate how blip activity can improve energy recovery and resolution and can aid in separation of flavor-exclusive $\nu_e$ charged current (CC) and flavor-inclusive neutrino-electron scattering (ES) channels.  


\subsection{Neutrino Energy Calorimetry Improvements}

To provide optimized low-energy neutrino energy reconstruction, one must perform calorimetry on all visible final-state particles.  
For supernova and solar neutrino interactions in argon, $\nu_e$ CC interactions represent the primary detection channel:
\begin{equation}
    \nu_e 
    + \prescript{40}{}{\text{Ar}} 
    \rightarrow e^- 
    + \prescript{40}{}{\text{K}}^*. \\
\label{eq:cc}
\end{equation}
Thus, for this channel, an optimal reconstruction of energy will include depositions not just from the final-state $e^-$, but also from the products of de-excitation of the $^{40}$K nucleus, primarily $\gamma$-rays and neutrons.  
While current tools will likely be capable of triggering on and reconstructing the former~\cite{dune_tdr2,ub_michel}, inclusion of the latter in calorimetry has not been closely studied in the literature.  
For the sub-dominant $\nu_x$-electron ES interaction process,  
\begin{equation}
    \nu_x + e^- \rightarrow \nu_x + e^-, \\
\label{eq:es}
\end{equation}
energy optimization will yield limited improvement due to the presence of the invisible final-state $\nu_x$; thus, the ES channel will be ignored in the present subsection.  

\begin{figure}
\includegraphics[trim = 0.0cm 0.0cm 0.0cm 1.0cm, clip=true,  width=0.9\columnwidth]{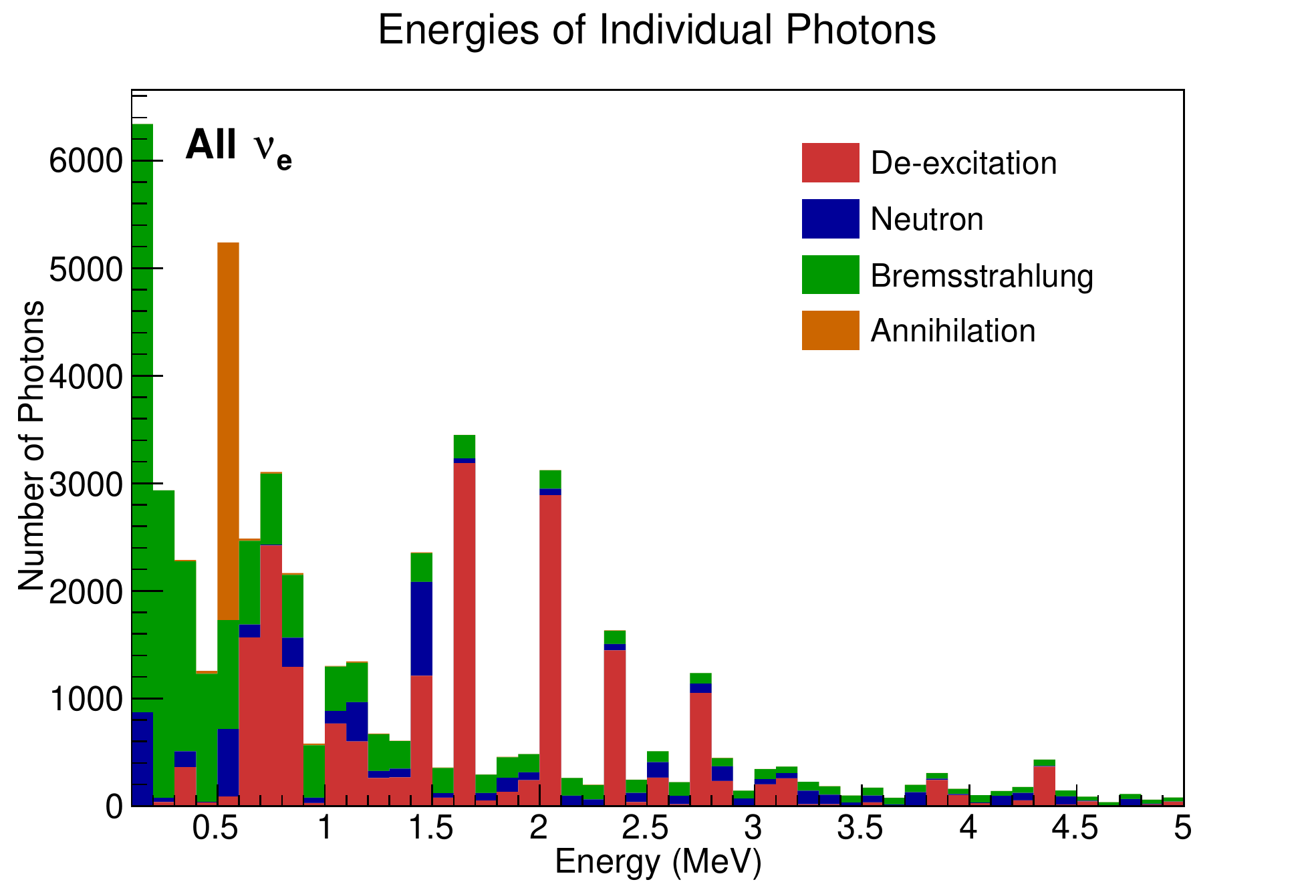}
\includegraphics[trim = 0.0cm 0.0cm 0.0cm 1.0cm, clip=true,  width=0.9\columnwidth]{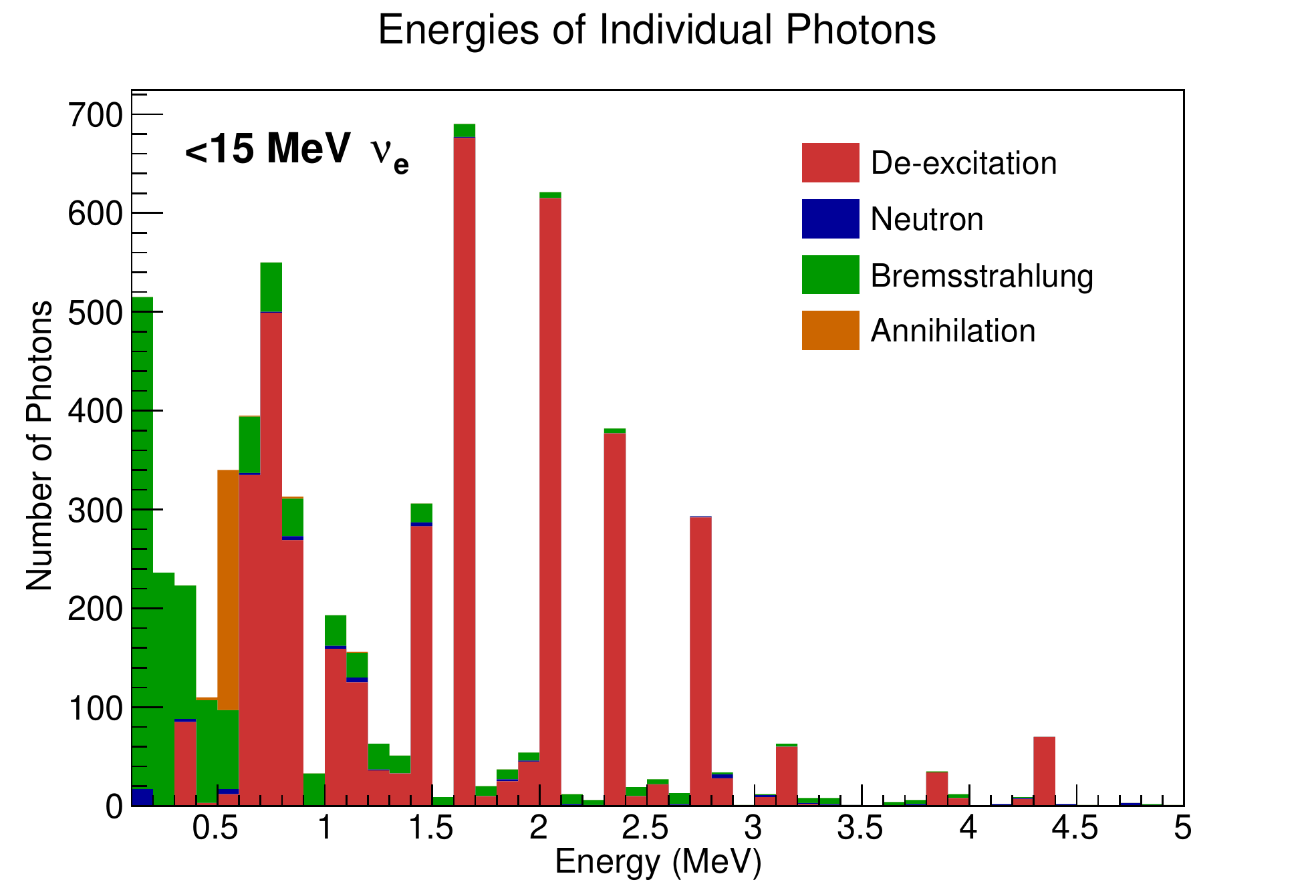}
\caption{Top: Energy distribution of $\gamma$-rays produced in $10^4$ MARLEY-generated $\nu_e$ CC supernova neutrino interactions.  Bottom: Subset of supernova CC events with $\nu_e$ energy below 15~MeV.}
\label{fig:sn_e}
\end{figure}

Figure~\ref{fig:sn_e} shows the energies of $\gamma$-rays produced in 10$^4$ $\nu_e$ CC interactions generated with MARLEY for the full supernova neutrino energy range.   
Also pictured is the same $\gamma$-ray energy spectrum for the subset of events with $\nu_e$ energies below 15~MeV, roughly matching the maximum energy limit for $^8$B solar neutrinos.
Separate accounting is done for photons produced by interactions of final-state de-excitation $\gamma$-rays, bremsstrahlung photons produced by the final-state electron, and de-excitation photons produced via inelastic scatters of final-state neutrons.  
It is clear that the average low-energy $\nu_e$ event will feature more than one MeV-scale $\gamma$-ray in the final state, each of which is quite likely to lead to multiple reconstructible blips.

\begin{figure}
\includegraphics[trim = 0.0cm 1.1cm 0.cm 1.1cm, clip=true, 
width=0.88\columnwidth]{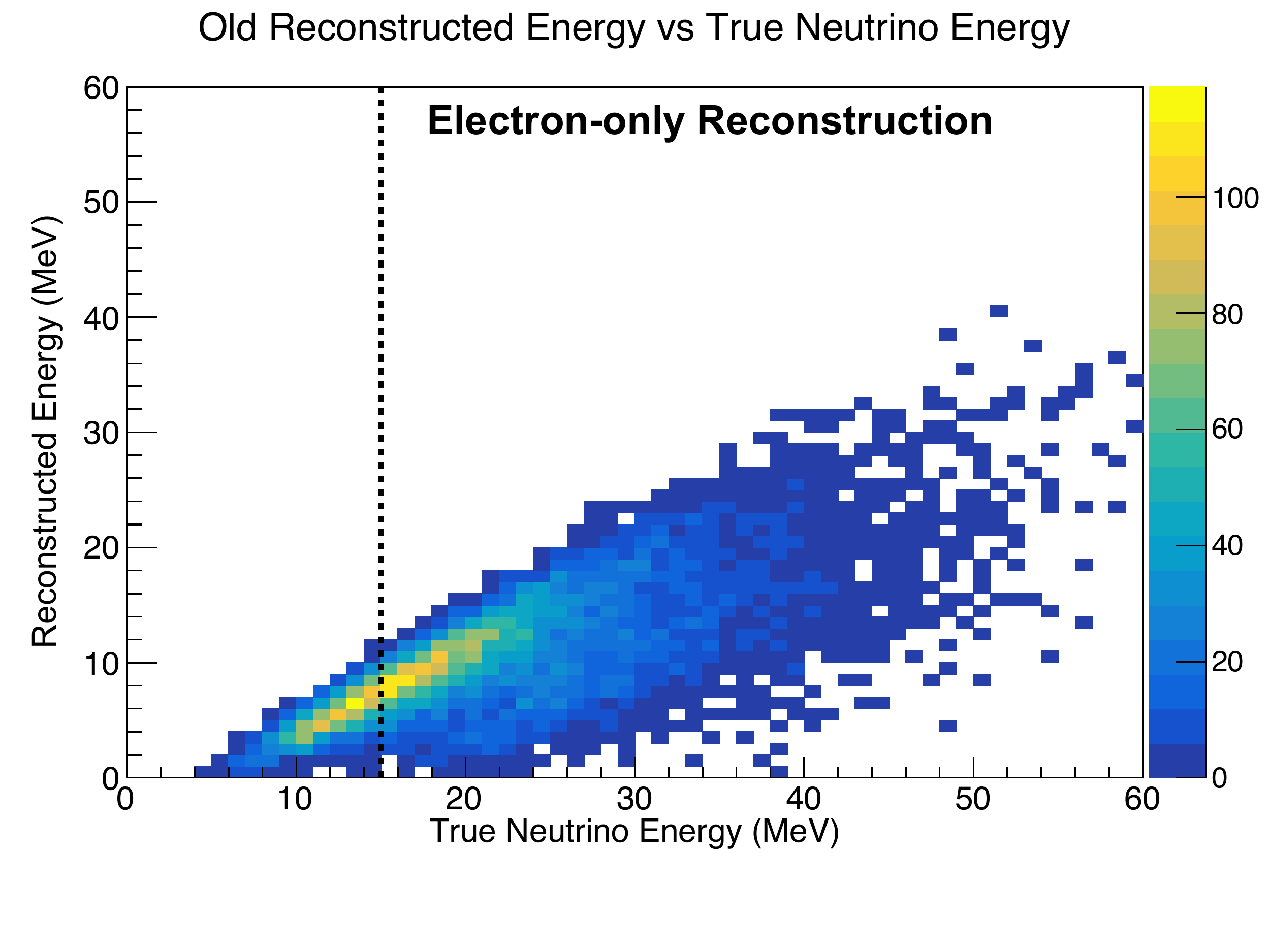}
\includegraphics[trim = 0.0cm 1.1cm 0.cm 1.1cm, clip=true, 
width=0.88\columnwidth]{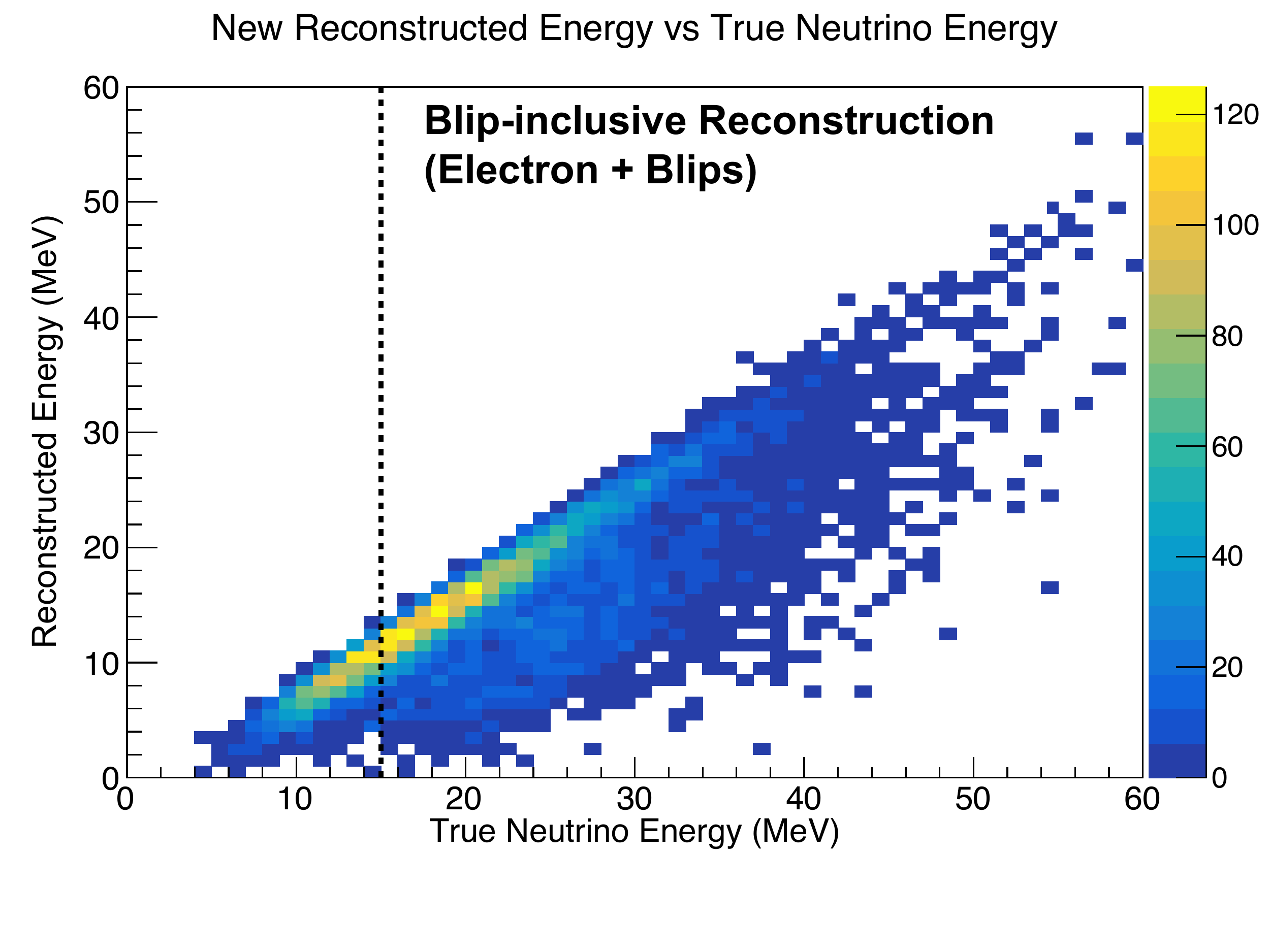}
\caption{Reconstructed vs. true energies for supernova $\nu_e$ undergoing CC interactions in liquid argon.  Top: reconstruction of only the primary final-state electron's trunk.  Bottom: improved reconstruction scenario in which blips above 75~keV and within 30~cm of the neutrino interaction vertex are included.  Events are generated using MARLEY simulations.  The vertical black dotted line at 15~MeV denotes the approximate endpoint of the ${}^8$B solar neutrino spectrum.}  
\label{fig:sn_budget}
\end{figure}
\begin{figure}
\includegraphics[trim = 0.0cm 1.2cm 0.0cm 1.2cm, clip=true,  width=0.88\columnwidth]{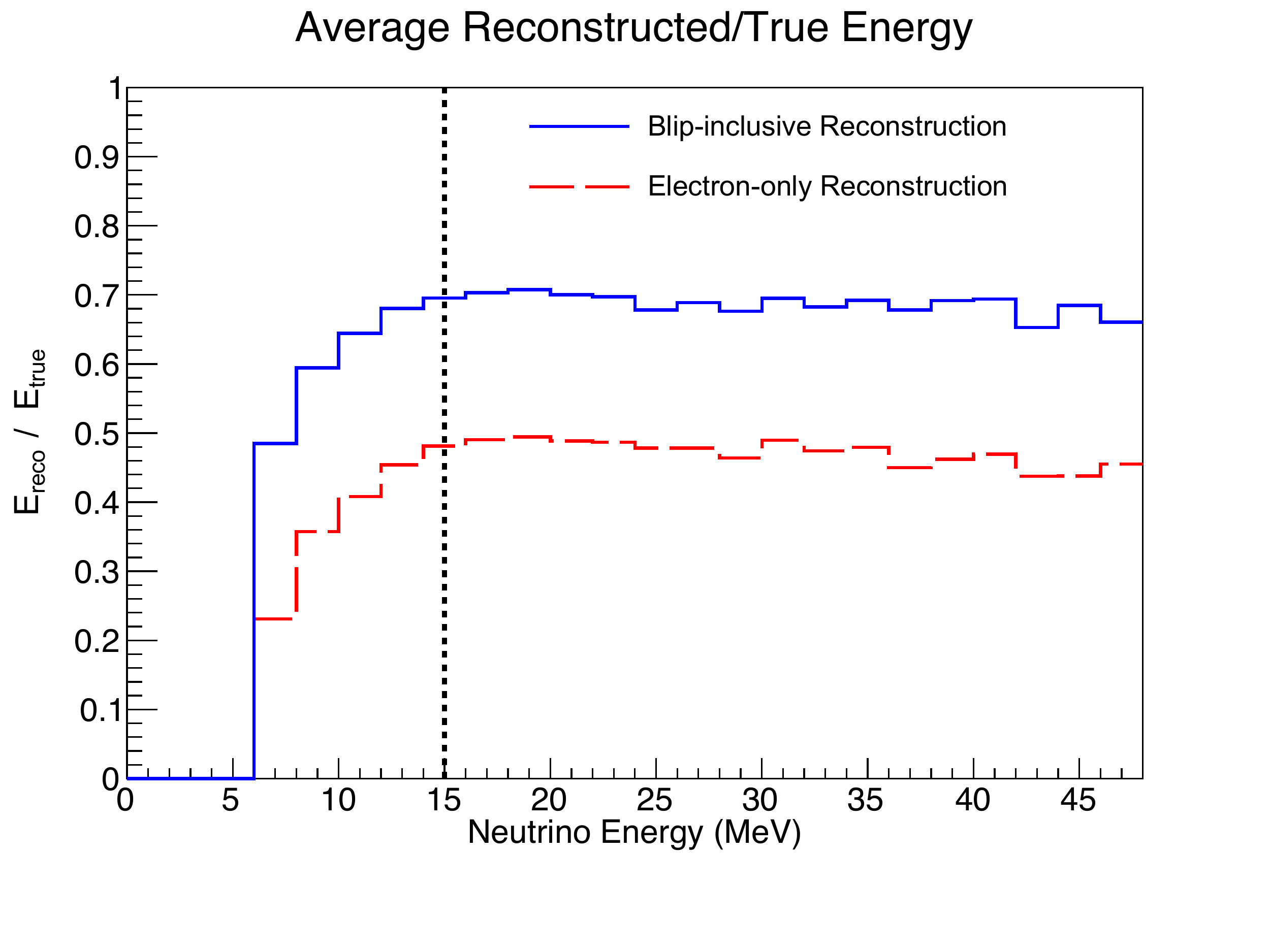}
\includegraphics[trim = 0.0cm 1.2cm 0.0cm 1.2cm, clip=true, 
width=0.88\columnwidth]{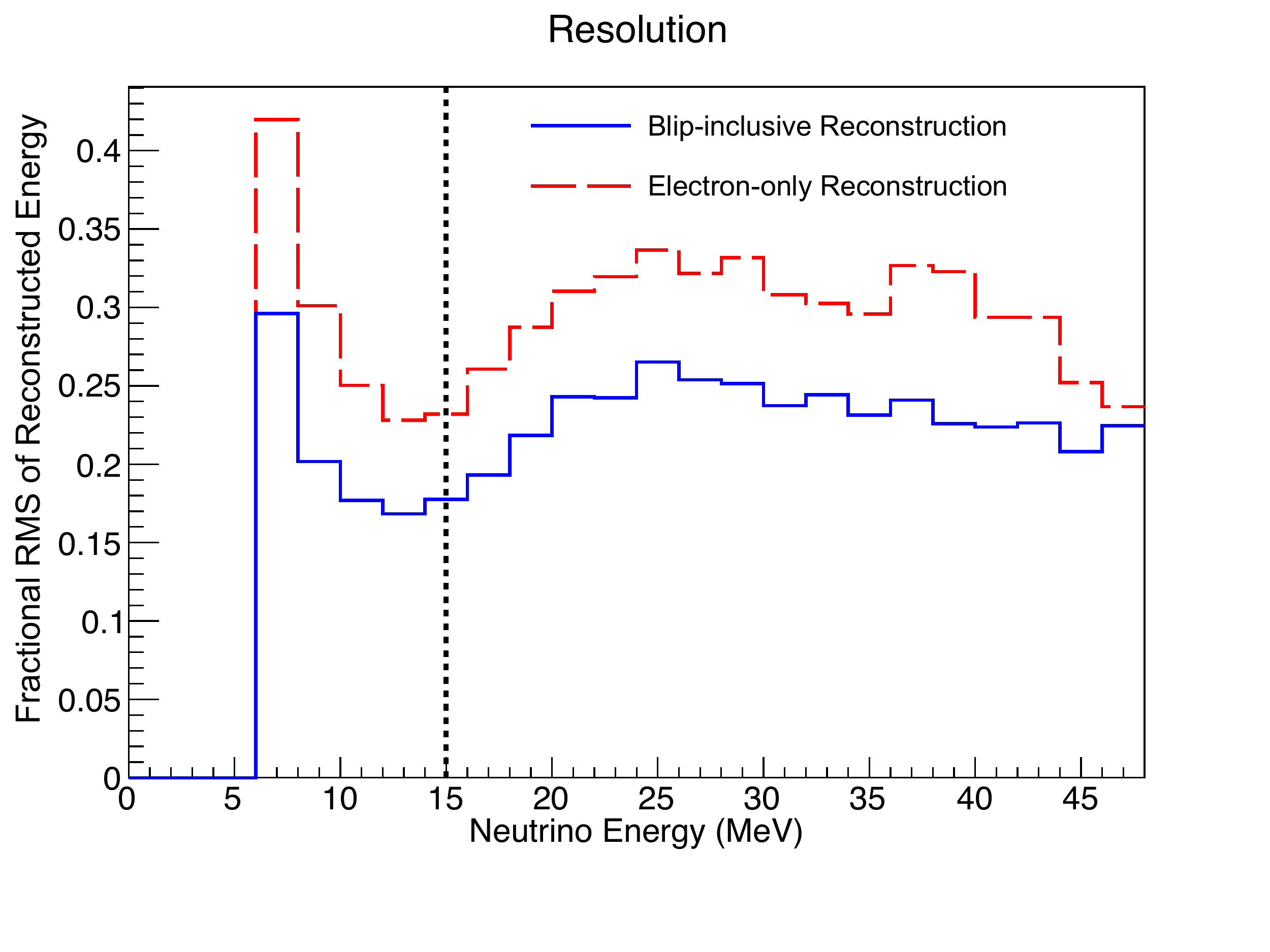}
\caption{Top: Ratio of the average reconstructed energy to the true neutrino energy for $\nu_e$ CC interactions using the electron-only (red) and blip-inclusive (blue) reconstruction methods shown in Figure~\ref{fig:sn_budget}. Bottom: The RMS of $E_\text{reco}$ for different bins of true neutrino energy divided by the average $E_\text{reco}$ in each bin.  The vertical black dotted line at 15~MeV denotes the approximate endpoint of the ${}^8$B solar neutrino spectrum.}
\label{fig:sn_reco_ratios}
\end{figure}

Figures~\ref{fig:sn_budget} and~\ref{fig:sn_reco_ratios} illustrate the impact of including blips when reconstructing the energy in $\nu_e$ CC events.  
The former presents the reconstructed and true $\nu_e$ energies for all events, while the latter presents the one-dimensional profiled mean and RMS reconstructed energy versus true $\nu_e$ energy.  
For the first scenario, we consider reconstruction of only the primary electron topological object, or `trunk'; this definition excludes energy lost by the primary electron to bremsstrahlung radiation.  
In the second scenario, we include reconstructed blips induced by the $\gamma$-rays in Figure~\ref{fig:sn_e}, as long as they occur within 30~cm of the neutrino interaction vertex.  
For the second scenario in Figure~\ref{fig:sn_budget}, a strong trend is visible just below the ``$E_\text{reco} = E_\text{true} - 2.8$~MeV'' diagonal representing the most complete possible extent of calorimetry.  
This diagonal trend is substantially more smeared for the case in which only the electron trunk is reconstructed -- particularly at higher energies, where bremsstrahlung interactions of the primary electron are more likely.  
Beginning at $\nu_e$ energy of approximately 18~MeV in both cases, a population of off-diagonal events is visible, arising primarily from binding energy losses associated with final-state neutron production.  
As will be discussed in Section~\ref{sec:neutrons}, it is possible (but unlikely) that this binding energy loss is recoverable in the reconstruction, given the challenge of positively identifying the presence of final-state neutrons in these interactions.  

For the `electron-only' case pictured in Figure~\ref{fig:sn_reco_ratios}, the reconstructed energy is significantly less than the true $\nu_e$ energy: on average, only 47\% of the neutrino's kinetic energy is reconstructed.  
When reconstructed blips are included, 70\% of the $\nu_e$ kinetic energy is recovered on average.  
Increasing the radius of blip inclusion from 30~cm to 60~cm improves this fraction to 79\%.  
Much of the remaining unaccounted energy in this case is due to the reaction threshold and binding energy losses.  
The former is constant with energy and thus does not result in broadened energy resolution.  The latter is energy-dependent, and does provide a substantial energy resolution contribution.  

Also illustrated in Figure~\ref{fig:sn_reco_ratios} is the fractional RMS of reconstructed energy, defined as the RMS of $E_\text{reco}$ divided by the average $E_\text{reco}$ within bins of $\nu_e$ energy.
The unusual shape of this curve -- specifically, the rise in fractional RMS beginning around 15~MeV -- is attributed to the broadening of the reconstructed energy caused by binding energy losses in events where a final-state neutron is produced~\cite{lar_res}.
The RMS curve for the blip-inclusive case remain within $\sim$15-25\% for $\nu_e$ energies above about 7.5~MeV, and an absolute improvement in RMS of roughly 5-10\% with respect to the electron-only case is visible at all true $\nu_e$ energies.  
If the proximity requirement is loosened to 60~cm, RMS is further improved by up to 4\%.
Thus, it is clear that blip reconstruction has the potential to improve calorimetry for both solar and supernova neutrino CC interactions.

We note that the blip-inclusive fractional RMS achieved in this study is comparable to the energy resolution estimated in the DUNE Technical Design Report (TDR) using existing charge signal reconstruction tools~\cite{dune_tdr2}.
Given that we implement an optimistic 75~keV blip threshold in this study, it seems unlikely that further improvements beyond that pictured in Figure~\ref{fig:sn_reco_ratios} and referenced above are possible with DUNE charge signals.  
Thus, our study appears to indicate that a resolution approaching the `physics limited' scenario in the DUNE TDR is likely not achievable.

\subsection{Interaction Channel Identification}

For low-energy neutrino signals, the multiplicity of topologically-isolated signatures in an event will increase as the assumed feature reconstruction threshold is decreased.  
This increase in multiplicity is likely to be sizeably different for $\nu_e$ CC and $\nu_x$ ES interactions of supernova and solar neutrinos, due to the lack of nuclear de-excitation activity in the latter case.

\begin{figure}
\includegraphics[width=0.9\columnwidth]{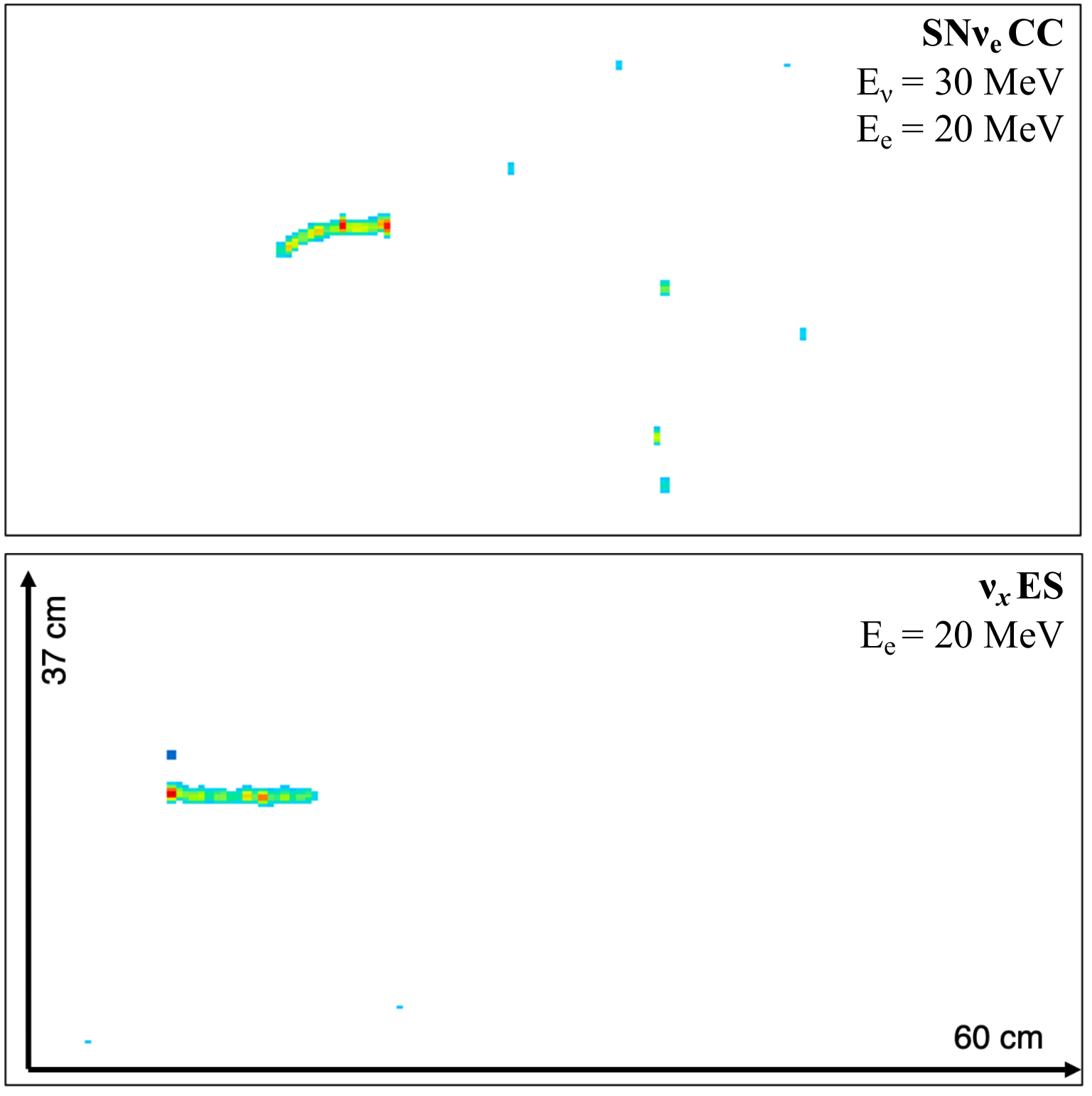}
\caption{LArTPC event displays showing a MARLEY-simulated supernova $\nu_e$ CC interaction (top) and a single electron (bottom) with an energy comparable to that expected from a supernova $\nu_x$ ES interaction.  
Both event displays are made with a simulated LArTPC incorporating wire-based charge readout, and only collection plane signals are shown.  Approximate dimensional scales based on the wire separation and electron drift time are shown. The color scale indicates the density of collected charge.}
\label{fig:sn_evd}
\end{figure}

Figure~\ref{fig:sn_evd} shows two simulated supernova neutrino interactions in a LArTPC.  
The top event consists of a 30~MeV $\nu_e$ CC event producing a 20~MeV electron and cascade of de-excitation photons. Compton scatters of these $\gamma$-rays produce the many blips seen in the event.  
The bottom event consists of a single 20~MeV electron, as could be produced by a 30~MeV $\nu_x$ ES interaction.  
We note that the extensive blip activity seen in the top event is absent in the bottom one.  
These topological differences suggest that a capability to distinguish between interaction channels is offered by blip reconstruction in LArTPCs.  
Such a capability would provide additional discrimination beyond that achieved by considering the forward-scattered kinematics of the final-state electron in the ES process~\cite{sno_flavor,dune_solar,dune_tdr2}.  

\begin{figure}
\includegraphics[trim = 0.0cm 1.2cm 0.0cm 1.1cm, clip=true, 
width=0.95\columnwidth]{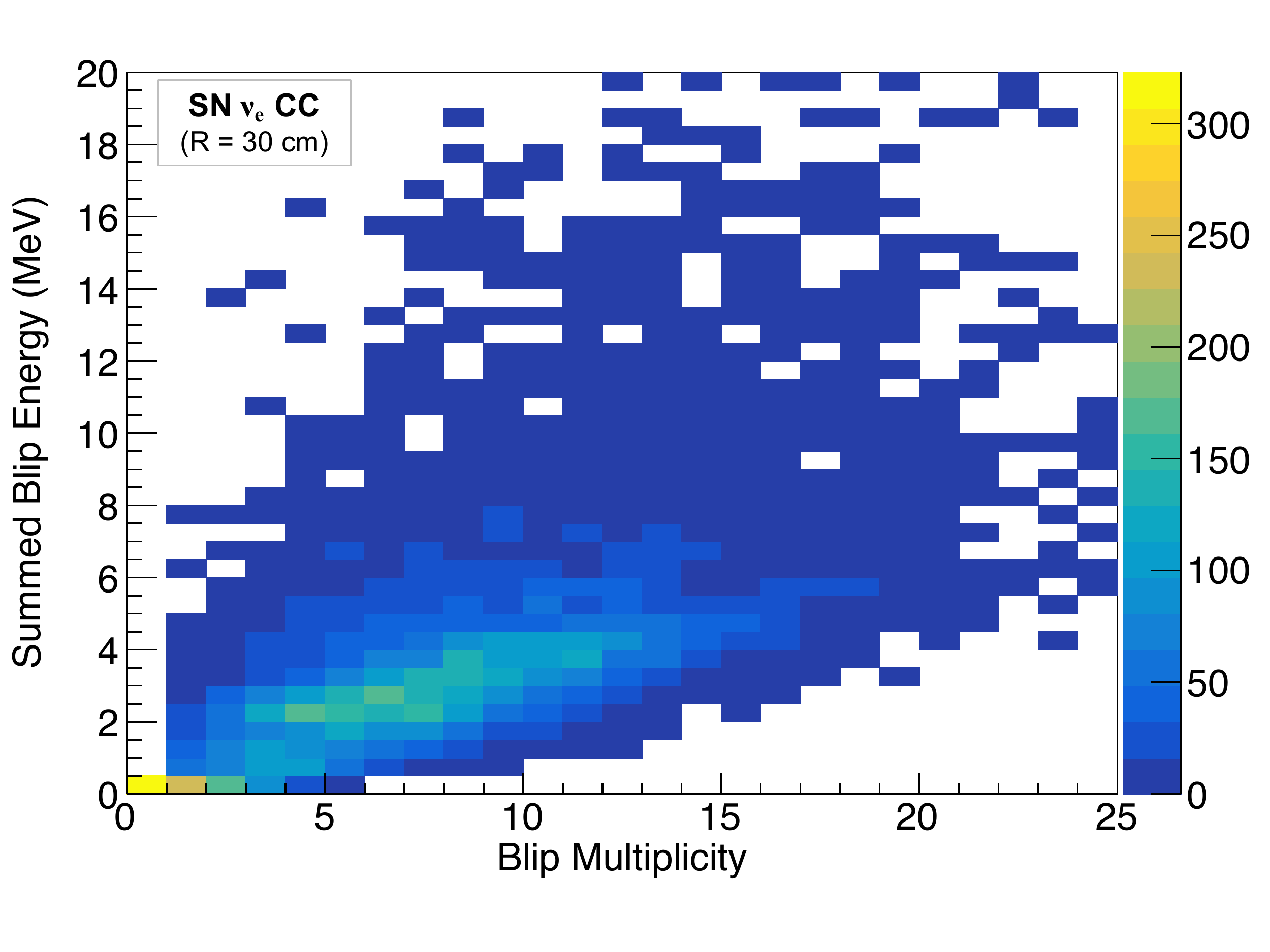}
\includegraphics[trim = 0.0cm 1.2cm 0.0cm 1.1cm, clip=true, 
width=0.95\columnwidth]{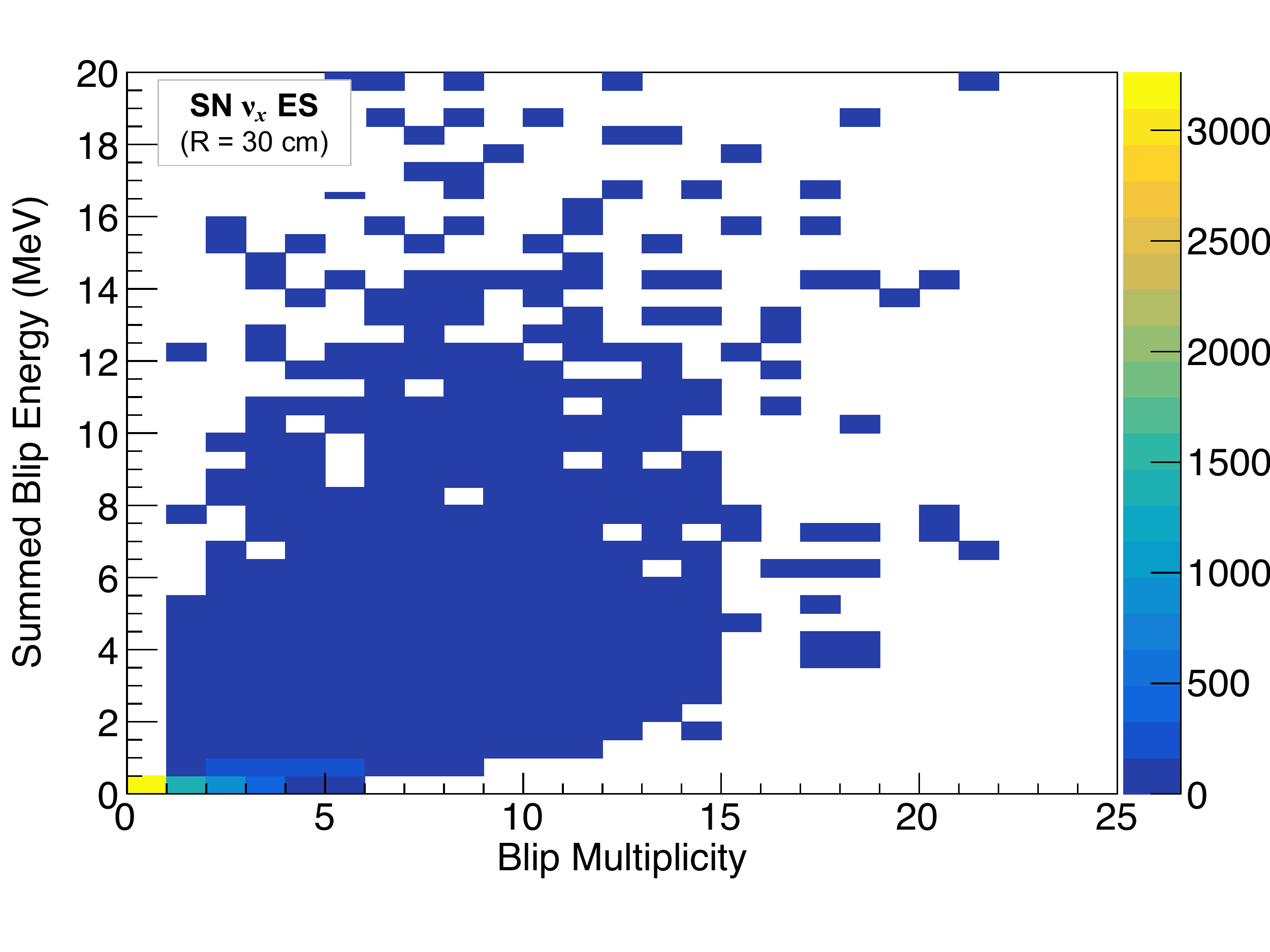}
\caption{Blip multiplicity versus summed energy of blips above 75~keV and within 30~cm of the primary electron start point for simulated $\nu_e$ CC (top) and $\nu_x$ ES (bottom) events.}
\label{fig:totalblipE_vs_multiplicity}
\end{figure}

Figure~\ref{fig:totalblipE_vs_multiplicity} illustrates the potential for this discrimination capability for supernova $\nu_e$ CC and $\nu_x$ ES events.  
The CC sample is the same as used in the previous section, while the ES electron sample is generated using the Geant4 particle gun with an input energy profile matching that expected from the default supernova energy parameterization described in the DUNE TDR~\cite{dune_tdr2}.  
This figure shows the multiplicity and summed energy of reconstructed blips which meet the default criteria described in previous sections.  
The two interaction types produce qualitatively different distributions, with generally higher blip multiplicities and summed energies in the $\nu_e$ CC sample. The use of total blip energy as a discriminating variable is motivated in part by the tendency of de-excitation photons, produced only in CC interactions, to be higher in energy compared to bremsstrahlung photons.

If we assume a CC:ES interaction ratio of $\sim$9:1 as given in the DUNE TDR~\cite{dune_tdr2}, we can apply a variety of cuts to simulate the extraction of ES events from the larger CC dataset.  
The efficiency and purity of these cuts are given in Table~\ref{tab:efficiency}, where these two metrics are defined as:
\begin{equation}
    \text{Efficiency} = \frac{(\text{\# ES selected})}{(\text{\# ES total})},
\end{equation}
\begin{equation}\label{eq:purity}
    \text{Purity} = \frac{(\text{\# ES selected})}{(\text{\# ES selected}) + \left(\frac{\text{CC}}{\text{ES}}\right) \times (\text{\# CC selected} )}.
\end{equation}

\noindent
The scaling factor CC/ES~(=9) in Equation~\ref{eq:purity} is used to account for the fact that the same number of events were simulated for each of the two interaction channels.

\begin{table}
    \centering
    \begin{tabular}{|c|c|c|c|c|c|}
    \hline
        Threshold & Sphere Radius & \# Blips & Energy & Efficiency & Purity \\
        \hline
        75~keV & $0.5-30$~cm & -- & $<1$ MeV 		& 72\% 	& 38\% \\
        75~keV & $0.5-30$~cm & $<3$ & -- 			& 61\% 	& 35\% \\
        75~keV & $0.5-30$~cm & $<3$ & $<1$ MeV 		& 58\% 	& 42\% \\
        \hline
        75~keV & $0.5-60$~cm & -- & $<1.5$ MeV 		& 70\% 	& 58\% \\
        75~keV & $0.5-60$~cm & $<4$ & --  			& 61\% 	& 47\% \\
        75~keV & $0.5-60$~cm & $<4$ & $<1.5$ MeV 	& 58\% 	& 62\% \\
        \hline
        300~keV & $0.5-60$~cm & -- & $<1$ MeV 		& 70\% & 57\% \\
        300~keV & $0.5-60$~cm & $<2$ & -- 			& 76\% & 40\% \\
        300~keV & $0.5-60$~cm & $<2$ & $<1$ MeV 	& 68\% & 59\% \\
        300~keV & $0.5-60$~cm & $<1$ & -- 			& 57\% & 66\% \\
        \hline
        \end{tabular}
        
    \caption{Efficiency and purity in selecting $\nu_x$ ES events from a larger sample of $\nu_e$ CC events using only cuts on reconstructed blip activity.  Efficiency and purity definitions are given in the text. For reference, when no cuts are applied, selection efficiency is 100\% and purity is 10\%.}
    \label{tab:efficiency}
\end{table}

A variety of cut scenarios are considered in Table~\ref{tab:efficiency}, with cut threshold values chosen based on noticeable differences in the distributions shown in Figure~\ref{fig:totalblipE_vs_multiplicity}.  
For the default blip selection case, combined cuts on multiplicity ($<$~3~blips) and summed blip energy ($<$ 1~MeV) produce a 42\% pure sample of ES events -- substantially increased from the pre-cut 10\% -- with an efficiency of 58\%.  
If blip proximity cuts are relaxed to 60~cm and multiplicity and energy cut values are adjusted to $<4$~blips and $<1.5$~MeV, respectively, purity is increased to 62\% while maintaining the same efficiency as before.
Interaction channel discrimination is not substantially degraded when considering a higher 300~keV blip energy threshold. 

These studies make it clear that even simplistic blip activity criteria have  power to separate $\nu_e$ CC and $\nu_x$ ES channels.  
It is likely that a more detailed study incorporating blip activity as well as additional variables, such as the reconstructed energy and directionality of the primary electron, would yield improved performance with respect to that given above. 

While only supernova neutrino fluxes have been considered here, we would expect similar levels of discrimination from solar neutrino fluxes, given the similar de-excitation $\gamma$-ray spectrum for higher- and lower-energy CC interactions (Figure~\ref{fig:sn_e}) as well as the reduced production of bremsstrahlung photons at lower solar $\nu_e$ energies. 


\section{Final-State Neutron Identification and Calorimetry}
\label{sec:neutrons}

Final-state and secondary neutrons play a key role in defining the energy budgets of GeV-scale accelerator neutrinos and antineutrinos -- and to a lesser extent, supernova neutrinos.  
With a neutron separation energy below 10 MeV for $^{40}$Ar, this should not be at all surprising.  
Accelerator neutrino physics experiments have recently begun considering signatures from  final-state neutrons in neutrino measurements~\cite{minerva_n, nova_osc}.  
Most neutrons produced as a result of supernova, solar, and beam neutrino interactions will have kinetic energies in the sub-MeV to 10s~of~MeV range.   
Below, we consider and summarize the role of final-state neutrons in defining MeV-scale and GeV-scale neutrino energy calorimetry in argon, and discuss the extent to which blip activity can aid in the recovery of neutron-related final-state information.  

\subsection{Neutron Signals in Liquid Argon}

\begin{figure}
\includegraphics[trim = 0.0cm 0.0cm 0.0cm 1.0cm, clip=true, 
width=0.47\textwidth]{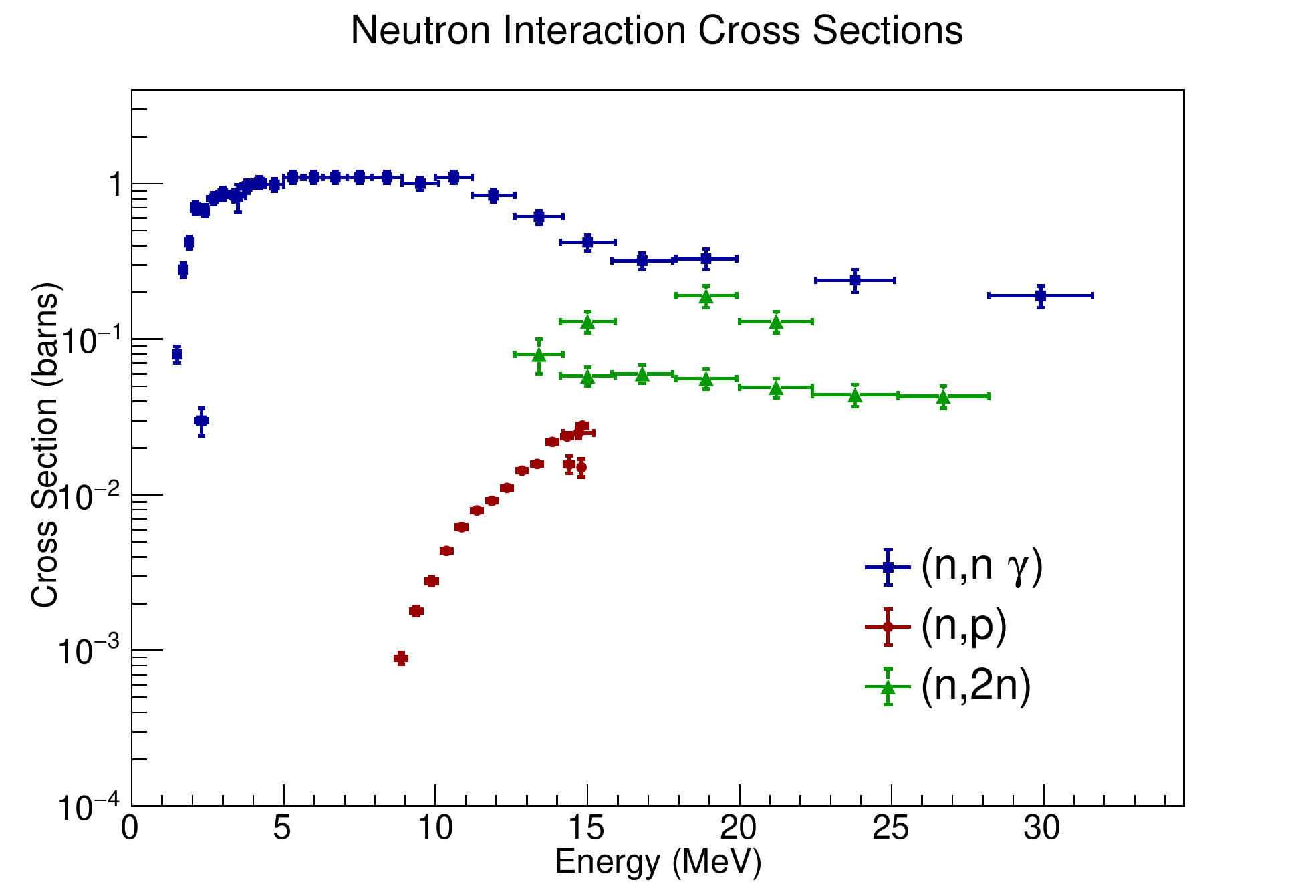}
\includegraphics[trim = 0.0cm 2.cm 0.0cm 1.0cm, clip=true, 
width=0.47\textwidth]{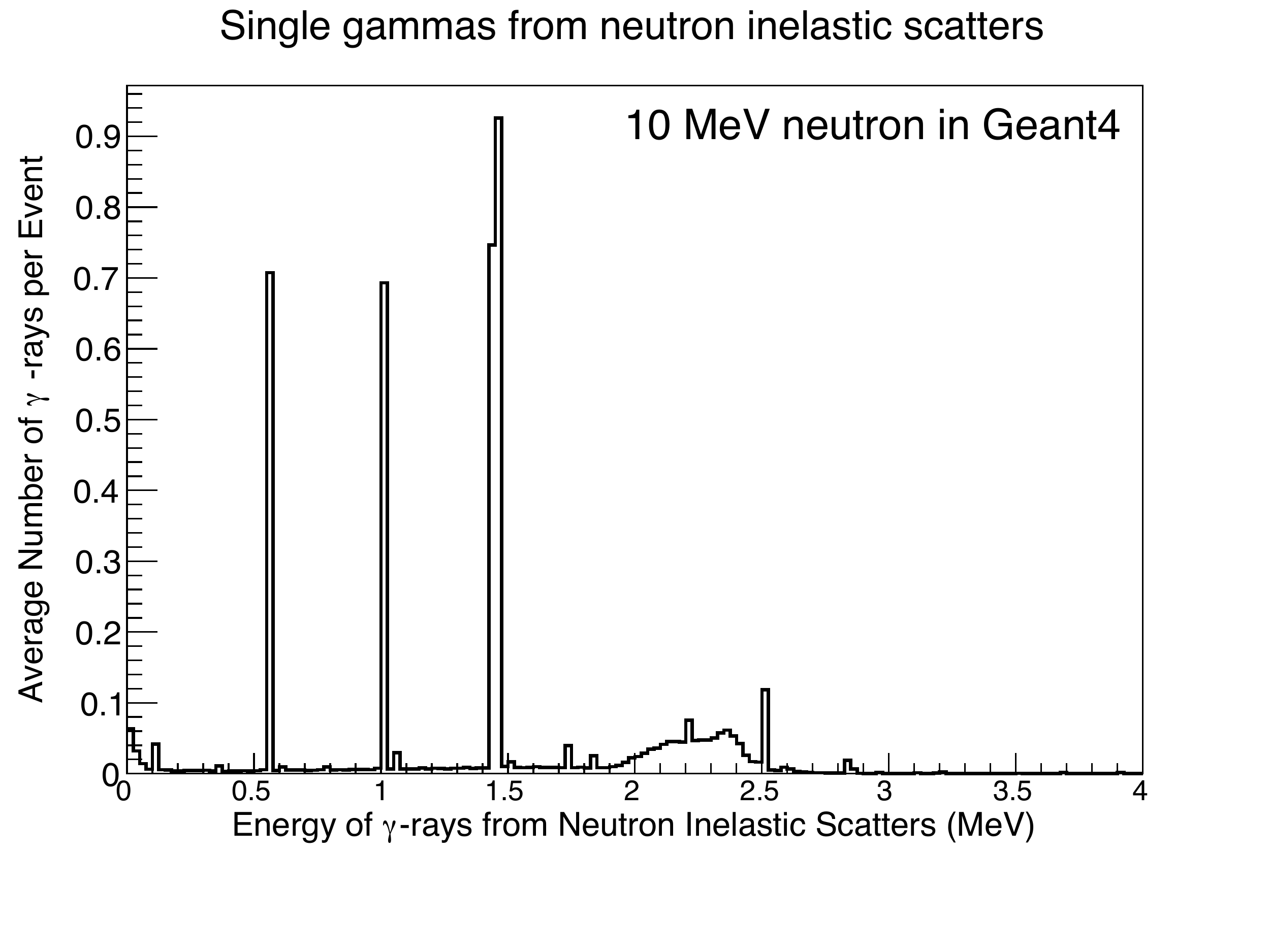}
\caption{Top: Cross sections of inelastic neutron scattering in LAr. Blue squares denote interactions with photon emission, green triangles denote neutron emission, and red circles denote proton emission. Data retrieved from~\cite{exfor} for Refs.~\cite{lar_ng,lar_np}.  Bottom: Geant4-reported true $\gamma$-ray energies created via inelastic scattering of neutrons in LAr, normalized to the number of primary 10~MeV neutrons simulated.} 
\label{fig:exfor}
\end{figure}

Recently-measured neutron interaction cross-sections on $^{40}$Ar in the $\sim$2-40~MeV energy regime~\cite{lar_ng,lar_np}, shown in Figure~\ref{fig:exfor}, are dominated by $\gamma$-producing inelastic scattering and by neutron-producing reactions -- the secondary neutrons from which will in turn lead to $\gamma$-producing inelastic scattering.  
These cross-sections correspond to effective neutron interaction lengths on the order of tens of~cm.  
$\gamma$-ray energies produced by de-excitation of $^{40}$Ar in response to inelastic interactions of 10~MeV neutrons are also given in Figure~\ref{fig:exfor}.  
These $\gamma$-rays are primarily in the 0-3~MeV range, roughly similar to those produced by supernova $\nu_e$ CC interaction final-state $^{40}$K$^*$ in Figure~\ref{fig:sn_e}.  
Thus, deposition of some final-state neutron energy will be reflected in LArTPC events as electron- and positron-produced blips.  
These blips will tend to be more concentrated in the general vicinity of the neutron's point of production.  
This suggests the ability to estimate the presence and/or energies of free neutrons in the final state of a neutrino interaction based on the presence or multiplicity of blips in its corresponding event.  
The remainder of this subsection will demonstrate this level of calorimetric capability for LArTPCs, as well as how this capability varies for different neutron energies.

\begin{figure}
\includegraphics[width=\columnwidth]{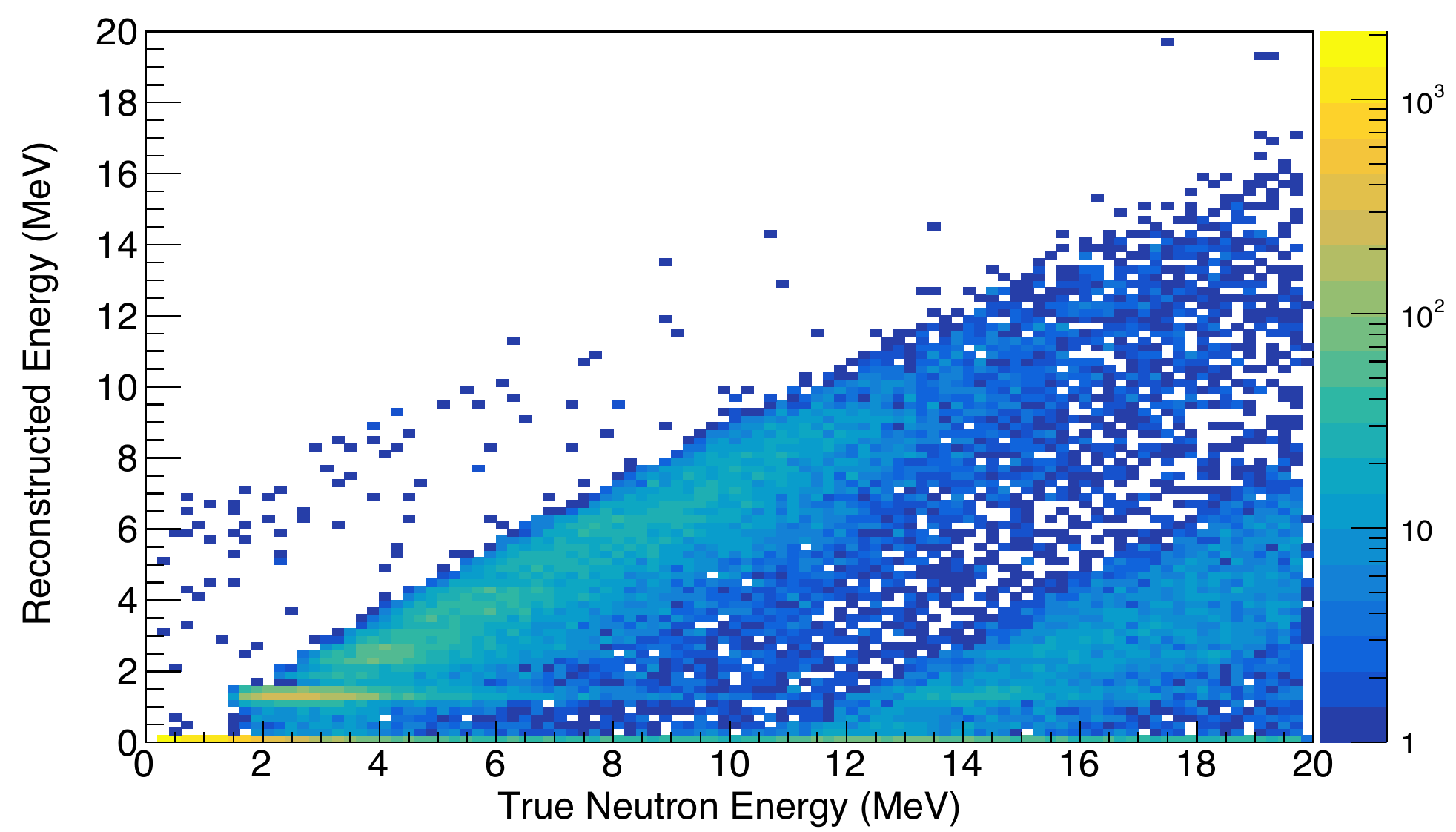}
\includegraphics[trim = 0.0cm 0.0cm 0.0cm 1.0cm, clip=true,  width=0.49\textwidth]{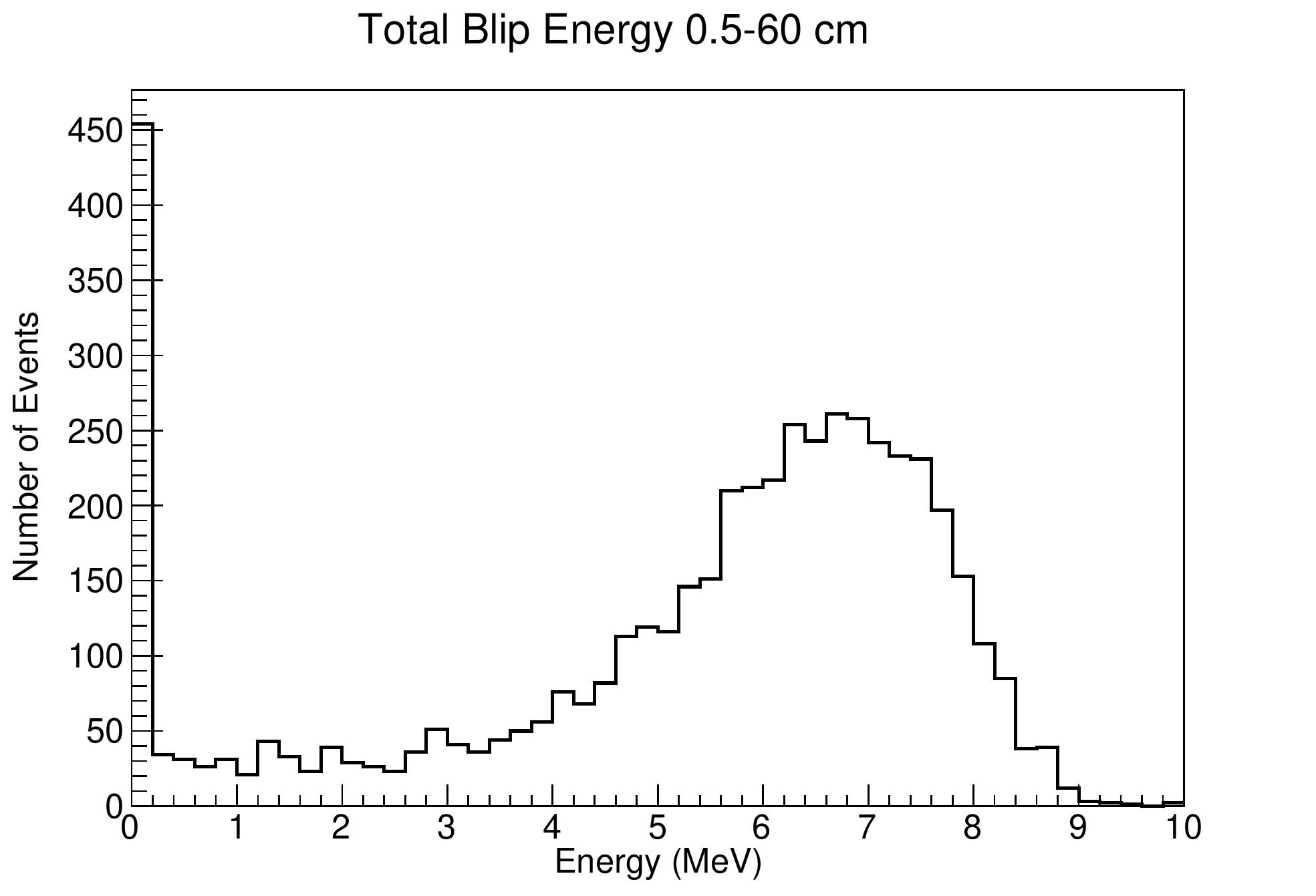}
\caption{Top: Summed energy of blips versus generated primary neutron energy. Bottom: Summed energy of blips produced by 10 MeV primary neutrons.  In both figures, only blips above 75~keV and within 60~cm of the neutron's production point are included.}
\label{fig:neutronblipE}
\end{figure}

At energies below a few~MeV, no excited states of $^{40}$Ar are accessible to the incident neutron and interactions are dominated by elastic scattering.  
Due to their low kinetic energy, these recoiling nuclei are not visible in large neutrino LArTPCs.  
This feature of LArTPC response to neutrons is demonstrated in Figure~\ref{fig:neutronblipE}, which shows the summed energy of blips surrounding primary neutrons generated with kinetic energy between 0 and 20~MeV (momentum between 0 and 195 MeV/c) using the procedures described in Section~\ref{sec:methods}.  
At the lowest energies pictured in this figure, almost no blip activity is visible.  

Very low-energy neutrons will produce MeV-scale $\gamma$-ray activity when they thermalize and capture on $^{40}$Ar.  
In Figure~\ref{fig:neutronblipE}, the few events containing blip activity from neutron captures exhibit a summed blip energy higher than the true kinetic energy of the produced neutron.  
Due to very low predicted interaction cross-sections in the 50-60~keV neutron energy range~\cite{endf}, these blips are likely to be produced tens of meters or more from the point of neutron production.
In addition, neutron captures in pure argon occur hundreds of $\mu$s following other event activity, which, due to charge drift effects in LArTPCs, is reflected in event displays by an additional spatial separation of order 20-30~cm. Therefore, even in a DUNE-sized detector, many neutrons are likely to escape the active LArTPC volume or data acquisition window.
In fact, only 12\% of 10~MeV neutrons simulated in a large volume of LAr resulted in a nuclear capture within a distance of 15~m. 
The large escaping neutron fraction is illustrated in Figure~\ref{fig:neutronblipE} by the extremely small number of events above the $E^\text{blip} = E_n^\text{true}$ diagonal.  
Due to the large distances between neutron production and capture locations, it will be difficult to use capture blip signals to provide more information about the neutron-producing interaction, such as final-state neutron multiplicities~\cite{nova_mult}.  
However, as will be discussed in Section~\ref{sec:spec}, neutron captures in LArTPCs can serve as a valuable source of monoenergetic MeV-scale energy depositions for detector calibration purposes.  

As neutron energies rise above a few~MeV, $\gamma$-producing inelastic scattering off of $^{40}$Ar becomes the dominant energy loss mechanism. 
Thus, it would be expected that, at this energy range and above, the summed energies of blips produced by a neutron should be proportional to that neutron's initial kinetic energy.  
This proportionality is demonstrated in Figure~\ref{fig:neutronblipE}. 
Summed blip energies include all neutron-derived blips passing the 75~MeV thresholding criterion; given the many-tens-of-cm~neutron interaction lengths involved, a proximity criterion of 60~cm with respect to the neutron generation vertex is applied. 
This choice is supported by Geant4 simulations which show the typical distance a neutron travels prior to its first inelastic interaction is relatively constant within the energy range of $\sim$2-20~MeV, averaging roughly 30~cm, with 67\% of neutrons interacting less than 30~cm and 86\% less than 60~cm. 
Starting at roughly 2~MeV, a proportionality is indeed visible a few MeV below  the $E^\text{blip} = E_n^\text{true}$ diagonal.  

Figure~\ref{fig:neutronblipE} provides further illustration of this trend by plotting the summed blip energy of a vertical slice of monoenergetic 10~MeV neutrons. 
The summed blip energy peak occurs at 6.6~MeV, with a resolution of approximately 1.0~MeV, or $\sim$15\% of the reconstructed peak.  
If we fit a linear trend to the mean of the main offset peak (Figure~\ref{fig:neutronblipE}) for each energy slice from 3 to 12~MeV, we find a slope of 0.75 MeV of summed blip energy per MeV of true neutron energy, with an intercept of about -1~MeV.  
The offset between blip and true neutron energies can primarily be explained by the cut-off of the ($n$,$n\gamma$) process near the MeV scale due to the lack of available excitable $^{40}$Ar states.  

The described linear relationship continues up to the $\sim$12-15~MeV neutron energy range, at which point an additional linear trend emerges at a much larger offset.  
While the slope of this higher-energy trend appears largely unchanged, the amplitude of the offset appears to be roughly 10~MeV larger than the 
fitted offset at lower energies, matching the 9.87~MeV neutron separation energy of $^{40}$Ar \cite{nndc_chart}.  
Without positive identification of an additional free final-state neutron, this reconstructed energy offset from binding energy loss will not be recoverable.  
Highly offset trends will continue with increasing neutron energy, with additional further-offset bands appearing as multiple nucleons are freed from the final-state nucleus.  

Neutrons of even higher energies than that described above, above 100~MeV, will be produced in interactions of GeV-scale beam neutrinos~\cite{minerva_n}.  
In liquid argon, these neutrons are much more likely to undergo proton- and/or neutron-releasing inelastic collisions with an argon nucleus, as suggested by Figure~\ref{fig:exfor} and discussed and demonstrated in Ref.~\cite{lar_captain}.  
The former case will result in the production of high-energy proton tracks that can be reconstructed using standard tools~\cite{ub_pandora} or high energy density blips in the general vicinity of the neutron production point.  
The latter case will result in a multiplication of neutrons and subsequent repetition of the various neutron scattering and binding energy loss processes, as discussed in Ref.~\cite{lar_res}.  

In summary, we have described the energy loss mechanisms of neutrons across all relevant energy ranges in large neutrino LArTPCs.  In particular: 
\begin{itemize}
\item For high-energy ($>100$~MeV) neutrons, existing large-feature reconstruction algorithms may be sufficient for performing final-state neutron calorimetry.  
\item Below 100~MeV, blip activity will play an essential role in determining the energy content of final state neutrons.  
\item Neutron energy recovery via blip identification will be most complete in the $\sim$2-12 MeV neutron energy range.  
\item Kinetic energy deposited in neutrino LArTPCs by $<$2~MeV neutrons is unlikely to be recoverable via blip identification or any other method.  
\item Final-state neutron multiplicity determination via blip activity will be extremely difficult, due to the highly displaced locations of neutron captures in LAr.  
\end{itemize}
Having outlined these neutron-related capabilities, we now consider how these capabilities can be leveraged for a few different neutrino energy ranges of interest.  

\subsection{Supernova and Solar Neutrino Neutrons}

\begin{figure}
\includegraphics[trim = 0.0cm 0.0cm 0.0cm 1.0cm, clip=true, width=0.49\textwidth]{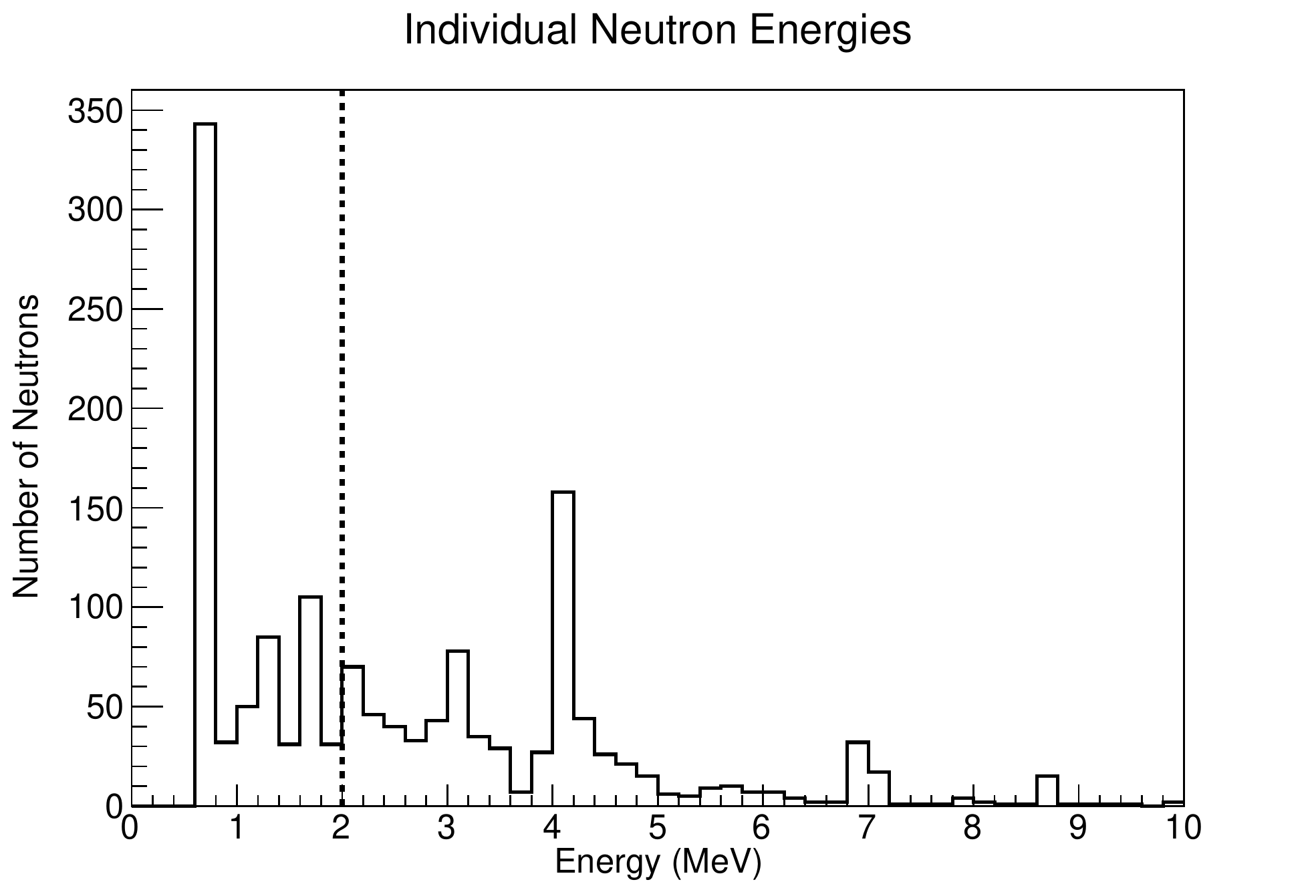}
\caption{
Energies of neutrons produced by supernova $\nu_e$ CC interactions. Vertical line indicates the energy at which neutrons no longer inelastically scatter.
}
\label{fig:sn_neutronE}
\end{figure}

Charged current supernova and solar neutrino interactions are kinematically constrained to produce no more than one or two final-state neutrons.  
These neutrons are sub-dominant contributors to the event-averaged neutrino energy accounting, as discussed in Section~\ref{sec:supernova}.  
In the case of the MARLEY $\nu_e$ CC events described in this section, we find that 15\% of all events have one or more produced neutrons, with a neutron energy spectrum as pictured in Figure~\ref{fig:sn_neutronE}.  
Of all $\nu_e$ CC interactions generated in Section~\ref{sec:supernova}, final-state neutron kinetic energy accounted for 1.7\% of the total kinetic energy of all interacting neutrinos.

All neutrons below $\sim$2~MeV initial energy, or 46\% of all those produced in Figure~\ref{fig:sn_neutronE}, will be invisible in LArTPC events.  
Those above $\sim$2~MeV initial energy will produce blips somewhat proportional in summed energy to the final-state neutron, as described in Figure~\ref{fig:neutronblipE}. 
For final-state $\gamma$-rays produced in our simulated $\nu_e$ CC dataset and depicted in Figure~\ref{fig:sn_e}, neutrons were responsible for 7.5\% of the total, comprising 13\% of the total energy of all pictured $\gamma$-rays.   
These numbers further illustrate the comparatively small calorimetric return from collecting neutron-produced blip activity.  

\begin{figure}
\includegraphics[trim = 0.0cm 1.0cm 0.0cm 1.5cm, clip=true, 
width=0.47\textwidth]{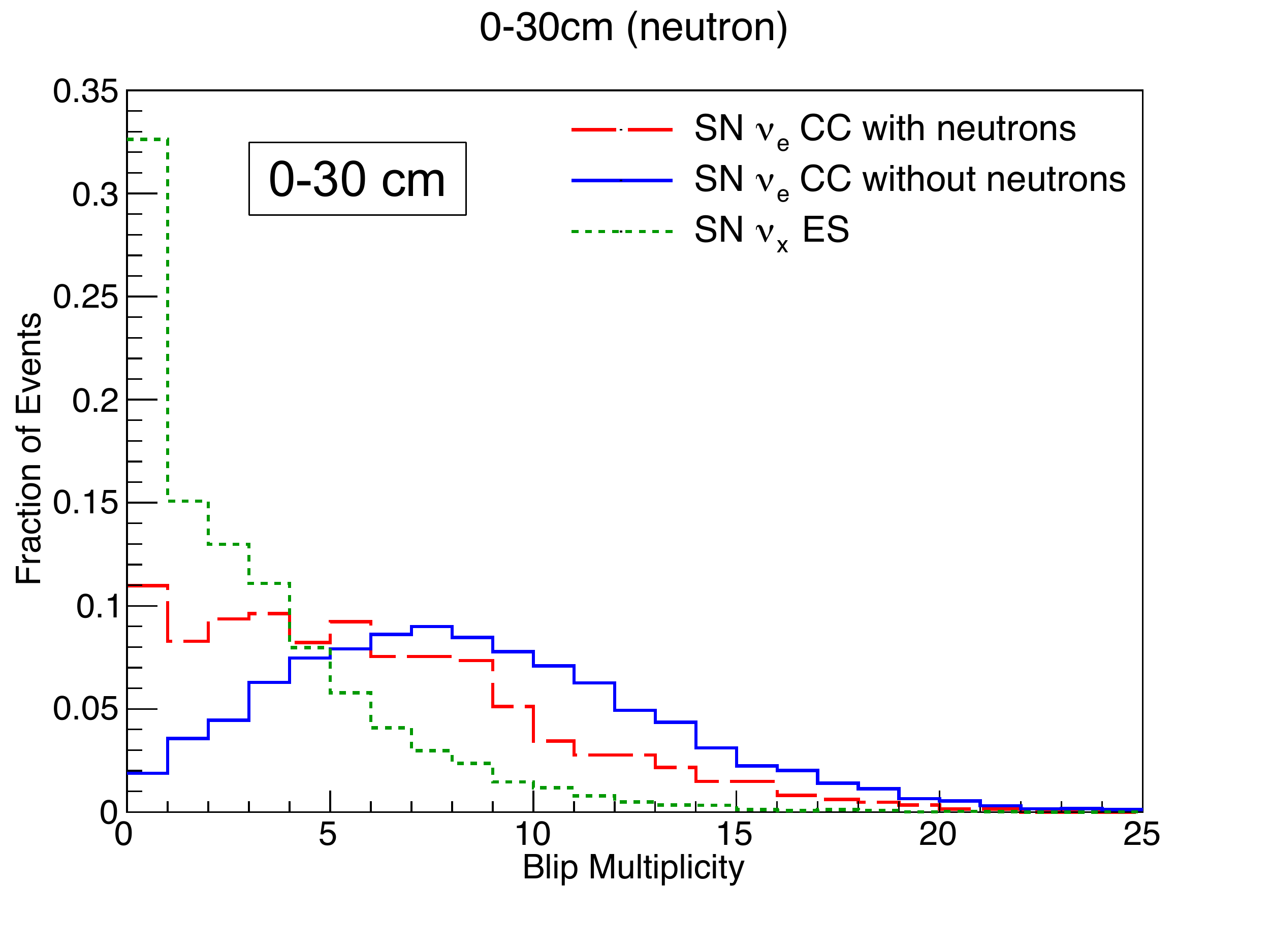}
\includegraphics[trim = 0.0cm 1.0cm 0.0cm 1.5cm, clip=true,  width=0.47\textwidth]{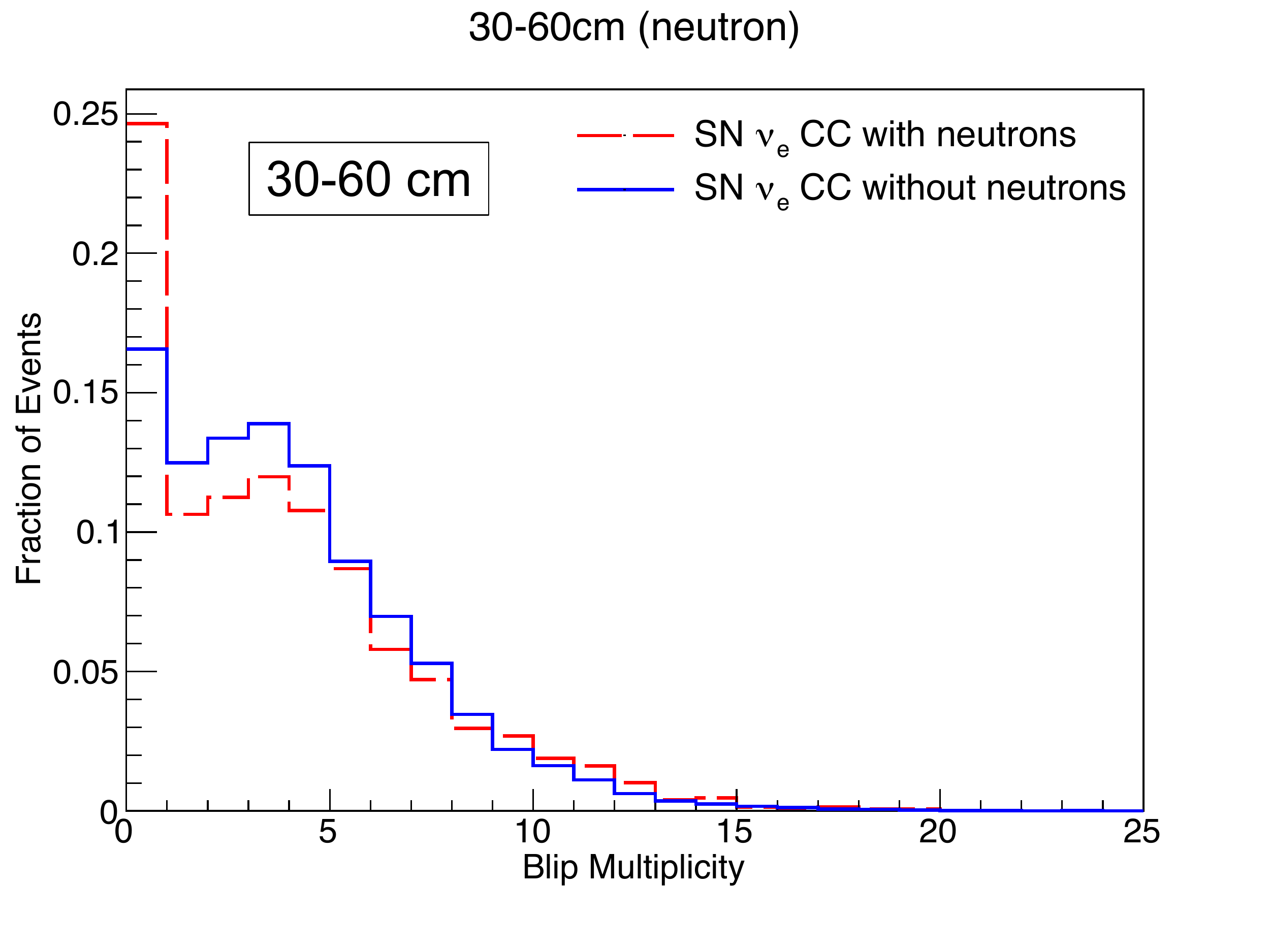}
\caption{Number of blips above 75~keV and within 30~cm (top) and between 30-60~cm (bottom) of the primary electron for supernova $\nu_e$ CC events with neutrons (red) and without neutrons (blue), and for $\nu_x$ ES events (green). The scattering contribution to the bottom panel is negligible, so it is not included. Distributions are area-normalized and integrated over all interacting neutrino energies.}
\label{fig:SN_nBlips}
\end{figure}

However, as illustrated in Figure~\ref{fig:sn_budget}, neutron-producing interactions are the source of a substantial increase in energy resolution due to the binding energy loss associated with the freed neutron.  
If 7.80~MeV of energy is required to free a neutron from $^{40}$K in 15\% of $\nu_e$ CC interactions, this corresponds to at least 5.1\% of the total kinetic energy of all interacting supernova neutrinos; this energy fraction is three times higher than that consumed by the kinetic energies of these neutrons.  
Thus, it would be beneficial to be able to positively identify the presence of a neutron in a $\nu_e$ CC interaction's final state.  
It is also worth noting that MARLEY may underestimate the true rate of $\nu_e$ CC neutron production, particularly at high $\nu_e$ energy~\cite{Friedland_marley}; if true, this enhances even further the calorimetric value of neutron tagging.
As mentioned in the previous subsection, identification via neutron capture tagging will be challenging, if not impossible, with existing LArTPC technology.  

As another method, one can attempt to exploit the relatively longer interaction length of final-state neutrons in liquid argon compared to final-state de-excitation $\gamma$-rays.  
This method is illustrated in Figure~\ref{fig:SN_nBlips} by showing total multiplicity for blips appearing within 30~cm of the neutrino interaction vertex, and within 30-60~cm of the interaction vertex.  
All CC and ES categories are normalized to one another, and are integrated over all interacting $\nu_e$ energies.  
As mentioned in Section~\ref{sec:supernova}, ES events are expected to have lower average multiplicity.  
For CC events, multiplicities at shorter vertex-blip distances are smaller for events containing final-state neutrons, as a larger portion of the excess energy of the final-state nucleus is spent in liberating the neutron.  
In contrast, neutron-containing events have a \emph{larger} proportion of high-multiplicity events at longer vertex-blip distances.  
It may be possible that a multivariate approach (e.g. a boosted decision tree) including this as well as other variables, such as individual blip energies and primary electron kinematics, may yield some discrimination and attendant improvement in neutrino energy recovery and resolution.

\subsection{Accelerator Neutrino Neutrons}

Final-state and secondary neutrons will also carry off a large portion of energy from interacting GeV-scale neutrinos, with neutron energies ranging in energy from the sub-MeV to hundreds of MeV scales.  
Hundreds of MeV of neutrino or antineutrino energy will regularly be lost or deposited in visible forms as a result of production or interaction of these primary and secondary neutrons in a LArTPC.  
Final-state neutrons in the lower-energy ($<$50~MeV) range will have properties similar to those described in the previous section, primarily producing $\gamma$-rays and subsequent blips via inelastic scattering.  
Final-state neutrons in the higher-energy range will have split energy depositions between $\gamma$-produced and proton-produced ionization, with the possibility of many follow-on generations of neutrons and subsequent nucleon binding energy losses.  

Ref.~\cite{lar_res} explains this neutron energy accounting in detail for GeV-scale neutrino interactions.  
Thus, for the purposes of completeness of our description, we will only briefly summarize some relevant conclusions in this paragraph, while encouraging the reader to carefully study that excellent paper.  
For FLUKA-simulated 4~GeV neutrino interactions in liquid argon, 30\% of hadronic energy is lost, on average, to the production (binding energy) or interaction (inelastic or elastic scattering) of final-state neutrons.  
Of this neutron-related budget, less than a third is likely to be identified using standard large-feature reconstruction tools, such as Pandora~\cite{ub_pandora}.  
One of the largest neutron-related energy loss categories is ionization below quoted DUNE CDR detection thresholds~\cite{dune_cdr}, i.e. electromagnetic and proton-produced blip activity.  
Proper identification and consideration of neutron-related blip activity 
can provide a relative improvement in energy resolution of order 25\% for both GeV-scale neutrinos and antineutrinos.  

Ref.~\cite{lar_res} also notes that binding energy represents the largest contributor to neutron-related energy losses.  
Thus, we might expect an additional substantial improvement in energy resolution from accurate determination of the number of final-state primary and secondary neutrons in an event.  
As mentioned in the previous section, precise capture-based neutron identification will be extremely challenging in LArTPCs.  
In addition, given the large number of average primary and secondary neutrons in a GeV neutrino event and the diffuseness of their produced activity, the blip proximity method introduced in the previous section also seems unlikely to provide easy insight into true neutron counts.  

Beyond the concretely defined improvements in energy accounting and resolution described above, blip multiplicities, energies, and positions represent a new source of data for constraining modeling of hadronic interactions and energy loss mechanisms in argon, as well as modeling of nuclear effects in neutrino-nucleus interactions.  
While information regarding final-state neutron multiplicities, such as that described recently in NOvA oscillation analyses~\cite{nova_mult}, may not be easily leveraged in LArTPCs, proxies for total neutron energy and the presence of high-energy neutrons will certainly be present in LArTPC events.  
A reduction in modeling systematics enabled by analysis of MeV-scale activity in beam neutrino events could have the potential to be more valuable than energy resolution reductions in maximizing the oscillation physics reach of DUNE.

\section{Electromagnetic Shower Reconstruction}
\label{sec:em}

While neutrino energy resolution improvements brought about by reconstruction of blip activity were demonstrated in previous sections, it is worth briefly defining calorimetric gains specifically for electromagnetic showers, such as those produced in interaction of solar, supernova, atmospheric, and beam $\nu_e$ and $\overline{\nu}_e$ in LArTPCs.  

Electromagnetic showers are composed of electrons and positrons produced by hard electron-electron scattering and bremsstrahlung photons.  
Many electrons produced by bremsstrahlung photons will have energies at or below the MeV-scale regime and may be lost from shower energy reconstruction in the absence of low thresholds and/or topologically loose feature collection criteria.  
These bremsstrahlung charge loss effects are stochastic, and contribute substantially to overall shower energy resolution in LArTPC reconstruction.  
These effects have been previously described in the literature:  
MicroBooNE reports Michel electron and $\pi^0$ electromagnetic shower resolutions of order 20\% over a range of energies, with much of this resolution arising from non-inclusion of charge below MicroBooNE hit-finding thresholds or outside of defined shower topologies~\cite{ub_michel,ub_pi0}. Similarly, LArIAT~\cite{lariat_detpaper,lariat_light} has demonstrated an energy resolution of approximately 10\% for the energy deposited by Michel electrons within their active volume, and an average overall energy resolution of about 3\% for fully-contained samples of simulated isolated electrons spanning a similar energy range. 

\begin{figure}
\includegraphics[trim = 0.0cm 0.0cm 0.0cm 1.1cm, clip=true, width=0.99\columnwidth]{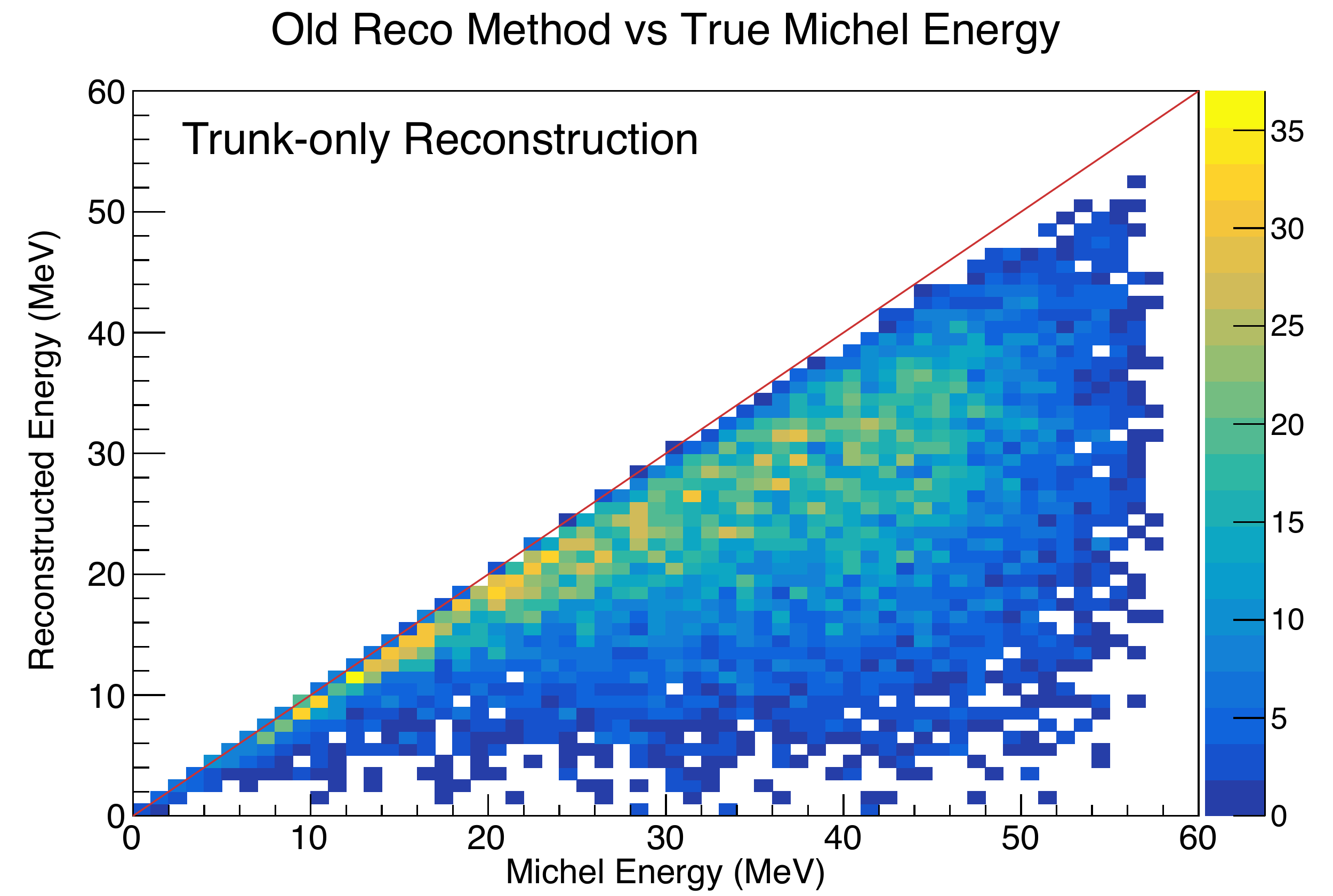}
\includegraphics[trim = 0.0cm 0.0cm 0.0cm 1.1cm, clip=true, width=0.99\columnwidth]{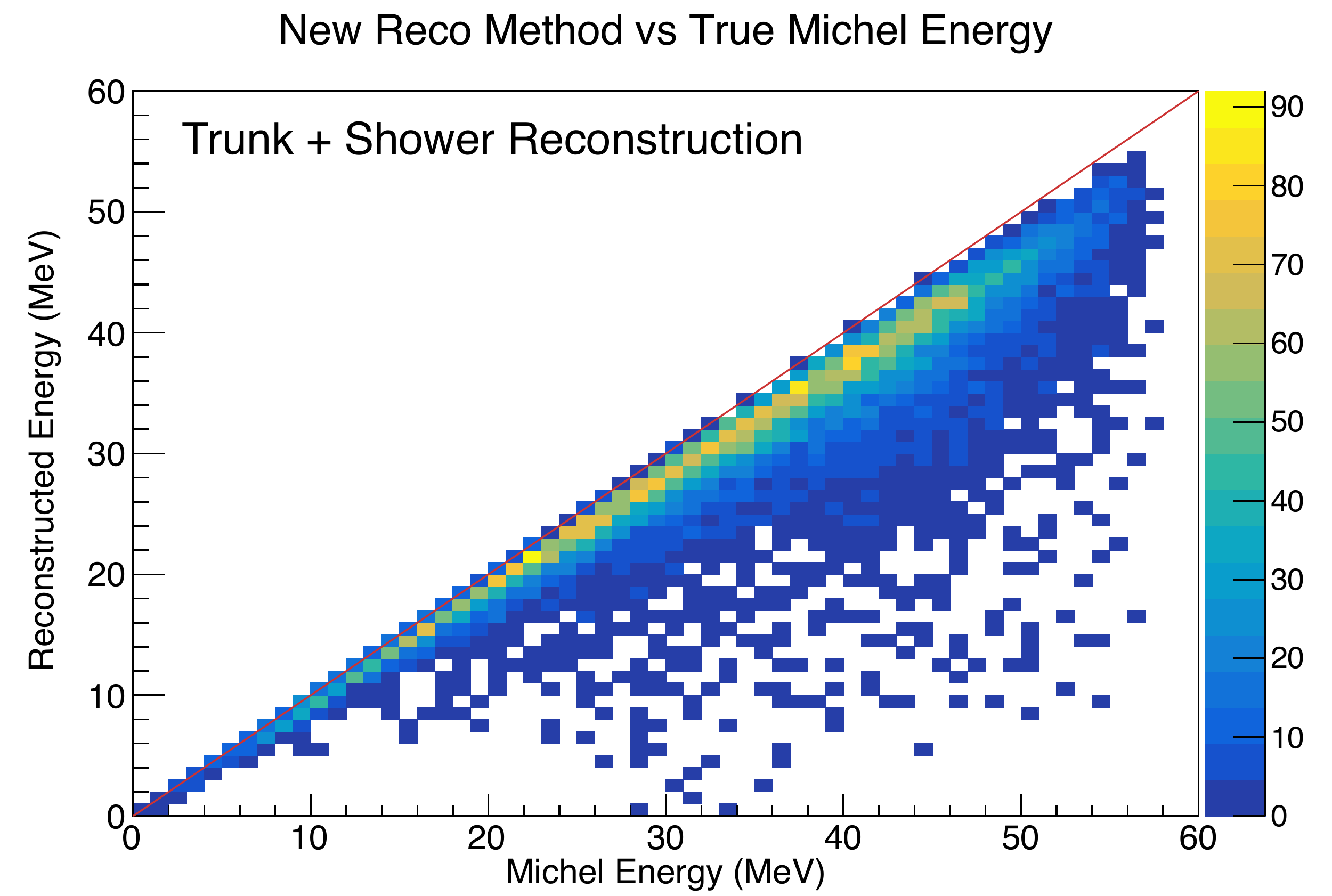}
\caption{Reconstructed energy plotted against the true energy for simulated Michel electrons. The top plot uses only the main electron trunk, while the bottom plot also incorporates blips within 60~cm of the electron's starting point. The red line indicates the expected trend for perfect reconstruction.}
\label{fig:michel_reco_true}
\end{figure}

Using the blip reconstruction procedure outlined in Section~\ref{sec:supernova}, we have conducted a similar study of reconstructed energy resolution for Geant4-generated 
electrons as done in Ref.~\cite{ub_michel} and Ref.~\cite{lariat_light}.  
The goal of this study is to demonstrate what the limits of electromagnetic shower resolution might be with the maximum achievable inclusion of charge (blip activity).  

\begin{figure}
\includegraphics[width=\columnwidth]{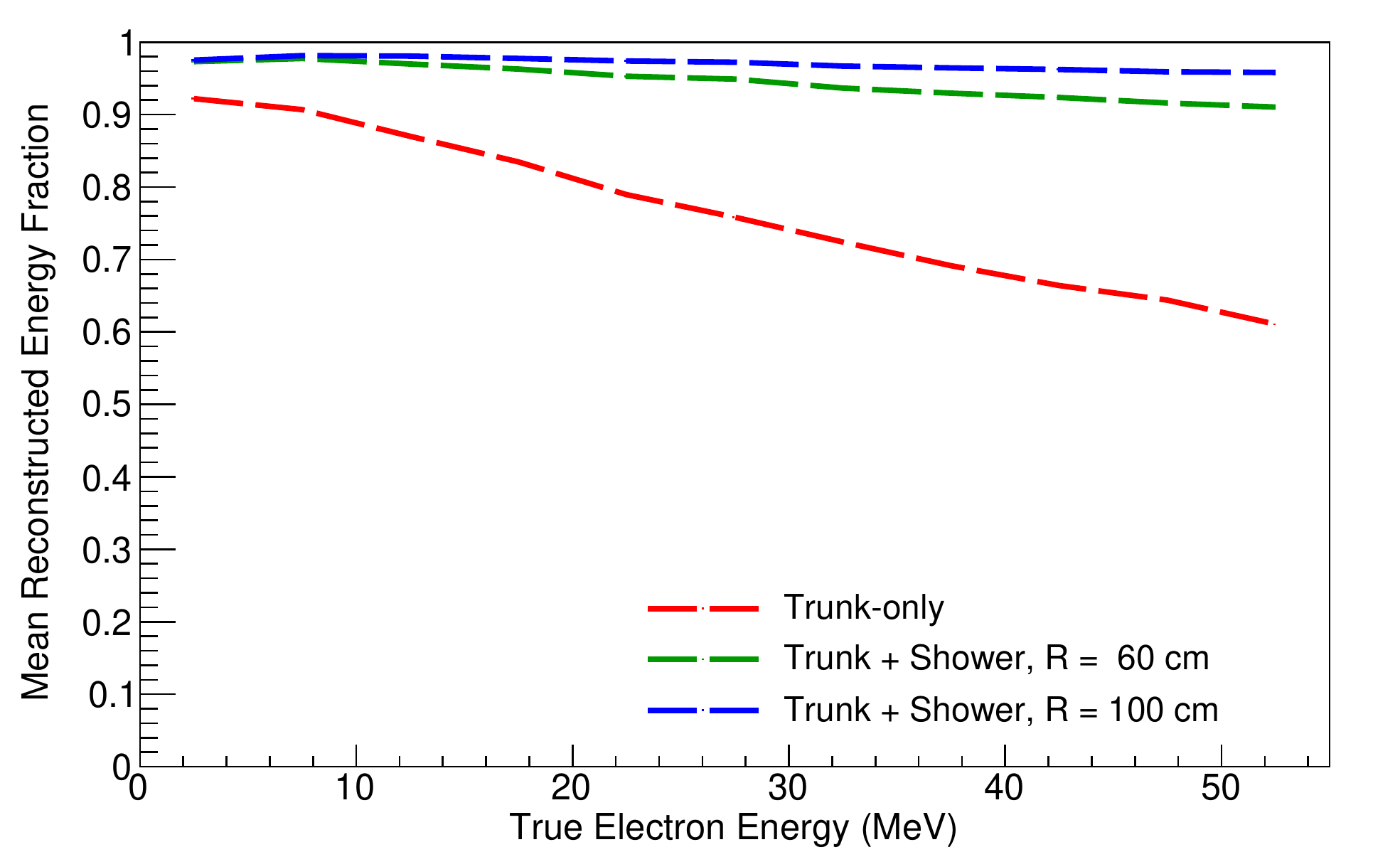}

\vspace{1ex}
\includegraphics[width=\columnwidth]{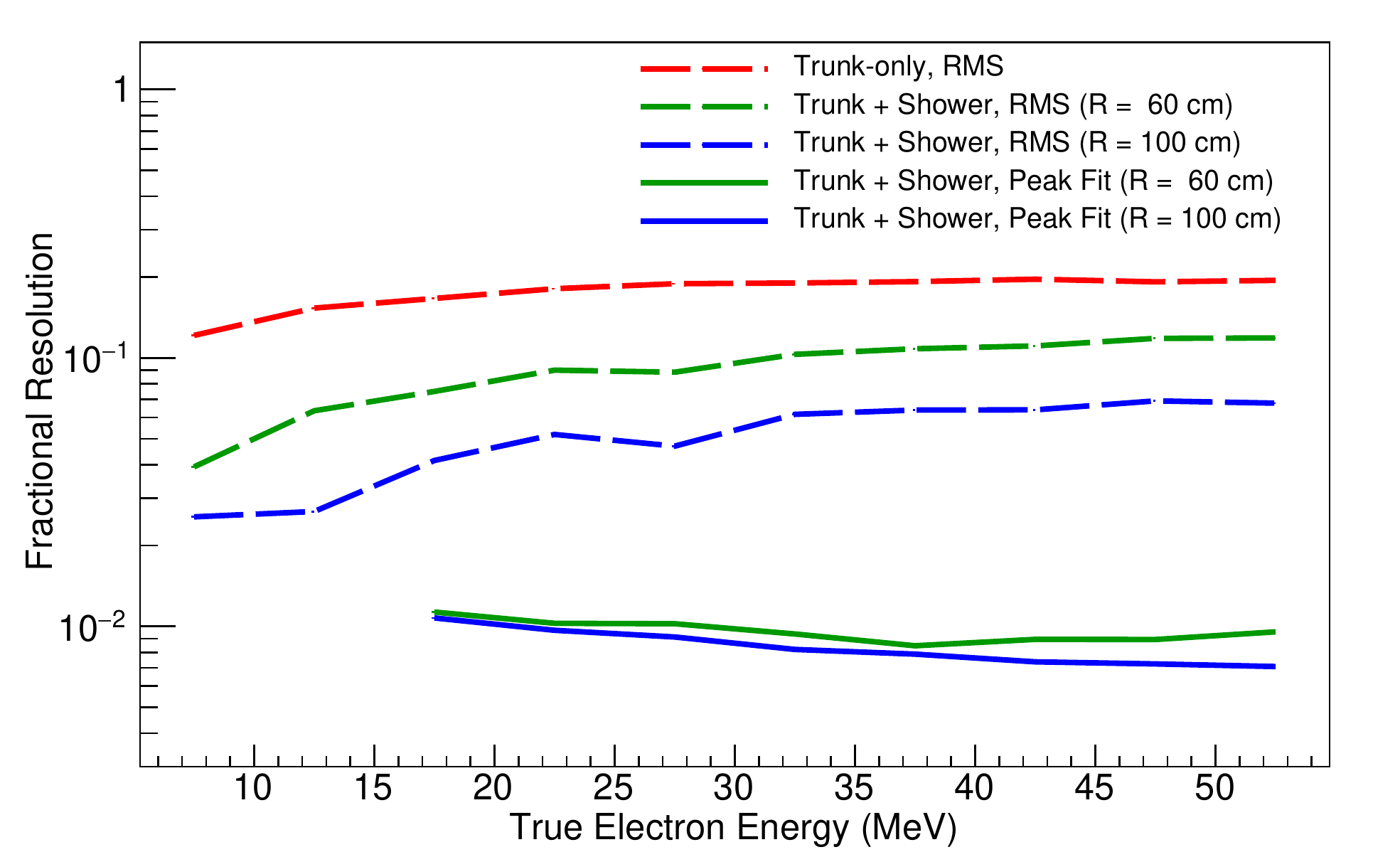}
\caption{Top: Average fraction of energy recovered versus the true initial electron energy when including  all reconstructed activity within either 100~cm or 60~cm of the electron's start point, as well as in the case when only the primary electron trunk is considered. Bottom: Fractional energy resolution plotted for these same three scenarios, measured using either the RMS or a Gaussian fit to the peak of the distribution of ($E_\text{reco} - E_\text{true})/E_\text{true}$ for different energy bins.}
\label{fig:michel_resolution}
\end{figure}

Figure~\ref{fig:michel_reco_true} shows reconstructed versus true energies for a sample of Michel electrons, which range in energy from 0 to 53~MeV. We consider both the `electron-only' reconstruction case (from Section~\ref{sec:supernova}) in which only ionization from the primary electron trunk is included, as well as the `blip-inclusive' which includes the electron trunk as well as the summed energy from all displaced bremsstrahlung-produced shower products and blips within 60~cm of the  electron start point.  
As expected, when incorporating displaced blip activity into the total energy reconstruction, a significant improvement is visible in both the accuracy and energy resolution.

For further illustration, the fraction of reconstructed energy and energy resolution are plotted in Figure~\ref{fig:michel_resolution} for a sample of isolated electrons spanning a range of 0-50~MeV. Resolution is calculated by composing distributions of the energy variance, $(E_\text{reco}-E_\text{true})/E_\text{true}$, across a range of true electron energy bins, and then taking either the RMS or the width parameter from a Gaussian fit to the peak of each distribution.
For the electron trunk-only reconstruction case, we see an energy loss ranging from only $\sim$10\% at the lowest energies to as much as $\sim$40\% near 50~MeV, with an RMS energy resolution in the 10-20\% range --- in reasonable agreement with that reported in Ref.~\cite{ub_michel} for a similar trunk-only case.    
If we instead consider a case in which we include all blip activity above 75~keV within a 60~cm (100~cm) radius of the electron start point, we achieve a relatively flat average energy loss of only $\sim$5\% ($\sim$3\%), with an RMS resolution ranging from 4\% to 12\% (3\% to 7\%).
When the resolution is calculated instead using the Gaussian fit, the total energy resolution for both the 60~cm and 100~cm radius scenarios drops to approximately 1\%.\footnote{For energies below about 15~MeV, electrons are too far below the critical energy to produce significant bremmstrahlung activity, and the peaks in their distributions of energy variance become highly non-Gaussian; data points from these cases are therefore excluded from Figure~\ref{fig:michel_resolution}.}
Thus, rather than substantially improving the resolution of the primary full-energy peak of the shower, increasing inclusion of blip activity serves primarily to reduce non-Gaussian off-diagonal energy smearing contributions.

From this study, we can conclude that for low-energy electromagnetic showers, the most optimistic calorimetric resolution that can be achieved in large neutrino LArTPCs is substantially better than 10\% for RMS-based resolution, and on the order of a percent when considering a Gaussian fit-based resolution. 
LArIAT has already demonstrated a charge-based reconstructed Gaussian energy resolution of $\sigma/E \approx$10\%$/\sqrt{E\text{[MeV]}}$~$\oplus$~2\% for 5-50~MeV electron showers, which equates to a resolution of $\sim$5\% at 5~MeV that drops to about 2.5\% at 50~MeV~\cite{lariat_light}. This $1/\sqrt{E}$ dependence arises from the presence of smearing introduced by the detector readout and by the reconstruction process. Our `best case' resolutions presented in this section do not include any simulated smearing, and are instead limited only by energy thresholding and by a simple blip proximity requirement; therefore, these resolutions are found to be largely flat for energies below 50~MeV.  
Further discussion of other detector-related resolution contributions is given in Section~\ref{sec:limitations}.  

For shower energies greater than 50~MeV, the primary electron and many in subsequent generations will be well above the electron critical energy, producing a large number of stochastic bremsstrahlung features. 
It is worth examining whether the trend in energy resolution observed for sub-critical electrons holds for this higher-energy regime.
For example, in Ref.~\cite{lar_res}, it is reported that for 4~GeV $\nu_e$ and $\overline{\nu}_e$, a 1.5\% electromagnetic shower energy resolution is produced by missed depositions below a 0.1~MeV blip identification threshold.
We observe a `best-case' Gaussian-fitted energy resolution of 0.3\% for simulated 200~MeV electrons using a 75~keV threshold and no proximity requirement. Increasing the electron energy to 500~MeV, we find the resolution improves even further to 0.2\%, though is then degraded to 1.2\% and 5.8\% when the energy deposition detection threshold is raised to 1~MeV and 10~MeV, respectively. 
For physics sensitivity studies, it therefore seems reasonable to assume an energy resolution modeling for high-energy electron showers that is substantially better than the 15\%/$\sqrt{\text{E [GeV]}}$ $\oplus$ 2\% assumed in previous literature, which translates to about 21\% at an electron energy of 500~MeV~\cite{dune_cdr,sbn}.

\section{Particle Discrimination Capabilities}
\label{sec:particleid}

We will now study the role that MeV-scale reconstruction can play in distinguishing the identity and charge of particles in LArTPC events.  
Focus will primarily be given to the role blip activity can play in distinguishing $\pi^+$, $\pi^-$, $\mu^+$, and $\mu^-$ from one another, given the limitations of existing tools for large neutrino LArTPCs.  

\subsection{Existing Particle Identification Methods in LArTPCs}

A variety of approaches have been advocated or demonstrated to use LArTPCs' excellent calorimetric capabilities and mm-scale resolution to provide discrimination between different types of particles.  Most prominently, energy deposition density helps discriminate between charged particles of substantially differing mass, such as protons, kaons, and pions~\cite{ub_cosmicreject,ub_cal,argo_pi,elena_phd}, as well as enabling discrimination between high-energy electrons and $\gamma$-rays~\cite{argo_nue,ub_pi0}.  

However, pions and muons are too close in mass to produce a highly-efficient density-based separation in a large LArTPC.  
If high purities are desired, discrimination must include other parameters.  
The presence of a Michel electron signature in either charge or light LArTPC data can be used to enrich a sample in specific muon or pion types~\cite{lariat_light}.   
Specifically, contained $\pi^-$ are far more likely to undergo nuclear capture than to decay to a Michel electron in a LArTPC via
\begin{equation}
\begin{aligned}
\pi^- \rightarrow~&\mu^- + \overline{\nu}_{\mu} \\
      &\drsh e^- + \nu_{\mu} + \overline{\nu}_{e},
\end{aligned}
\end{equation}
while all $\mu^+$ will decay to a Michel electron, 
\begin{equation}
\mu^+ \rightarrow e^+ + \nu_e + \overline{\nu}_{\mu}.
\end{equation}
The presence of hard or inelastic scatters along the path of a track is more indicative of a strongly-interacting pion~\cite{elena_phd}.  
When both a muon and a pion are possibly present and share a vertex, relative track length is also used as a proxy for particle identification~\cite{ub_cc,ub_hnl}.   
Sign selection is also an important consideration in understanding neutrino interaction images, particularly in antineutrino-mode accelerator neutrino data, which has an outsized wrong-sign contamination.  
In this case, Michel electron identification can also be considered as a possible tool in LArTPCs for muon or pion sign determination, for the reasons stated above.  
For the case of muons, some charge-sign discrimination may also be gained by exploiting the longer characteristic decay time of $\mu^+$ in liquid argon with respect to $\mu^-$~\cite{lariat_light}.  


\begin{table}[]
    \centering
    \begin{tabular}{|c|c|c|c|c|}
        \hline
        Particle & \makecell{Decay\\(in flight)} & \makecell{Decay\\(at rest)} & Capture & Other \\ \hline  
        $\pi^+$ & 3\%& 69\%    & 0\%     & 28\% \\ 
        $\pi^-$ & 3\%	& 0\%     & 63\%    & 34\% \\ 
        $\mu^+$ & 0\% & 100\%   & 0\%     & 0\% \\ 
        $\mu^-$ & 0\% & 26\%    & 74\%    & 0\% \\ 
        \hline
    \end{tabular}
    \caption{End-of-life processes for 100~MeV positively and negatively charged muons and pions as simulated in Geant4.  Decay processes will produce Michel electrons, while capture and other processes (such as absorption and charge exchange) will not.}
    \label{tab:michel}
\end{table}

As an example of the difficulty of disambiguating pion and muon identities, Table~\ref{tab:michel} illustrates the level of pion-muon or charge-sign purity that can be achieved by Michel electron identification.  
This table considers the specific case of primary particles with kinetic energy of 100~MeV generated in an effectively infinite volume of liquid argon using LArSoft and the \texttt{QGSP\_BERT\_HP} high-energy hadronic library in Geant4.  
This kinetic energy corresponds to that commonly observed for pions produced in interactions of GeV-scale neutrino interactions~\cite{minerva_pi,t2k_pi}.  
Most of these pions and muons will end their lives either decaying or capturing at rest.  
Even assuming 100\% efficient charge-based Michel electron identification, the contamination issues are apparent.  
For selection of muons (pions) only, a requirement of finding one (zero) Michel still accepts 72\% (74\%) of contaminating $\pi^+$ ($\mu^-$).  
For exclusive charge-sign selection, a required Michel count of one will produce a $\pi^+$ sample nearly free of $\pi^-$, but a $\mu^+$ sample containing 26\% of contaminating $\mu^-$.  
A Michel count requirement of zero will produce a pure $\mu^-$ sample, but a sign-contaminated $\pi^-$ sample, due to the non-negligible nuclear absorption of 100~MeV $\pi^+$ in flight.
In the case of less than 100\% efficient Michel electron identification, the described separations above will be further degraded.  

\subsection{Discriminating $\pi^{\pm}$ and $\mu^{\pm}$ End States Using Blip Activity}

Pion-muon discrimination should in principle be possible based on the level of MeV-scale activity at or near the end of a candidate pion or muon track.  
Pion nuclear capture leaves the entirety of the pion's rest mass energy in the capturing nucleus, which will be released in the form of final-state protons, neutrons, and de-excitation $\gamma$-rays.  
In addition to having less rest mass energy to begin with, a capturing muon will convert a substantial portion its rest mass into invisible final-state $\nu_{\mu}$ kinetic energy.  
Thus, muon capture should be expected to produce comparatively less proton, neutron, and $\gamma$-ray activity around its capture point.  
As pion and muon decay involve no direct nuclear interaction at all, the only visible activity at the particle end point should be a Michel electron and its attendant bremsstrahlung photons.  

While a complete theoretical description of final-state energy accounting for the case of pion and muon capture is very complex and does not exist, there are many existing nuclear physics measurements of these processes~\cite{cap_measday,cap_meyer}.  
In Geant4, final-state pion and muon capture on argon are modeled primarily using parameterizations based on existing measurements on lighter and heavier nuclei.  
We use Geant4 simulations of 1~MeV $\pi^-$ and $\mu^-$ to demonstrate the relevant truth-level and reconstructed blip activity differences.  

\begin{figure}
\includegraphics[trim = 0.0cm 0.0cm 0.0cm 1.0cm, clip=true, width=0.45\textwidth]{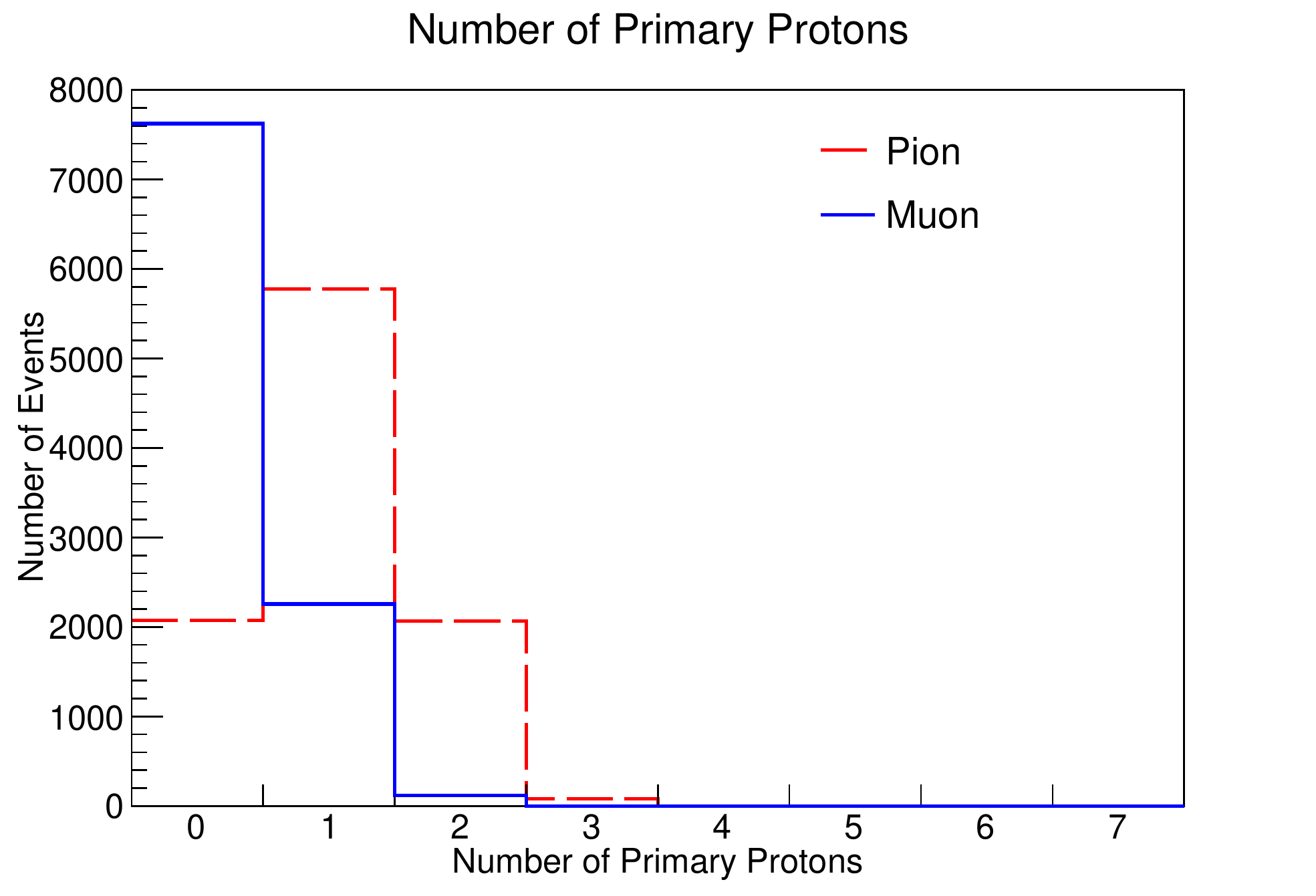}
\includegraphics[trim = 0.0cm 0.0cm 0.0cm 1.0cm, clip=true, width=0.45\textwidth]{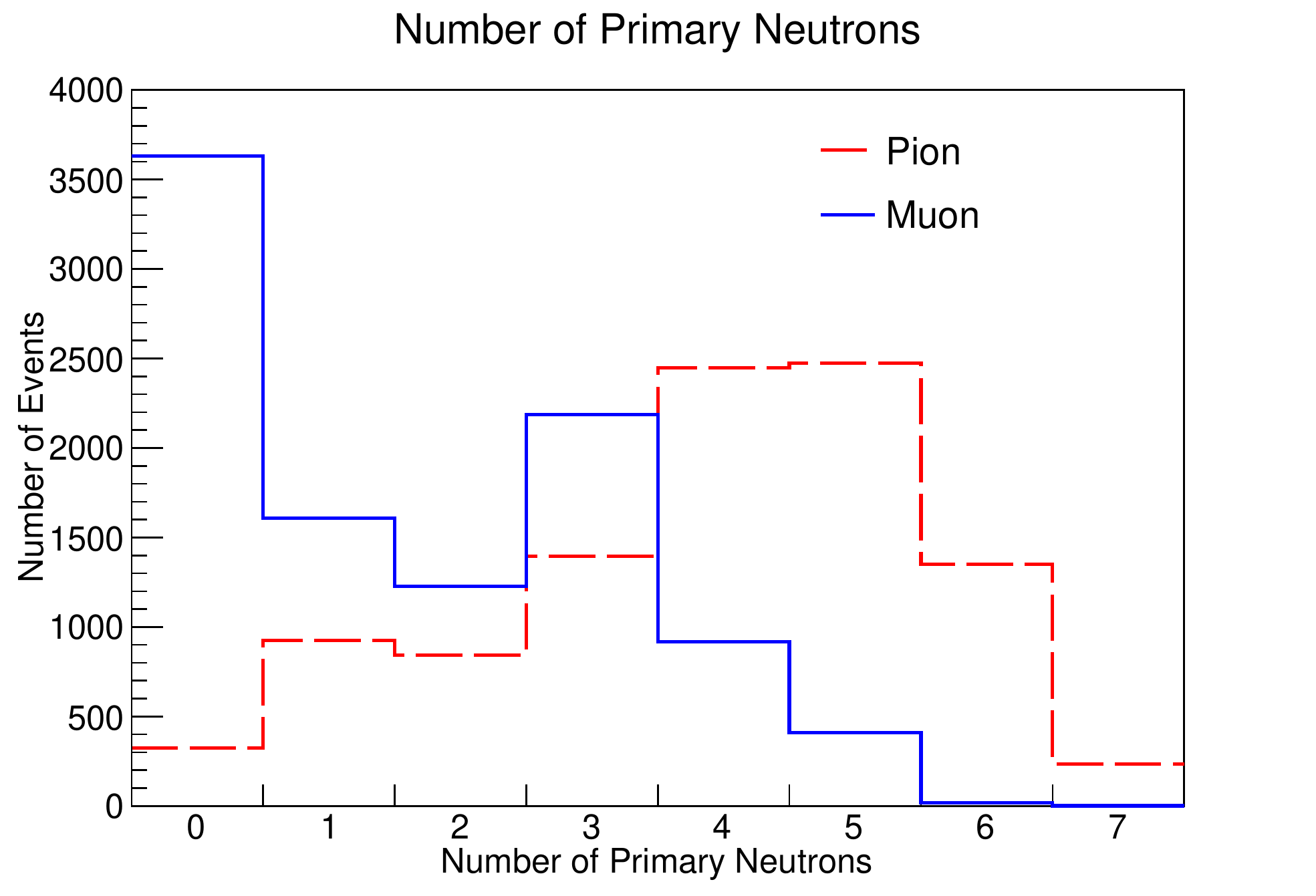}
\caption{Number of protons (top) and neutrons (bottom) emitted from  $\pi^-$ captures at rest (red dashed line) and  $\mu^-$ captures at rest (blue solid line).  Distributions are area normalized.}
\label{fig:capture_primary_particles}
\end{figure}

Figure~\ref{fig:capture_primary_particles} shows final-state proton and neutron multiplicities for muons and pions that undergo nuclear capture on Ar; for decaying pions and muons, obviously these multiplicities are zero.  
We find that pion capture emits more protons and neutrons than muon capture. In about 75\% of muon nuclear captures, no proton is emitted, while this occurs only about 20\% of the time in pion nuclear capture.  
The substantial difference in average neutron multiplicity indicates an expected difference in blip multiplicities and summed blip energies for these various cases.  
In addition, de-excitation $\gamma$-ray production at the capture vertex may also differ between pion and muon capture as different daughter nuclei are produced in different excited configurations. 

\begin{figure}
\includegraphics[trim = 0.0cm 1.0cm 0.0cm 1.2cm, clip=true, width=0.45\textwidth]{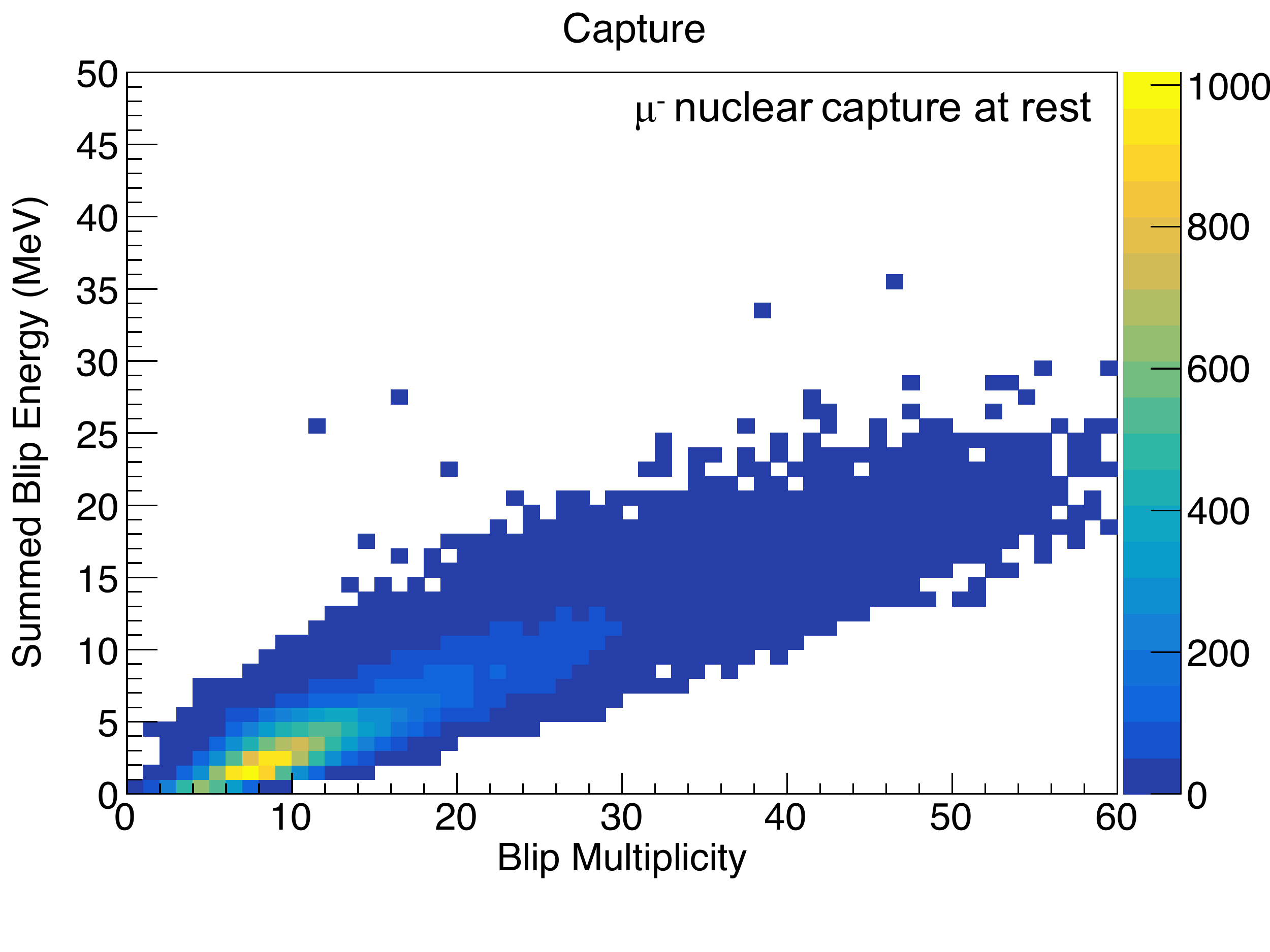}
\includegraphics[trim = 0.0cm 1.0cm 0.0cm 1.2cm, clip=true, width=0.45\textwidth]{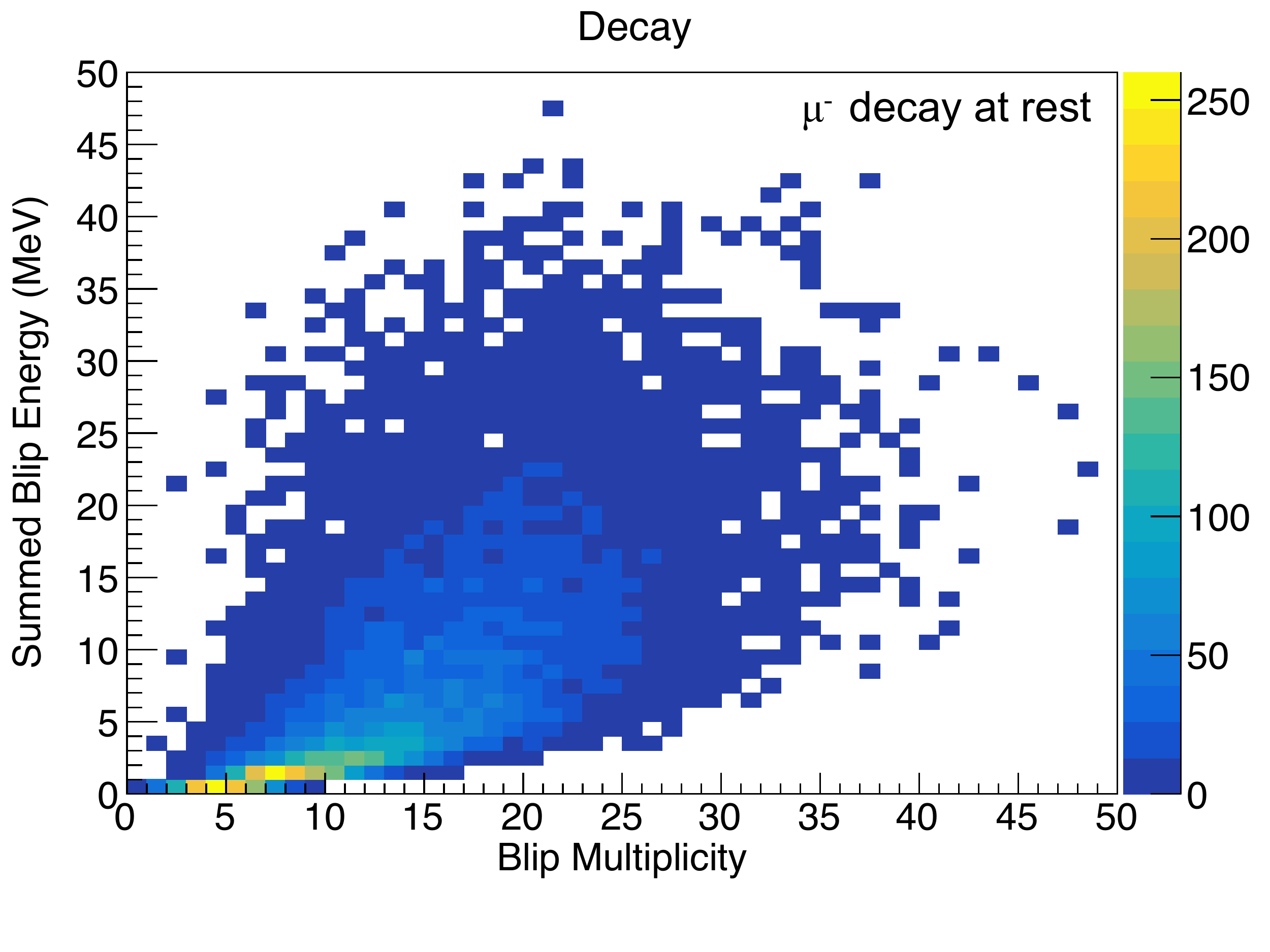}
\includegraphics[trim = 0.0cm 1.0cm 0.0cm 1.2cm, clip=true, width=0.45\textwidth]{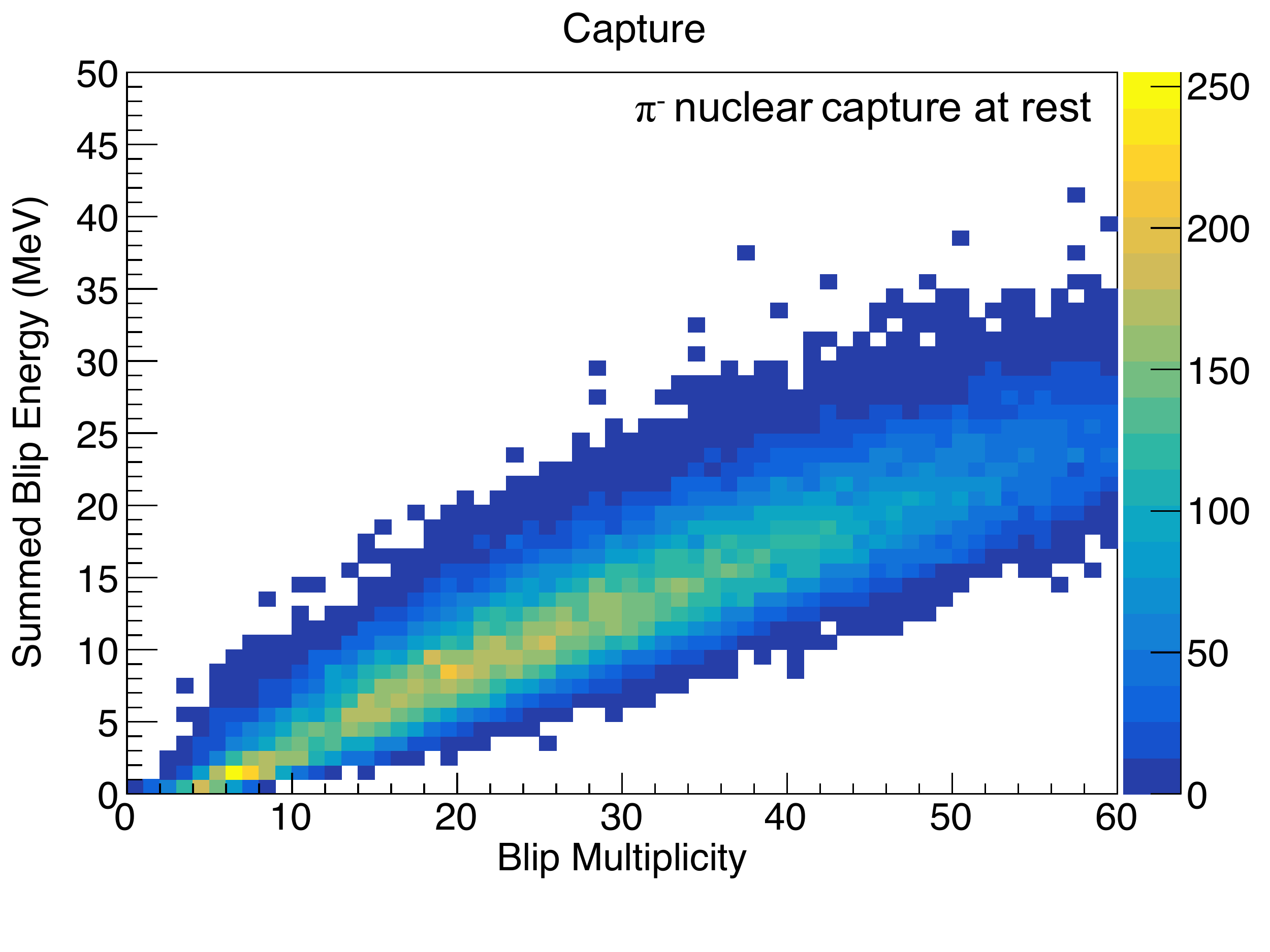}
\caption{Summed blip energy versus blip multiplicity within 60~cm of the capture/decay point for $\mu^-$ captures at rest (top), decaying $\mu^-$ (middle), and $\pi^-$ captures at rest (bottom).}
\label{fig:muonpion_ENumber_blips}
\end{figure}

To judge the level of discrimination provided by blip signals, we again consider the signal blip metrics examined in previous sections.  
Figure~\ref{fig:muonpion_ENumber_blips} shows the 2D joint distribution of blip multiplicity and total energy as in Section~\ref{sec:supernova}, but for $\mu^-$ and $\pi^-$ capture, and for  $\mu^-$ decay.  
Distributions are shown for blips within 60~cm of the primary particle end point.
Since a $\pi^+$ at rest decays to a $\mu^+$ of approximately 4~MeV, which travels only a negligible distance before subsequently decaying to a Michel electron, its final state will look very similar to that of $\mu$ decay. Therefore, while neither $\mu^+$ or $\pi^+$ undergo nuclear capture, their distributions of blip multiplicity and energy for decay at rest are well-represented by the $\mu^-$ decay distribution shown in Figure~\ref{fig:muonpion_ENumber_blips}.
%
We see that, of the three possible end-state processes, pion nuclear capture at rest produces the most blip activity within 60~cm of the capture point.  
In addition, there are noticeable differences in the distributions for $\mu-$ capture (consisting only of de-excitation gammas) and decay (consisting of a final-state electron that can produce bremsstrahlung-induced blip activity).
Thus, blip activity can provide new discrimination capability independent of whether other discrimination methods, such as Michel electron identification, are employed.

\begin{table}
\centering
	\begin{tabular}{|c|c|c|c|c|c|c|}
    \hline
   	Radius & N$_\text{blip}$ & E$_\text{blip}$ & E$_\text{vert}$ &  $\mu$ CAR & $\mu$ Decay & $\pi$ CAR \\  \hline
   	30~cm 	& $>7$ 	& -- 			& -- & 52\% & 65\% & 85\%  \\
   	30~cm 	& -- 	& $\geq 4$ MeV 	& -- & 33\% & 47\% & 77\%  \\
    30~cm 	& $>7$ 	& $\geq 4$ MeV 	& -- & 30\% & 42\% & 75\%  \\
    \hline
    60~cm 	& $>14$ & -- 			& -- & 34\% & 46\% & 85\%  \\
    60~cm 	& -- 	& $\geq 8$ MeV 	& -- & 22\% & 44\% & 78\%  \\
    60~cm 	& $>14$ & $\geq 8$ MeV 	& -- & 21\% & 33\% & 77\%  \\
    \hline
    60~cm 	& --  	& -- 			& $>5$ MeV & 18\% & 0\% & 74\%  \\
    60~cm 	& $>14$ & $\geq 8$ MeV 	& $>5$ MeV & 6.3\%& 0\% & 53\%  \\
    \hline
    \end{tabular}
    
\caption{ 
Selection efficiency for various blip activity and vertex activity cuts for $\mu^-$ captures at rest (CAR), decaying $\mu^-$, and $\pi^-$ CAR. 
The vertex region is defined by a 5~mm radius sphere centered at the particle's decay or capture point; only blips found outside of this region are considered.}
\label{tab:muonpion}
\end{table}

To demonstrate more quantitatively the level of pion-muon and sign discrimination possible using blip information, we place a variety of cuts on blip multiplicity and summed blip energy.  
In Table~\ref{tab:muonpion}, one can see the capabilities of these cuts alone to distinguish the pion capture, muon capture, and decay end processes.  
To further illustrate, we focus on a hypothetical identification of $\pi^-$ in LArTPC events.  
For the case where blips within 60~cm of the track endopoint, we find that by placing a cut of $\geq 8$ MeV ($>$14~blips) on summed blip energy (multiplicity), we are able to correctly identify a $\pi^-$ 78\% (85\%) of the time, while rejecting all but 44\% (46\%) of decaying muons and 22\% (34\%) of capturing muons.  
If these two cuts are combined, we reject 79\% (67\%) of all capturing (decaying) muons, with a 77\% $\pi^-$ selection efficiency.  
Since pion blips are primarily neutron-generated, a similar selection based on a 30~cm proximity requirement, also given in Table~\ref{tab:muonpion}, performs substantially less well, particularly in discriminating pion and muon nuclear capture.  
It should also be noted that the `other' category of pion end-states in Table~\ref{tab:michel} is dominated by nuclear absorption in flight, which will produce even more blip activity than nuclear capture at rest, due to the additional absorbed pion kinetic energy.  
This high $\pi^-$ selection efficiency should thus be realizable at kinetic energies higher than the simplified 1~MeV case simulated here. 

For comparison to the 60~cm proximity blip selection described above, a Michel-rejecting selection with perfect Michel tagging would yield 97\% $\pi^-$ selection efficiency while rejecting 0\% (100\%) of capturing (decaying) muons.  
In this case, blip-based discrimination excels where the Michel-based selection performs less well, and vice-versa.  
This emphasizes the value of combining blip-based discrimination with the other forms described in the previous subsection; in this case, the combination of methods would yield a substantial reduction in muon contamination.  
A similar level of discrimination as that described above should be achievable when considering an exclusive selection of $\mu^+$.

To demonstrate sign selection capability, we use the results of Table~\ref{tab:muonpion} to consider the case of distinguishing $\pi^+$ from $\pi^-$ for pions that have ranged out and are approximately at rest. Nearly 100\% of these $\pi^-$ will undergo nuclear capture, and thus be rejected at a rate of 77\% if blip multiplicity and energy cuts for a 60~cm proximity requirement are inverted -- i.e., rejecting events that have both $>$~14 blips and $\geq$~8~MeV summed blip energy. On the other hand, $\pi^+$ will exclusively decay to a low-energy $\mu^+$, producing a muon decay final-state signature and thus leading to a sign-selection efficiency of 77\%, assuming again that cuts are inverted.
However, the situation gets more complicated when one instead considers a sample of 100~MeV pions. From Table~\ref{tab:michel}, we find that 34\% of $\pi^-$ at this energy will undergo an inelastic scatter (which we reject for the purposes of this exercise), while 63\% will capture at rest. 
After also considering the small contribution from $\pi^-$ decay, we obtain $\sim$83\% reduction of wrong-sign $\pi^-$ background with a $\pi^+$ identification efficiency of roughly 49\% for 100~MeV pions using only blip-related metrics.
It is likely this purification can be substantially improved with a Michel electron requirement.

For wrong-sign purification of $\mu^+$, the situation is simpler since muons almost always range out and come to a rest, rarely undergoing inelastic interactions. However, blip-related metrics alone are not as effective at discriminating between $\mu^+$ and $\mu^-$ as they are for pions, given the similar extent of the distributions between muon capture and decay. Using only total blip energy in a 60~cm sphere, the 74\% of $\mu^-$ that capture will be rejected at a rate of 78\%, while the other 26\% that decay will be rejected at a rate of 56\%, giving us a total wrong-sign $\mu^-$ rejection rate of 72\%, but with an accompanying $\mu^+$ selection efficiency of only 44\%.

While outside the realm of blip reconstruction so far considered in this paper, the proton final-state multiplicities in Figure~\ref{fig:capture_primary_particles} are also worth considering in the context of particle discrimination.  
In particular, pion capture will produce final-state protons, which will produce either tracks or excess ionization at the capture vertex beyond that expected from a pion or muon Bragg peak.  
To estimate the discrimination power provided by these protons, we consider the ambitious case where we are able to positively identify the presence of a proton with kinetic energy in excess of 5~MeV~\cite{uB_dedx}.  
For this case, we see that a cut on $>$5~MeV vertex energy produces a complete rejection of decaying pions and muons while rejecting 82\% of capturing muons, with a 74\% selection efficiency for capturing pions.  
A combination of both blip-based and vertex cuts further reduce muon capture contamination to the 6.3\% level while maintaining better than 50\% pion capture efficiency. 
\section{BSM Physics Capabilities}
\label{sec:bsm}

A variety of BSM searches in large neutrino LArTPCs can benefit from the identification of blip activity and classification of events based on the presence or absence of these features.  
We will briefly highlight a few promising scenarios here and encourage the performance of more quantitative studies in the future using full simulations of the BSM processes and final-state distributions in question.  

Many BSM physics processes discussed as possible points of focus within the SBN and DUNE physics programs can be categorized by the variety of distinctive particle combinations they produce~\cite{dune_tdr2,bsm_white,sbnd_phys}.  
For example, high-energy di-lepton pairs can be expected from Standard Model neutrino trident interactions in liquid argon~\cite{dune_trident,dune_trident2}, which have the potential to uncover new physics --- such as heavy sterile neutrinos~\cite{bsm_heavy}, a dark neutrino sector~\cite{bsm_heavy}, or dark Higgs~\cite{bsm_higgs,bsm_mpd} --- if rates are divergent from Standard Model predictions.  
Other specific particle combinations have also been hypothesized: for example, pion-muon pairs would be expected from decays of heavy neutral leptons~\cite{ub_hnl} produced in accelerator neutrino experiments, while pion-pion pairs could be produced in these experiments by decays of dark Higgs bosons~\cite{bsm_higgs} or up-scattered dark neutrinos~\cite{bsm_dark}, respectively.  

It is expected that the primary backgrounds to these dedicated BSM searches are different final-state particle combinations produced by common Standard Model neutrino interactions.  
For example, Ref.~\cite{dune_trident} provides an excellent overview of the various expected background channels to neutrino trident $\mu^+$-$\mu^-$ production, particularly 1$\mu$-1$\pi$ final states from $\nu_{\mu}$ CC interactions.  
The pion-muon discrimination capability delivered by analysis of reconstructed blips, as described in Section~\ref{sec:particleid} above, may be a useful additional tool in reducing backgrounds for this and other BSM analyses in large neutrino LArTPCs.


Other BSM signatures can be characterized primarily by the unique topological distribution of blip signals they produce in LArTPC events.  
An obvious example is searches for millicharged particles produced in neutrino beams, as recently demonstrated by the ArgoNeuT experiment on the NuMI beamline~\cite{argo_mcp}.  
The track of weak ionization produced by these particles would be visible in a LArTPC as two or more reconstructed blips that can be connected by a line pointing back to the neutrino beam's target~\cite{harnik_mcp}.  
We note that lowered LArTPC blip thresholds in these searches leads directly to improvements in sensitivity.  
Other hidden sector particle interactions in LArTPCs, such as up-scattering and decaying heavy neutral leptons~\cite{sbnd_phys}, can produce two displaced event vertices, one of which consists of a de-exciting nucleus exhibiting primarily or exclusively reconstructed blip activity.  
Thus, identification of these secondary low-activity vertices is likely only possible through the use of blip identification capabilities.  

Hidden sector physics scenarios may also be characteristic in the total level, rather than the spatial distribution, of MeV-scale activity present in events.  
For example, decays of hidden sector particles in LArTPCs, such as heavy neutral leptons, dark photons, or dark Higgs, need not include substantial momentum exchange with an argon nucleus. 
These decay vertices, unlike neutrino-argon interactions, will not include the neutron and photon products of final-state nuclear de-excitation, resulting in an event with little or no blip activity near the interaction vertex.  
This lack of blip activity is another possible input for reducing neutrino-induced backgrounds to these BSM scenarios.

\section{Single $\gamma$-Ray Calibration and Spectroscopy}
\label{sec:spec}

Previous sections describing the utility of MeV-scale LArTPC signals have focused on a handful of metrics related to total blip energy or multiplicity.  
In addition, LArTPC physics analyses may be enhanced by considering individual blips or blip sub-groups within an event.  
In this section, we will focus primarily on the benefits of blip sub-grouping for performing MeV-scale single $\gamma$-ray spectroscopy in LArTPC events.  
This technique could be valuable for different purposes, such as low-energy LArTPC calibration~\cite{dune_tdr2} or tagging of final-state nuclei produced in neutrino or BSM interactions~\cite{t2k_spec}. 


We have attempted to resolve $\gamma$-ray spectrum features in an event by iteratively forming sub-groups of blips produced by electrons that are daughters of the same parent $\gamma$-ray.  
Proximity is our sole metric in determination of common parentage, with grouping achieved by the following algorithm.  
First, we identify all of the blips in an event by tagging electrons that deposit at least 75~keV of energy.
Then, a candidate `reconstructed $\gamma$-ray' is formed by grouping all identified blips 
 located within a 30~cm spherical radius centered around the highest-energy blip. 
The blips present in this reconstructed $\gamma$-ray candidate are then removed from consideration, and the process is repeated using the remaining blip of highest energy.  
Formation of reconstructed $\gamma$-rays continues until no blips above our energy threshold of 75~keV remain in the event.  
The primary reconstructed $\gamma$-ray metric investigated here will be total energy.  

To generate reconstructed $\gamma$-ray energy spectra more closely resembling those attainable from a LArTPC, we apply a 50~keV energy smearing to each blip's energy to simulate the impact of electronics noise on LArTPC ADC signals; this energy smearing choice is guided by measurements of raw wire waveform noise in MicroBooNE~\cite{ub_noise,uB_ar39}. Further discussion of the limitations presented by electronics noise in blip analyses will be given in Section~\ref{sec:limitations}.  

It is likely that a more detailed analysis of blip sub-grouping will yield algorithms with improved spectroscopic performance.  
In particular, optimal grouping criteria are likely to be dependent on the exact signal in question.  
It also seems likely that additional spectroscopic information can be gleaned through combined consideration of reconstructed $\gamma$-rays' total energies and blip multiplicities~\cite{lar_neutron}.  
Nonetheless, here we forego these considerations, as the method described above is sufficient to demonstrate the value of blip activity in performing MeV-scale $\gamma$-ray spectroscopy in LArTPCs.  


The benefits of blip sub-group metrics are illustrated by applying the blip reconstruction technique from Section~\ref{sec:methods} and the blip grouping algorithm described above to LArSoft simulation of individual $\gamma$ ray and neutron samples of various types.  For each sample, 10$^5$ total events are produced.  

\begin{itemize}
    \item{\uline{A single 1.46 MeV $\gamma$-ray}: This sample can be used to directly characterize the impact of thresholding on $\gamma$-ray calorimetric capabilities at the MeV scale.   This energy reflects that of $\gamma$-rays preferentially produced in neutron inelastic scattering off $^{40}$Ar, as visible in Figure~\ref{fig:exfor}}.
    \item{\uline{Two 1.46 MeV $\gamma$-rays generated in the vicinity of one another}: $\gamma$-rays are simulated 30~cm apart, traveling at randomized angles.  This sample enables us to investigate the ability to separate secondary electrons from different $\gamma$-rays, and to examine the effects of blip pile-up on reconstructed $\gamma$-ray spectra, resolutions and biases.}  
    \item{\uline{A single neutron capture in liquid argon}: A largely monoenergetic 6.1~MeV signal from capture on $^{40}$Ar, comprised of a cascade of $\gamma$-rays of varying energy~\cite{lar_aced}, likely to be observed in large LArTPCs like DUNE.}
    \item{\uline{A single 10~MeV kinetic energy neutron}.  These events, as described in Section~\ref{sec:neutrons}, will generate many $\gamma$-rays with a variety of true energies.  We can attempt to reconstruct spectral features within this realistic mass of overlapping Compton electron activity}.
\end{itemize}

\begin{figure}[htb!]
\includegraphics[trim = 20 0 20 0, clip=true, width=0.95\columnwidth]{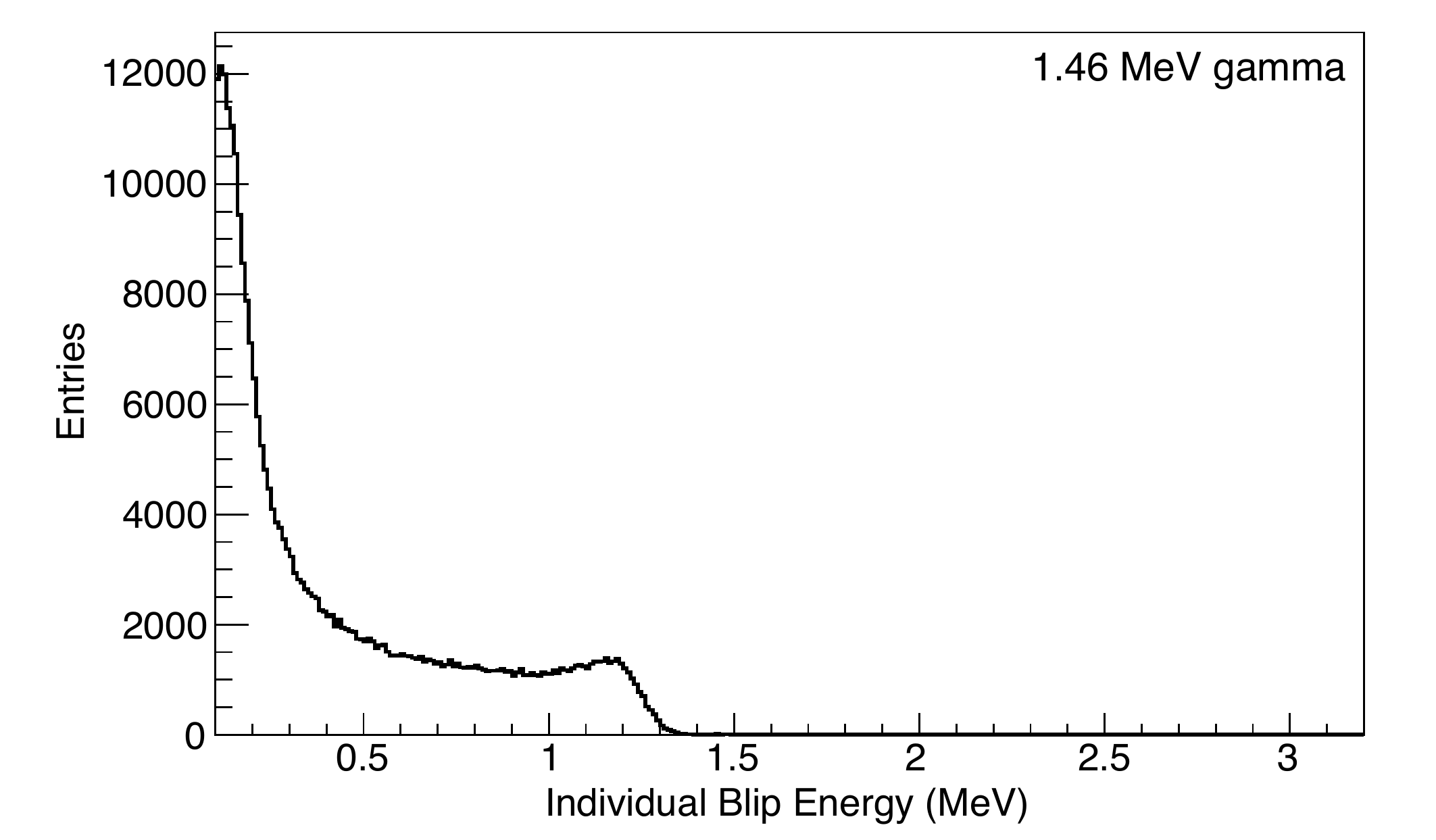}
\includegraphics[trim = 20 0 20 0, clip=true, width=0.95\columnwidth]{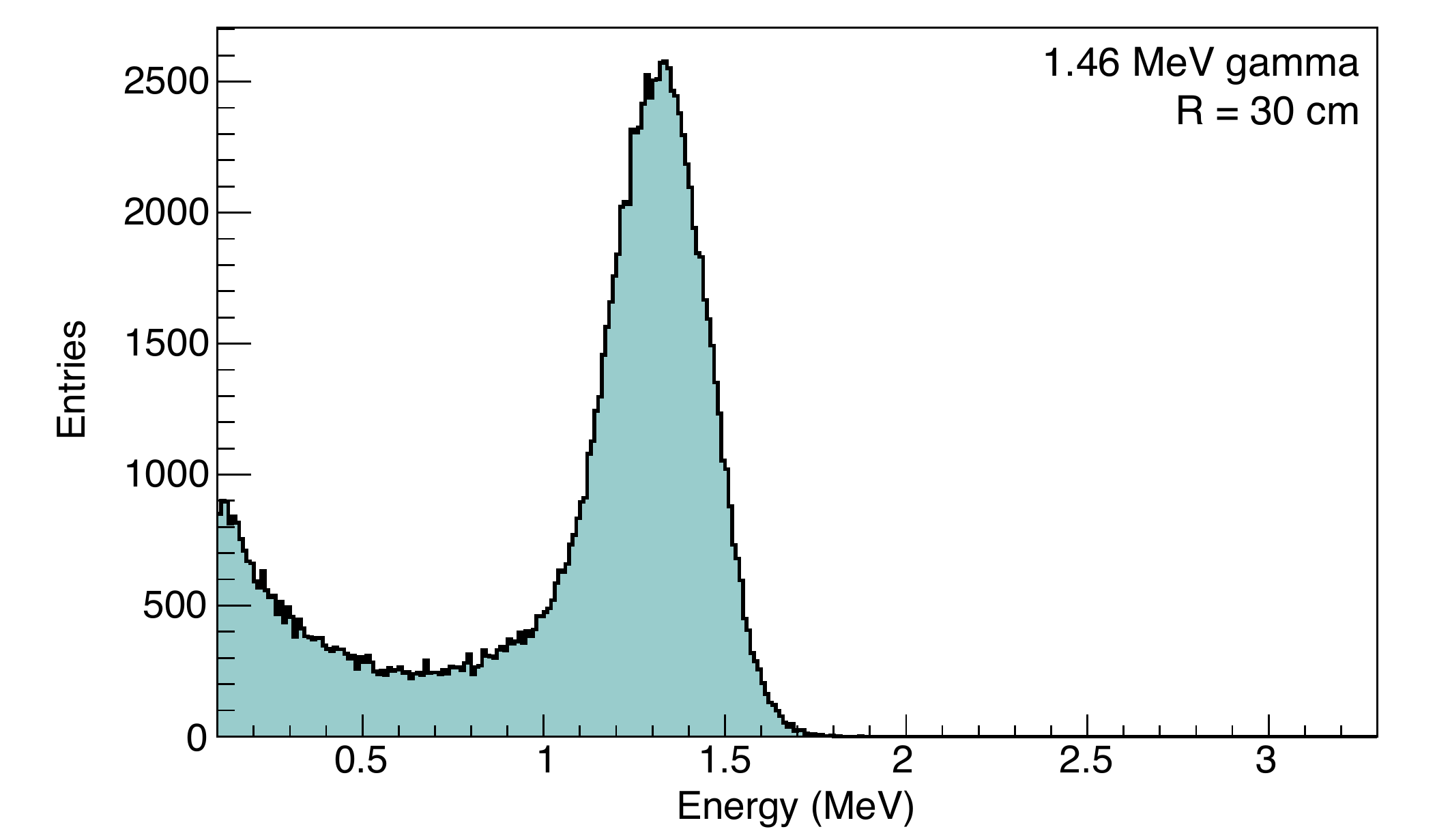}
\caption{Reconstructed energies of individual blips (top) and grouped-blip reconstructed $\gamma$-ray energies produced from a LArSoft simulation of single 1.46~MeV $\gamma$ rays (bottom).  The $\gamma$-ray's Compton shoulder is visible in the single-blip spectrum, while the full-energy peak is the most prominent feature in the reconstructed $\gamma$-ray spectrum.}
\label{fig:spec_single}

\vspace{2ex}

\includegraphics[trim = 20 0 20 0, clip=true, width=0.95\columnwidth]{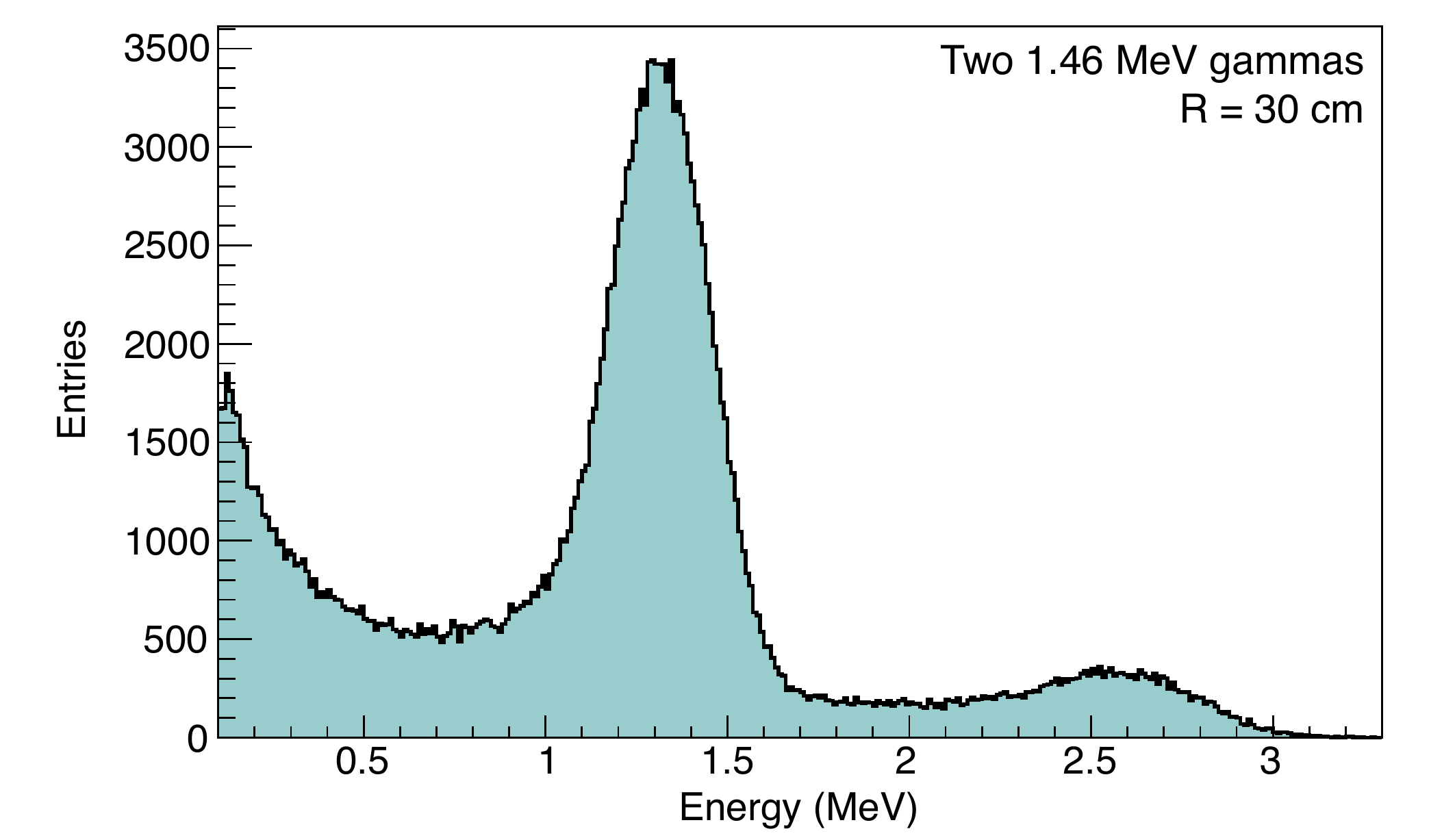}[htb!]
\caption{Reconstructed $\gamma$-ray energies produced from a LArSoft simulation of two 1.46 MeV $\gamma$-rays generated at random angles at a separation distance of 30~cm.  The full-energy peak and the pile-up peak containing energies of both $\gamma$-rays are most prominent features in the spectrum.}
\label{fig:spec_double}

\end{figure}

The energies of individual blips in events containing a single simulated 1.46~MeV $\gamma$-ray, as well as energies of reconstructed $\gamma$-rays using the iterative sphere-based grouping method described above, are shown in Figure~\ref{fig:spec_single}.  
In the individual blip spectrum, a Compton edge is observed at roughly 1.25~MeV, as would be expected from a 1.46~MeV incident $\gamma$-ray.  
This edge is accompanied by dramatically increasing blip counts at lower energies.  
As blip sub-groups are formed, this low-energy tail is decreased in magnitude as a prominent peak emerges at an energy above that of the Compton edge in the single-blip spectrum.  
The latter feature represents the reconstructed $\gamma$-ray full-energy peak, which is biased downward from the true $\gamma$-ray energy of 1.46~MeV due to the non-collection of energy in below-threshold blips.  
The full-energy peak is fit to a Gaussian function with a linear background component to account for the overlap from the distribution of incomplete $\gamma$-ray candidate energies to the left of the peak.
The Gaussian fit provides a mean of 1.32~MeV, biased -9.5\% from the true energy, as well as a 1$\sigma$ resolution of 0.13~MeV, or 9.5\%.
Taking the integral of the Gaussian component of the fit and dividing by the total number of simulated events, we calculate a `peak efficiency' of 75\%.  
This indicates that the existing algorithm is relatively efficient in its grouping of blips originating from a common $\gamma$-ray.  
These performance metrics are summarized in Table~\ref{tab:spec_metrics}.
It should be noted that since energy peaks reported on in this section are non-Gaussian to varying degrees, values reported in this table fluctuate at the few-percent level based on exact fitting assumptions and ranges.   

Similar metrics are provided for cases where altered blip grouping settings have been applied.  
If thresholds are raised to 150~keV, the resulting full-energy peak bias, resolution, and efficiency come out to -18\%, 7.8\%, and 55\%, respectively.  
For this case it appears that total energy is biased further downward from the expected true energy, while peak efficiency is also degraded significantly.  
Meanwhile, if we return to the 75~keV energy threshold but reduce the sub-group proximity to 20~cm, these metrics are altered to -11\%, 10.4\%, and 62\%, respectively.   
In this case, the energy bias and resolution remain relatively stable compared to the 30~cm radius scenario, while peak efficiency is worsened.  


\begin{table}[]
    \centering
    \begin{tabular}{|l|c|c|c|c|}
        \hline
        Sample \& Sphere Radius
        & \makecell[c]{$E_\gamma$ Bias \\(\%)} 
        & \makecell[c]{1$\sigma$ Res. \\(\%)} 
        & \makecell[c]{Peak Eff. \\ (\%)} 
        & \makecell{Pile-Up \\(\%)} \\ \hline  
        
        1 $\gamma$, 30~cm           & -9.5    & 9.5   & 75    & - \\ 
        1 $\gamma$, 30~cm (150~keV) & -18    & 7.8   & 55    & - \\ 
        1 $\gamma$, 20~cm           & -11    & 10.4  & 62    & - \\ \hline
        2 $\gamma$, 30~cm           & -9.5    & 10.3  & 107   & 28 \\  
        2 $\gamma$, 20~cm           & -11    & 11.2  & 123   & 7.0 \\ \hline 
        $n$-$^{40}$Ar capture, 60~cm  & -8.2    & 5.0   & 58    & 1.8 \\  \hline 
        10~MeV $n$, 30~cm           & -10.5    & 10.2  & 27    & 144 \\ 
        10~MeV $n$, 20~cm           & -11.8    & 11.2  & 37    & 146 \\ 
        \hline
    \end{tabular}
    \caption{Total energy bias, resolution, and efficiency metrics for different $\gamma$-ray samples, each containing 100k simulated events, using a blip energy threshold of 75~keV and proximity requirement of 30~cm.  A variety of alternate threshold and proximity cases are also shown.} 
    \label{tab:spec_metrics}
\end{table}


The energies of reconstructed $\gamma$-ray candidates identified in simulated events containing \emph{two} mono-energetic $\gamma$-rays are shown in Figure~\ref{fig:spec_double}, with performance metrics also outlined in~Table~\ref{tab:spec_metrics}.  
The full-energy peak of the single $\gamma$-ray from Figure~\ref{fig:spec_single} is again apparent in this sample's energy spectrum, with a similar bias and resolution: -9.5\% and 10.3\%.  
Thus, single $\gamma$-ray spectroscopy can still be performed even when signals from multiple $\gamma$-rays are present in the same event region.
A peak efficiency of 107\% is produced, indicating that, on average, one of the two simulated $\gamma$-rays will have its energy properly reconstructed.  

%

We also note the additional peak at roughly twice the energy of the first peak; this feature is the result of grouping energies from the two different $\gamma$-rays into one reconstructed $\gamma$-ray.  
We characterize the size of this effect by counting the number of reconstructed $\gamma$-rays $>$3$\sigma$ above the single full-energy peak and dividing by the total number of simulated two $\gamma$-ray events; this metric is referred to as the `pile-up fraction.'  
Applying the default blip reconstruction and sphere-based grouping methods to this two $\gamma$-ray sample, we observe a pile-up fraction of 28\%.  
If the smaller 20~cm group proximity requirement is used on the two $\gamma$-ray sample, the single $\gamma$-ray energy resolution is again modestly degraded as in the one $\gamma$-ray case.  
However, higher fidelity is achieved in $\gamma$-ray energy grouping: peak efficiency increases to 123\%, while the pile-up fraction reduces to 7.0\%.  

\begin{figure}
\includegraphics[trim = 20 0 20 0, clip=true, width=0.95\columnwidth]{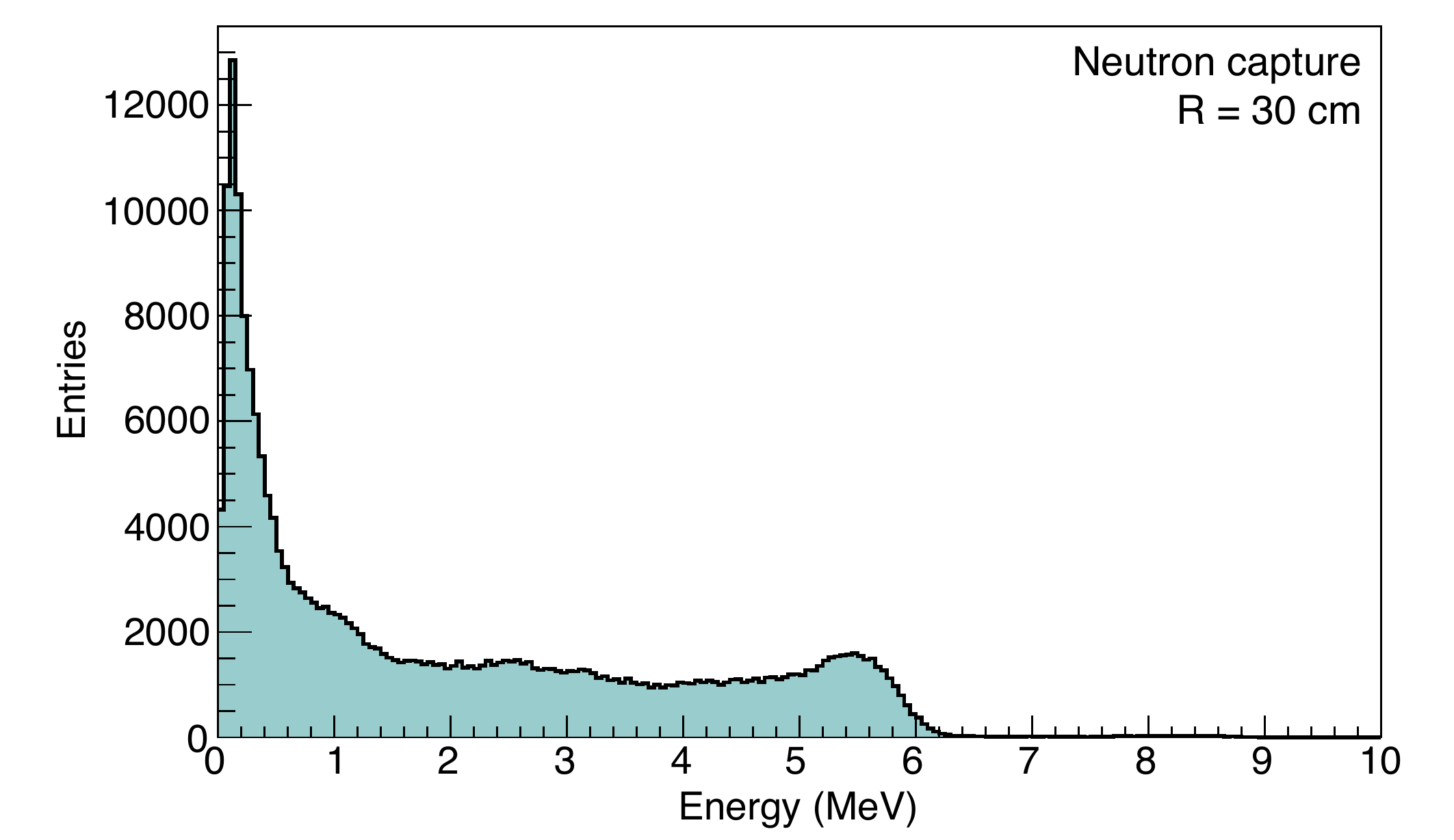}
\includegraphics[trim = 20 0 20 0, clip=true, width=0.95\columnwidth]{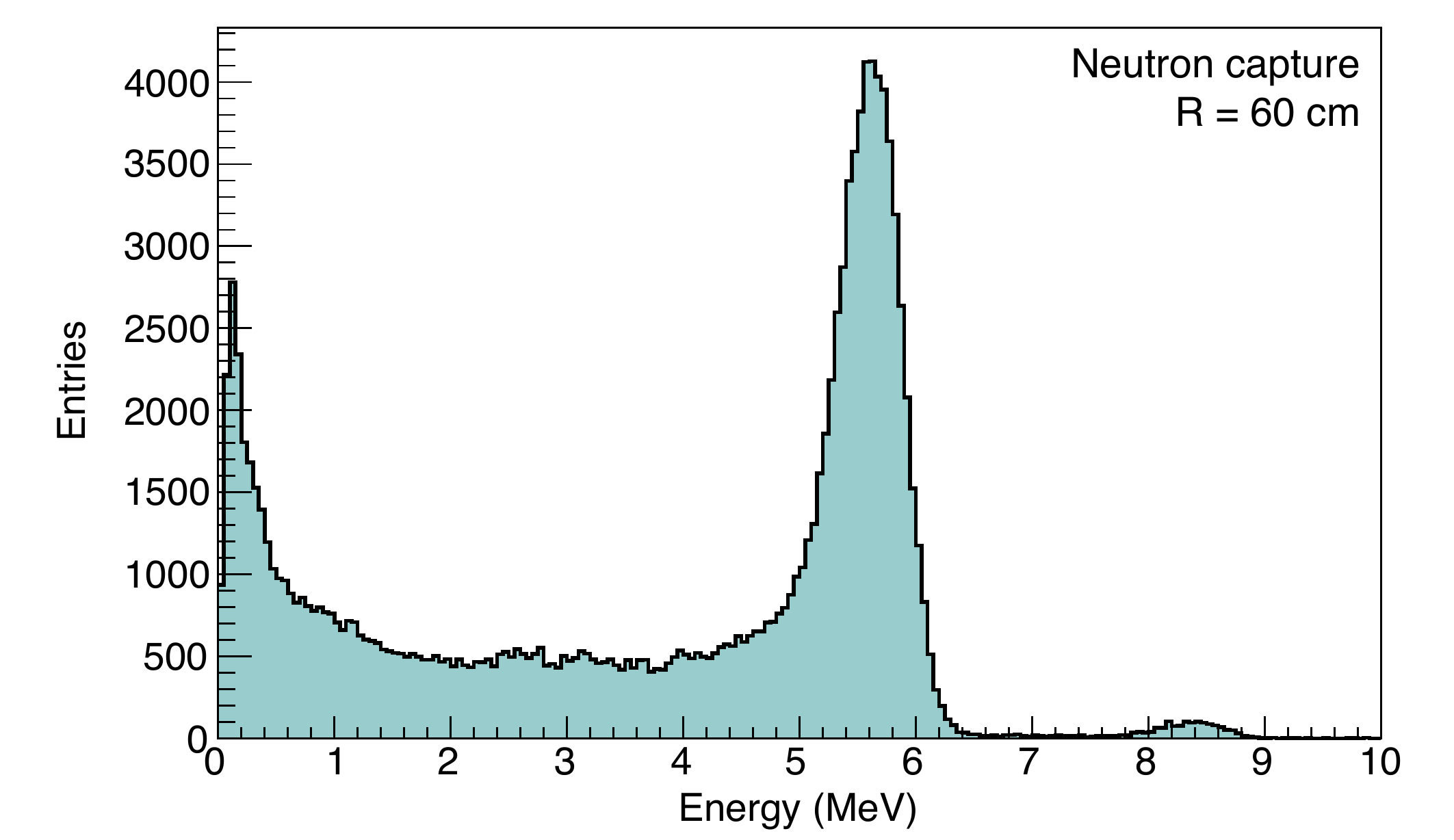}
\caption{Reconstructed $\gamma$-ray energies produced from LArSoft simulation of captures of 1~eV primary neutrons, using either a 30~cm (top) or 60~cm (bottom) proximity requirement.  For the 60~cm case, the 6.1~MeV peak from capture on $^{40}$Ar is clearly visible.}
\label{fig:spec_cap}
\end{figure}

The reconstructed $\gamma$-ray energy spectra following the simulation of single thermal neutrons are shown in Figure~\ref{fig:spec_cap} for both a 30~cm and 60~cm proximity requirement, with performance metrics outlined in~Table~\ref{tab:spec_metrics}.  
For this sample, the neutrons are not energetic enough to produce the 1.46~MeV $\gamma$-rays that are characteristic of our other samples; instead, our full-energy peak corresponds to 6.1~MeV, the total energy of  $\gamma$-rays emitted during neutron capture on $^{40}$Ar.
Applying the default $\gamma$-ray reconstruction to this sample yields a distribution that remains largely flat above 1~MeV, with a muted full-energy peak.  
Thus, it appears that the default 30~cm blip proximity requirement is better tuned to the identification of individual $\gamma$-rays as shown in previous samples, but it is insufficiently wide to capture the energy of all $\gamma$-rays from the neutron capture cascade.  
At the same time, the spectrum does not reflect the rich underlying forest of true monoenergetic $\gamma$-rays produced by the Geant4 simulation, highlighting the combined limitations of our technique and inherent LArTPC capabilities.  
If the proximity requirement is loosened to 60~cm, a clear peak is visible just below 6.1~MeV.  
This peak exhibits a bias of -8.2\%, a resolution of 5.0\%, and a peak efficiency of 58\%. We also note the existence of a much smaller peak just below 8.8~MeV due to neutron capture on $^{36}$Ar.

\begin{figure}
\includegraphics[trim = 20 0 20 0, clip=true, width=0.95\columnwidth]{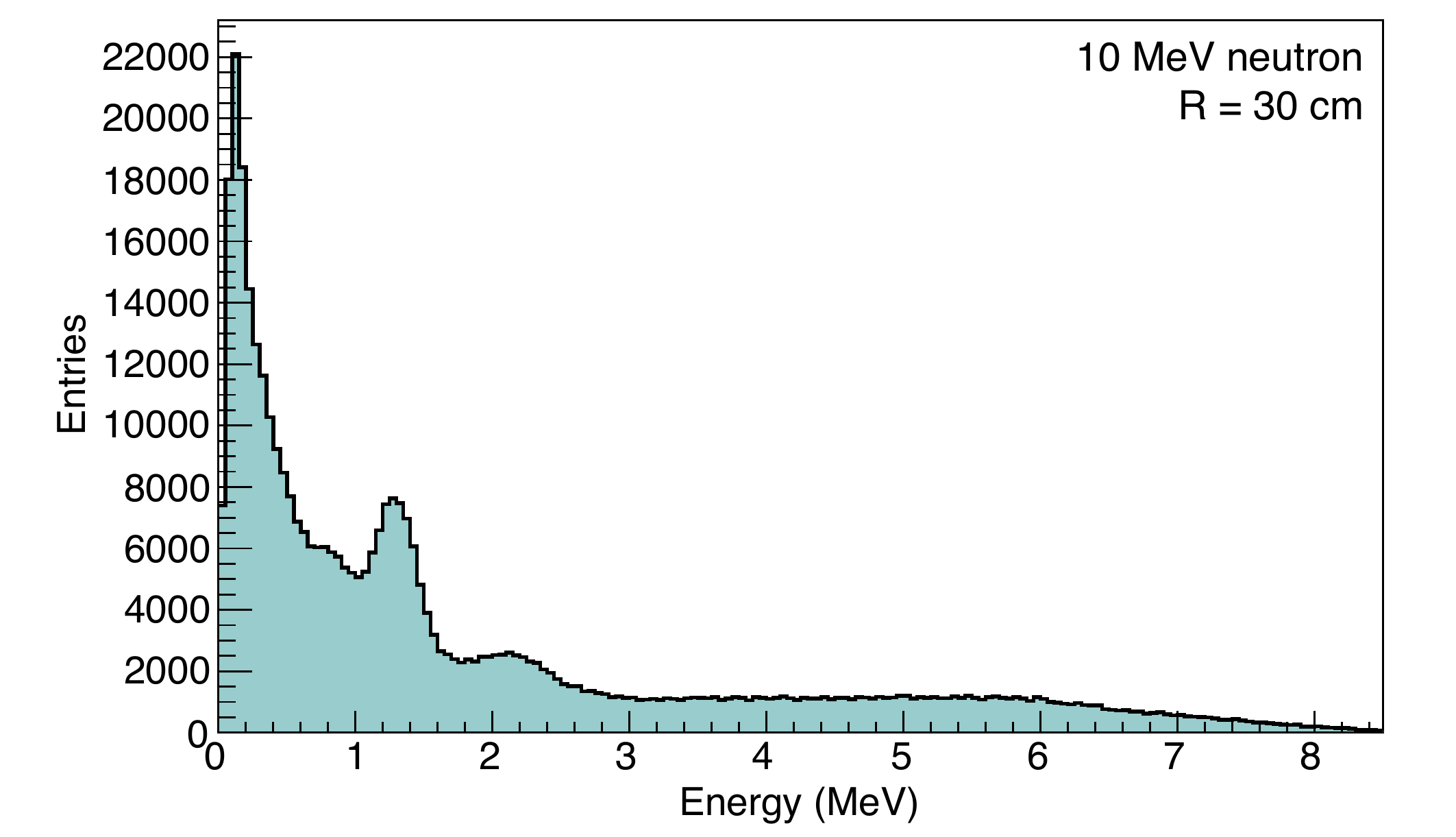}
\caption{Reconstructed $\gamma$-ray energy spectrum produced from LArSoft simulation of 10~MeV primary neutrons, using a 30~cm blip proximity requirement.  The 1.46~MeV peak corresponding to the first excited state of $^{40}$Ar is clearly visible.}
\label{fig:spec_fn}
\end{figure}

In Figure~\ref{fig:spec_fn}, we plot reconstructed $\gamma$-ray energies for monoenergetic 10~MeV fast neutrons, as might be produced by neutrino interactions, nuclear interactions of final-state heavy charged particles, or on-surface cosmic rays.  
An exponentially-decreasing spectrum is observed with a clear peak in the vicinity of the 1.46~MeV first excited state of $^{40}$Ar.  
Using the default reconstruction, a combined Gaussian plus linear fit yields a 1.46~MeV peak bias of -10.5\%, and a resolution of 10.2\%, comparable to that obtained from the single 1.46~MeV $\gamma$-ray sample.  
Based on the area of the fitted Gaussian, we see that for every simulated 10~MeV fast neutron, we identify 0.27 well-reconstructed 1.46~MeV $\gamma$-rays.  
The large pile-up fraction for this dataset is produced by overlap of $\gamma$-ray signals from multiple inelastic neutron scatters and from de-excitations of higher-lying states of $^{40}$Ar.  
The latter is likely responsible for the additional energy peak appearing at roughly 2.2~MeV.  

Both capture and inelastic scatter $\gamma$-ray signals will be naturally produced during operation of all LArTPC detectors, whether in signal neutrino interactions, or in background radiogenic and cosmogenic processes.  
Thus, these can serve alongside $^{39}$Ar as additional naturally-occurring low-energy  calibration signals in existing and future large LArTPCs.

\section{Limiting Factors in LArTPC MeV-Scale Reconstruction}
\label{sec:limitations}

In summarizing the uses of blip activity in the previous sections, we have deliberately overlooked the discussion of some of the possible limitations of this method.  
In this section we will summarize what we see as the most obvious possible limitations, and will then either quantitatively assess their impact or suggest avenues for future assessment.  

\subsection{$^{39}$Ar Contamination}

Blip activity is or will be ubiquitous in the event displays of all current and future planned large LArTPCs due to the natural presence of 1~Bq/kg specific activity of $^{39}$Ar in the liquid argon~\cite{lar_ar39}.  
While there are some benefits to the presence of this signal for detector response calibration~\cite{dune_tdr2}, its $\beta$ decay electrons are an irreducible background for the purposes of uncovering physics with non-radiogenic blip activity.  
Due to our knowledge of $^{39}$Ar specific activity in LArTPCs, however, it is straightforward to estimate the impact.  

To do so, we have used the existing radioactivity generator in LArSoft to simulate $^{39}$Ar $\beta$ decays in an otherwise-empty LArTPC. The active volume has a height of 12~m and a length of 14~m, with a full drift distance of 3.5~m, matching what is expected in DUNE. Decays are simulated throughout the volume for a length of time corresponding to one full drift period, about 2.2~ms at a nominal electric field of 500~V/cm~\cite{dune_tdr2}. 
For each event, we then choose a series of random points to serve as candidate vertices, requiring that each is least 150~cm from any active volume boundary. Blips contained within `spheres' surrounding each chosen vertex are then selected according to the requirements described in Section~\ref{sec:methods}. The extent of the proximity requirement, or sphere radius, is varied from 10~cm to 150~cm.

\begin{figure}[htb!]
\includegraphics[width=0.43\textwidth, trim = 0. 0. 1cm 1cm, clip]{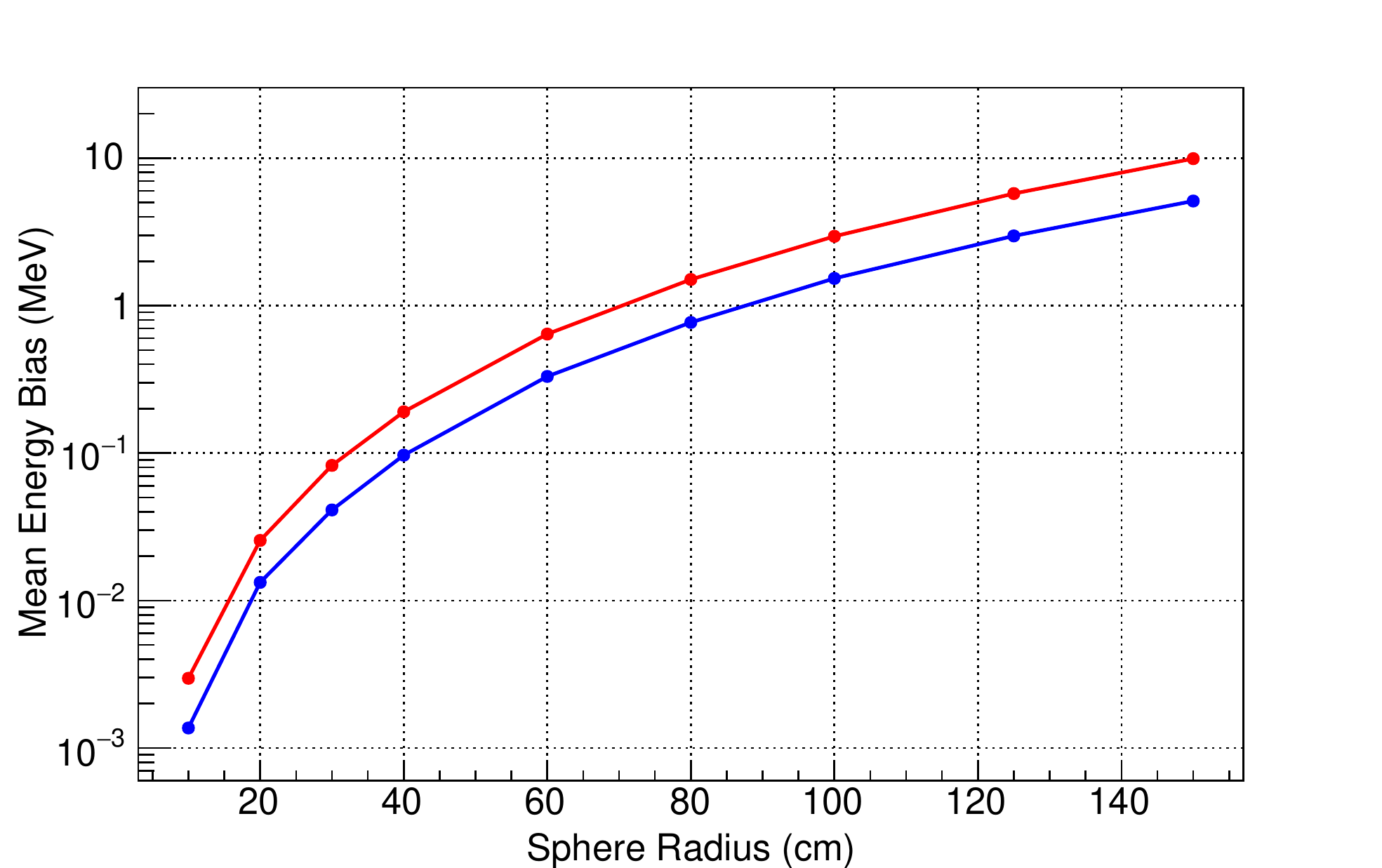}
\includegraphics[width=0.43\textwidth, trim = 0. 0. 1cm 1cm, clip]{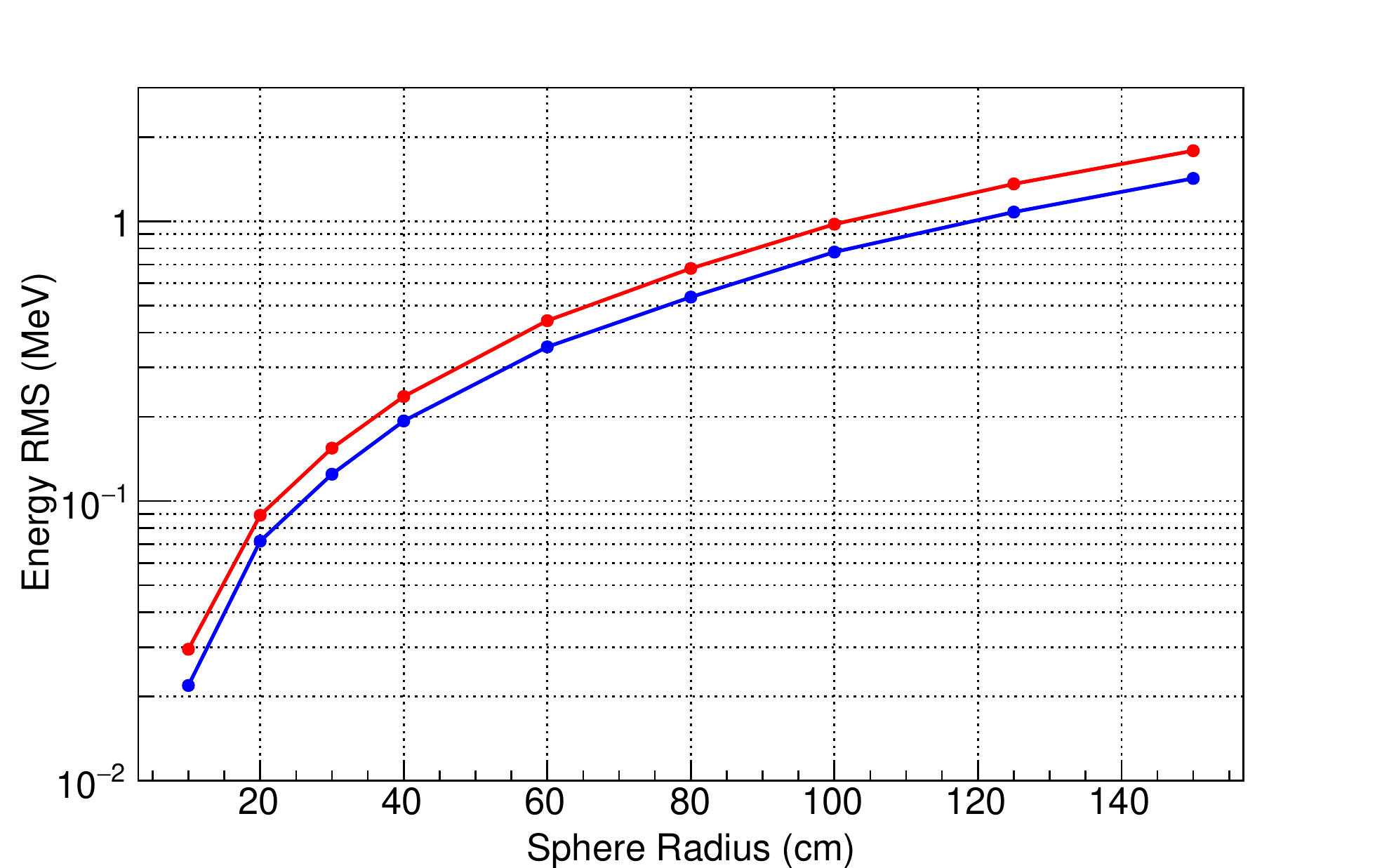}
\includegraphics[width=0.43\textwidth, trim = 0. 0. 1cm 1cm, clip]{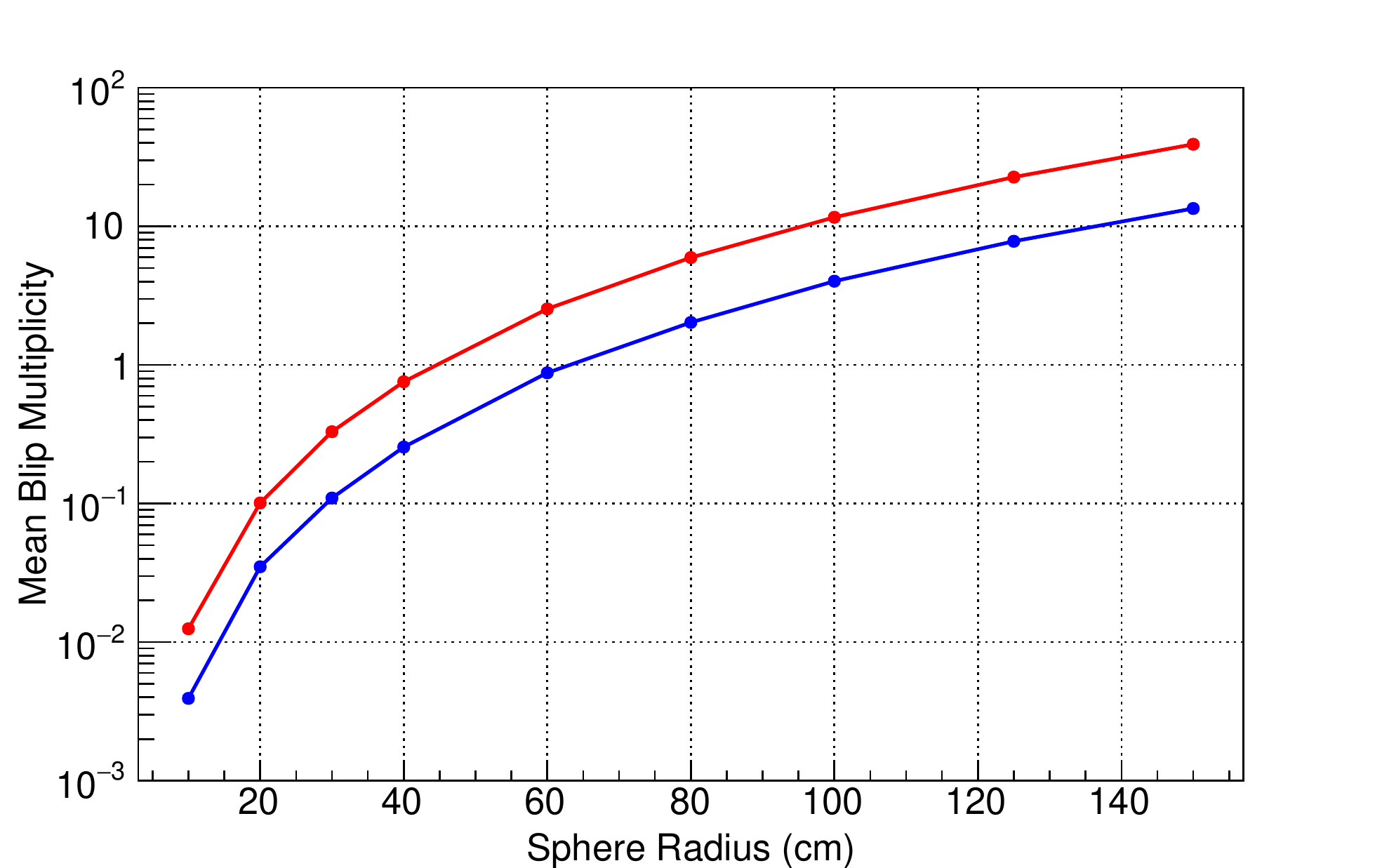}
\includegraphics[width=0.43\textwidth, trim = 0. 0. 1cm 1cm, clip]{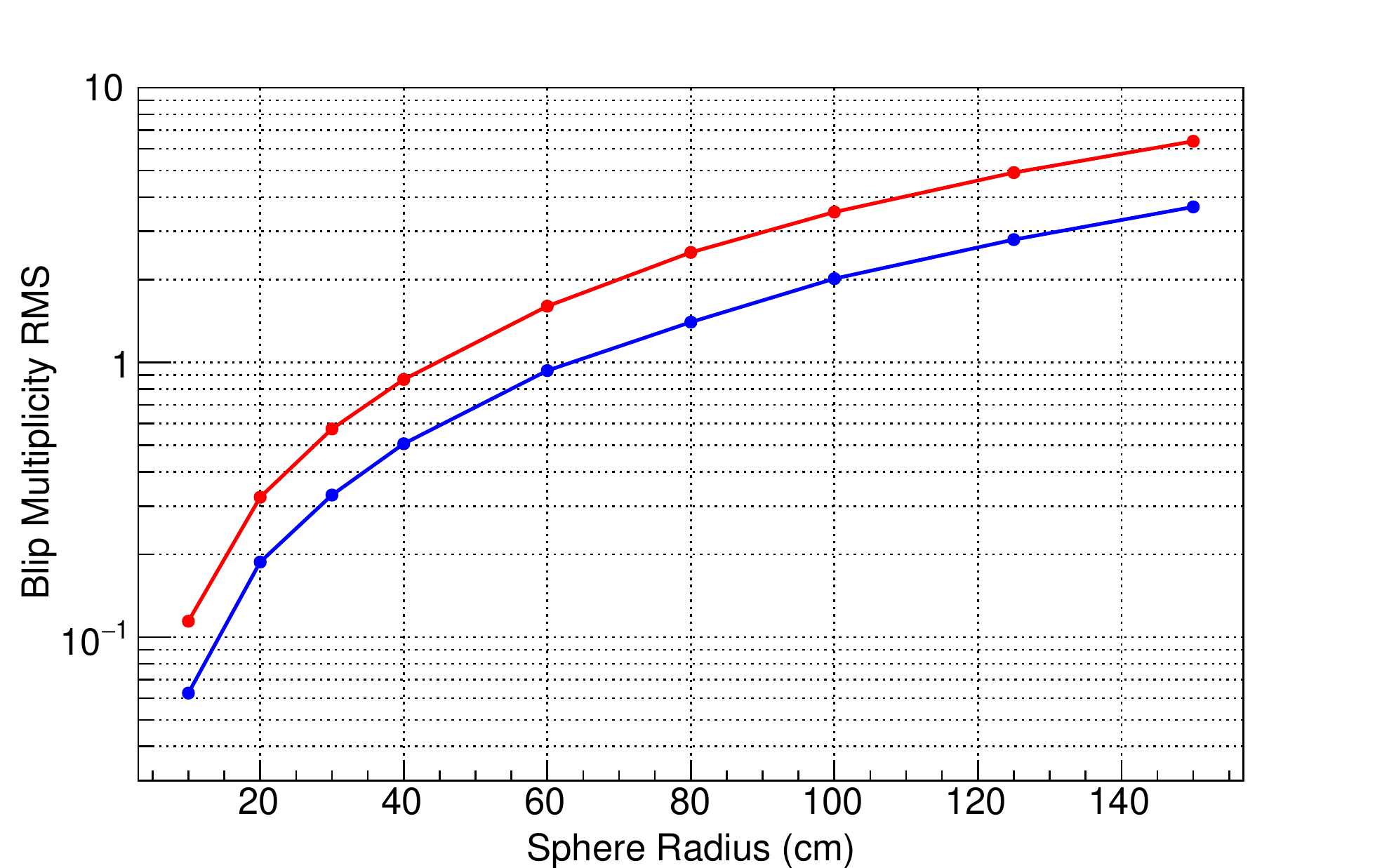}
\caption{The contribution to summed blip energy, energy RMS, blip multiplicity, and blip multiplicity RMS due to the presence of blips produced by background \ce{^{39}Ar} $\beta$ decays for varying proximity requirements and blip energy thresholds.}
\label{fig:ar39_res_bias}
\end{figure}

Blip activity metrics obtained using this procedure and dataset are shown in Figure~\ref{fig:ar39_res_bias}. 
As would be expected, as the volume considered for blip reconstruction increases, blip multiplicity, summed energy, and RMS energy spread of $^{39}$Ar blips increase.  
When a proximity of more than a meter is considered, energy biases of order 2~MeV are produced, with RMS spreads in energy contribution nearing $\sim$1~MeV.  
For this reason, we have considered only sub-meter proximity in the physics analyses shown above.  
As the Q-value of the $^{39}$Ar $\beta$ decay is 0.565~MeV, these contributions are only modestly reduced at a higher threshold of 300~keV, as shown in Figure~\ref{fig:ar39_res_bias}.

With a blip energy threshold of 75~keV and a proximity requirement of 30~cm (60~cm), an average energy bias of roughly 0.1~MeV (0.65~MeV) is expected, with an RMS spread of 0.15~MeV (0.45~MeV).  
In particular, this RMS spread can be compared directly to the calorimetric resolutions and distributions reported in the previous sections.  
In Sections~\ref{sec:supernova},~\ref{sec:neutrons}, and~\ref{sec:em}, reported energy resolutions are found to be larger than this: 
for supernova neutrinos, resolution is $\sim$20\% with a 30~cm proximity cut; for 10~MeV neutrons, resolution is 1~MeV with a 60~cm cut; and for low-energy electromagnetic showers, resolutions are $\sim$5\% with a 60~cm cut.  
In the case of single $\gamma$-ray spectroscopy, all described sources had full-energy peak resolutions of order 0.15-0.25~MeV.  
Thus, for physics processes depositing above roughly 5~MeV, $^{39}$Ar activity can negligibly impact calorimetric capabilities; blip thresholding plays a much more important role for these event classes.  
For MeV-scale physics processes, such as very-low-energy supernova neutrinos, solar neutrinos, and single $\gamma$-ray spectroscopy, $^{39}$Ar blips will likely play a non-negligible role, and should be considered when modeling achievable energy resolution.  

Using the same energy threshold and proximity requirement of 30~cm (60~cm), only about 0.3~blips (2.5~blips) are found on average, with an RMS spread in multiplicity of 0.6~blips (1.6~blips). In Sections~\ref{sec:supernova} and~\ref{sec:particleid}, blip multiplicity and summed energy were used for neutrino interaction channel and particle discrimination.
Given the relatively large cut values used for muon and pion particle discrimination in Table~\ref{tab:muonpion}, the contribution due to $^{39}$Ar is unlikely to have a significant effect on these selection efficiencies.
However, for the neutrino interaction channel identification studies, the cut values are relatively small and the effect of $^{39}$Ar cannot be easily dismissed. To gauge the impact, the values in Table~\ref{tab:efficiency} were recalculated while biasing and smearing the values of summed energy and multiplicity on an event-by-event basis to mimic the effects of $^{39}$Ar contamination found above. For the 30~cm proximity requirement, selection efficiencies were reduced by $\sim$5\% while maintaining comparable purity. A similar drop of $\sim$5\% was seen in both efficiency and purity for the 60~cm case, though the nominal cut values for multiplicity and energy were shifted by +2 and +1~MeV, respectively, to account for the sizeable bias caused by $^{39}$Ar in a sphere of this size.

\subsection{Electronics Noise}

While we have implicitly acknowledged in this study that electronics noise will define achievable blip reconstruction low-energy thresholds, we have in most cases not addressed the contribution of electronics noise to the energy resolution of reconstructed blip information.  
On the contrary, we have assumed perfect correspondence between reconstructed blip energy and true electron energy deposition.  

For wire-based charge readout systems, an equivalent noise charge (ENC) of $<$400 $e^-$ and $<$700 $e^-$ has been achieved on all wire planes in MicroBooNE and ProtoDUNE, respectively~\cite{ub_noise,dune_tdr2}.  
Noise floor performance is expected to be enhanced with a pixel-based charge readout system, with an ENC of $\sim$275~$e^-$\cite{Dwyer_2018}, while somewhat degraded in a dual-phase DUNE module, with $\sim$1100~$e^-$ ENC expected \cite{dune_idr3}.
These values should be compared to an expected muon minimum ionizing particle ENC of order 15,000~$e^-$ to 20,000~$e^-$ for MicroBooNE and DUNE.  
Based on these numbers and a simple scaling argument, one would expect electronics noise levels in various large LArTPCs to range from approximately 20-80~keV.  
As mentioned earlier, when considering integrated MicroBooNE waveforms over a wire-time tick area comparable to that expected from $^{39}$Ar blips, an average noise level of $\sim$50~keV is observed~\cite{uB_ar39}.  

A direct comparison of this 50~keV single-blip noise level to the results in previous sections indicates that noise contributions are likely to be sub-dominant for many of the calorimetric and discrimination use cases discussed above.  
For example, the 1~MeV calorimetric resolution for 10~MeV neutrons shown in Figure~\ref{fig:neutronblipE} is more than an order of magnitude larger than this estimated noise level per blip; noise levels appear similarly small compared to the 10+\% supernova neutrino energy resolutions shown in Figure~\ref{fig:sn_reco_ratios}.  
To provide context for interaction and particle discrimination capabilities, we note that  Figure~\ref{fig:totalblipE_vs_multiplicity} and Figure~\ref{fig:muonpion_ENumber_blips} are binned in 1~MeV increments, much more coarsely than any additional smearing one might expect from noise.  

The lowest observed resolutions discussed in this paper appear in Sections~\ref{sec:em} and Section~\ref{sec:spec}, where shower calorimetry and single-gamma spectroscopy are discussed, respectively.  
These sections show full-energy peak Gaussian resolutions of order 50-500~keV (Figure~\ref{fig:michel_resolution}) and 120-300~keV (Table~\ref{tab:spec_metrics}), respectively,  much closer to expected single-blip noise levels.  
Thus, in these cases, it seems likely that simulation of noise effects will be important in determining realistic capabilities.  

\begin{figure}
    \centering
    \includegraphics[width=\columnwidth]{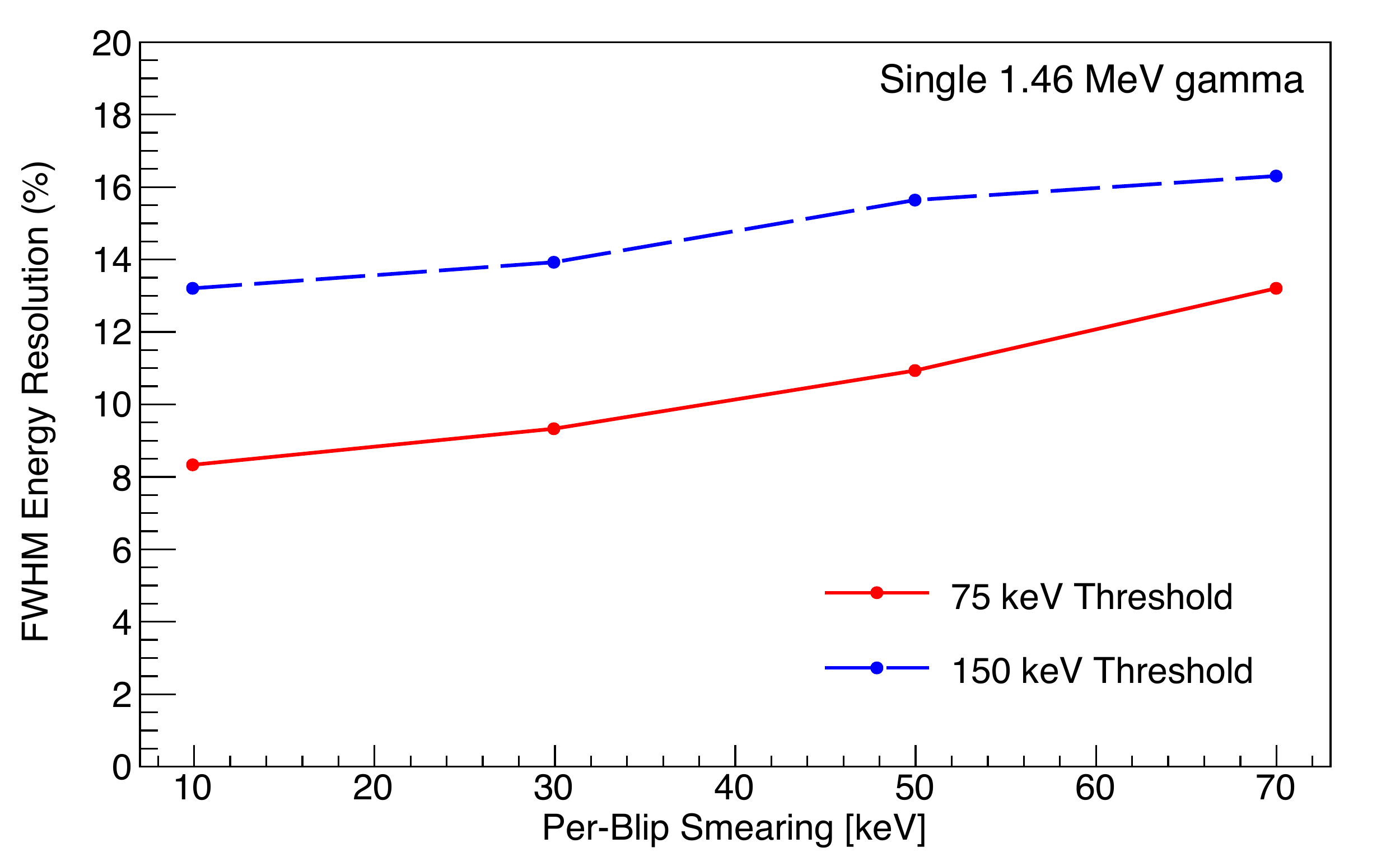}
    \caption{
    The resolution of the full-energy peak for simulated 1.46~MeV $\gamma$-rays, over a range of different blip smearing levels, for both 75~keV and 150~keV energy thresholds. A proximity requirement of 30~cm is used. Resolution is calculated based on the FWHM of the peak using the relationship to standard deviation: $\sigma = \text{FWHM}/(2\sqrt{2\ln{2}})$.
    }
    \label{fig:noise_vs_res}
\end{figure}

As a demonstration of the impact of noise, we show in Figure~\ref{fig:noise_vs_res} the variations of observed full-energy peak resolution for the single 1.46~MeV $\gamma$-ray dataset described in the previous section as per-blip noise smearing is varied from 10~keV to 70~keV.  
To reduce fitting dependencies for this relatively clean peak, we calculate the fractional FWHM resolution with respect to the FWHM window midpoint; then, to enable a more direct comparison to Table~\ref{tab:spec_metrics}, we scale the result by the relation between FWHM and the 1$\sigma$ width expected from a Gaussian distribution: $\sigma = \text{FWHM}/(2\sqrt{2\ln{2}})$.  
We remind the reader that a default per-blip noise smearing of 50~keV was applied to produce the results in Table~\ref{tab:spec_metrics}.  
In Figure~\ref{fig:noise_vs_res}, resolution is observed to decrease modestly as the applied noise smearing is decreased: when smearing is reduced to 10~keV, resolution is improved from 11.5\% to roughly 8.5\%.  
A similarly-sized decrease is observed if a higher blip threshold of 150~keV is chosen.  
Thus, for the purposes of gamma spectroscopy, it is clear that per-blip noise smearing has a modest but non-negligible impact on achieved resolutions.  

Using our current truth-based simulation method, it is more difficult to convey the impact of noise on electromagnetic showers. 
For this signal type, the average energy per topological feature (trunk or blips) is substantially higher than the previous single-gamma case, meaning that many features will consist of more than one or two above-threshold charge collection elements (wires or pads).  
In these cases, each element (rather than each blip) generates a fixed noise contribution.  
As we do not simulate individual charge collection elements, we are not able to comment accurately on this case: given that it will generate far more hit elements than blips, noise smearing contributions are certain to be higher than what would be estimated using our truth-based methods.  
As a remedy, we would encourage readers to examine the discussion of noise contributions to low-energy electron showers from LArIAT and ICARUS given in Refs.~\cite{lariat_light} and~\cite{icarus_michel}.  

\subsection{Other Detector Response Features}

Our truth-level study procedures also do not account for a variety of other features of LArTPC detector response.  
Due to the variability of many of these features between LArTPCs or to their indirect relationship to the blip physics studies presented here, we will only comment briefly on them.  

Triggering LArTPCs to capture primary electron and blip signals from low-energy neutrino events is a challenge that must be addressed by future LArTPC experiments.  
DUNE supernova and solar studies have identified triggering scenarios producing high ($>$90\%) efficiency for detection of individual neutrinos above roughly 10~MeV kinetic energy~\cite{dune_tdr2}.  
Triggering is unlikely to be a concern when considering blip activity in higher-energy particle interactions.  
For both high- and low-energy event data acquisition schemes invoking zero suppression for data reduction~\cite{ub_trig,uB_sn2}, care should be taken to choose suppression thresholds low enough to ensure acquisition and storage of blip signals.  
Impacts of blip signal smearing from ionization charge drift diffusion should also be closely considered when defining these thresholds~\cite{lar_diff1,lar_diff2}.  

Due to the small energy and size of blips and to the limited reconstruction capabilities of light-based readout systems in existing and future LArTPCs, it is likely that blip signals will not have well-defined light signatures matched to them.  
Thus, it is possible that blip signatures will be smeared in amplitude (energy) due to the unknown level of ionization electron drift losses they experience.  
While in some applications, blip drift losses can be estimated based on the relative readout time of blips with respect to larger topological features, this is not true for all cases considered in this paper.  
Thankfully, electron lifetimes achieved in current LArTPCs~\cite{ub_cal} and aimed for in future large LArTPCs~\cite{dune_tdr2} are expected to produce drift losses at the few-percent level, leaving intact the blip-related calorimetric capabilities described above.  
Variations in drift charge diffusion between blip signals, mentioned in the previous paragraph, may also impact calorimetric precision to some degree.  

\subsection{Pile-Up of Blip Activity From Many Physics Sources}

Most of the physics capabilities afforded by blip reconstruction described in this paper have been demonstrated in otherwise empty LArTPC environments.  
In reality, this will rarely be the case.  
For on-surface LArTPCs, cosmic rays will be a source of constant activity totally unrelated to any interesting physics events, including blip activity.  For example, Refs.~\cite{argo_mcp} and~\cite{uB_ar39} provide measurements of cosmogenic-related blip activity for ArgoNeuT and MicroBooNE, respectively, before and/or after applying various forms of track proximity-related blip rejection.  
Even absent cosmic ray activity, whether by being deep underground or by using offline data filtering, high-energy physics processes will produce multiple final-state particles producing different kinds of blip activity in different locations.   
These different populations can overlap, an effect that has the potential to make targeted calorimetry and identification tasks much more difficult.  

Unlike the first two limitations discussed, the level to which this effect limits the utility of blip reconstruction is completely dependent upon the application being considered.  
Thus, we do not attempt to quantify this limitation for all scenarios, and instead highlight two cases that represent the large range of possible impacts.  
For the case of supernova neutrino or solar neutrino detection, pile-up from separate physics events (i.e. different supernova or solar neutrino interactions) should have essentially no impact on the calorimetric or interaction channel identification information discussed in Section~\ref{sec:supernova}.  
For a 10~kpc distance supernova, even during the moment of highest interaction vertex density at the arrival of the neutronization burst flux, no more than a few dozen interactions are expected in an entire event for the 40~kt DUNE detector.  
On the other hand, consider the case of using blip activity to perform pion-muon discrimination on final-state tracks from GeV-scale neutrino interactions.  
Blip activity is likely to provide the most utility here for low-energy pions, which have a higher probability of capturing at rest and producing an appearance very similar to a stopping muon.  
In these cases, the pion will often end its life within 30~cm of the neutrino interaction vertex, resulting in a large degree of blip activity overlap from $\gamma$-rays and neutrons produced both at the pion end point \emph{and} the neutrino interaction vertex.  
Detailed simulation and study of these substantially-overlapped cases will be essential to understanding the usefulness of blip-based information in them.  

\subsection{Imperfect Nuclear Physics Simulation in Argon}

For many of the studies in the paper, we have relied on Geant4 and MARLEY modeling of final-state neutron and $\gamma$-ray multiplicities and energies for complex nuclear interactions on argon.  
With the exception of neutron scattering and capture, many of the processes discussed have never been measured in argon.  
Thus, we stress that our studies are meant to highlight the potential for using blip-related information, rather than to provide authoritative predictions of attainable capabilities.  
Before full use of some of these methods for high-level physics analysis, it would be prudent to assess Geant4 and neutrino generator modeling of these final-state products with dedicated measurements or systematics studies using LArTPC test beam experiments, meson decay-at-rest neutrino LArTPC experiments, and neutrino beam LArTPC experiments.

\section{Summary}
\label{sec:conc}

Using truth-level MC simulations in a generic liquid argon volume, we have demonstrated how the unique combination of excellent position resolution and low energy thresholds can be leveraged to provide new information about particle interactions from the MeV to GeV scale in large neutrino LArTPCs.  
The reconstructed positions and energies of compact, topologically isolated energy depositions of MeV-scale electrons, or blips, have been shown to enable better understanding of the identities and energies of the ancestors that created them.  
This paper has outlined the following uses for blip activity in large neutrino LArTPCs: 
\begin{itemize}
\item{Improved calorimetry and interaction channel discrimination for supernova and solar neutrino interactions}
\item{Calorimetry of final-state uncharged particles (such as $\gamma$-rays and neutrons) produced in high-energy interactions of neutrinos and other particles with argon nuclei}
\item{Improved calorimetry for electromagnetic showers}
\item{Improved discrimination and sign selection capabilities for pions and muons}
\item{Improved sensitivity for BSM searches by enabling improved background rejection and/or identification of interaction-specific topological features.}
\item{Spectroscopy of single MeV-scale $\gamma$-rays.}
\end{itemize}

This list of use cases is certainly non-exhaustive: we foresee broad applications including using nuclear decays and final state nucleus tagging, low-energy particle identification, and detector calibration, and anticipate further possibilities will be identified in the future.  
These capabilities should be generally applicable to all existing and future LArTPC experiments, such as the SBND, MicroBooNE, and ICARUS experiments in the Fermilab SBN Program and the ProtoDUNE and DUNE experiments.  
Many of these concepts and use cases are equally relevant to other particle detector technologies possessing excellent positioning and threshold combinations, such as opaque scintillator detectors~\cite{liquido_phys} or optical TPCs~\cite{o_tpc}.  

In demonstrating these capabilities, we have also identified notable limitations of blip-based information.  
While calorimetry of final-state neutrons in LArTPCs is enabled by blip reconstruction, this capability is degraded by the likely unrecoverable loss of primary and secondary neutron binding energy; further, final-state neutron multiplicity determination will be difficult, if not impossible.  
The ubiquitous presence of $^{39}$Ar decays in argon limits the scope of blip reconstruction; fortunately, for all but the lowest considered MeV-scale energies, $^{39}$Ar blip contamination is likely to play a sub-dominant role with respect to to blip energy thresholding.  
Finally, for some physics analysis scenarios, overlap of blip activity from different physics processes is likely to degrade the capabilities described above.  

Analyses focused on blip activity have already been performed using the ArgoNeuT LArTPC~\cite{argo_mev,argo_mcp}, and studies are now also underway in other LArTPC experiments, such as MicroBooNE and ProtoDUNE.  
However, blip reconstruction should not be relegated solely to the realm of dedicated studies.  
Hopefully, we have convinced the reader that blip activity can play a role in many of the centerpiece LArTPC physics analyses expected in the coming decades, such as beam and atmospheric oscillation measurements, BSM searches, and solar and supernova neutrino studies.  
Indeed, blip activity provides valuable information often orthogonal to that provided by the larger topological features in LArTPC events.  

As we see it, there are two technical roadblocks that slow a more complete implementation of blip activity in LArTPC physics analyses.  
The first is the lack of a standard software toolkit focused on reconstruction of low-energy features in LArTPCs.  
Such a toolkit could standardize low-level thresholding, identification, and position/energy reconstruction tasks that should be relatively common across LArTPC experiments, and provide for the end user blip physics objects in a format similar to that currently provided for tracks/showers/particles by Pandora~\cite{ub_pandora}.  
This will lower the barrier to entry for new analysis by making low-energy features as easily accessible as high-energy ones.  
Inclusion of blips in mainline physics analyses is also hampered by the uncertainty in the underlying nuclear modeling that determines the appearance of blip activity in most of the use cases considered.  
As mentioned in the previous section, this limitation must be resolved via dedicated measurements and subsequent model tuning, as has been done recently for argon-neutron  interactions~\cite{lar_aced,lar_captain}, argon-pion hadronic interactions~\cite{elena_phd}, and more generally for neutrino interactions on heavy nuclei~\cite{nustec, mahn}.

\begin{acknowledgments}
This work was supported by DOE Office of Science, under award No. DE-SC0008347, as well as by the IIT College of Science and Rutgers University--New Brunswick School of Arts and Sciences.  
We thank Russell Betts, Avinay Bhat, Erin Conley,  Ryan Dorrill, Roni Harnik, Zhen Liu, and Pedro Machado for insightful discussions that helped motivate this study, and especially thank David Caratelli, Steven Gardiner, Shirley Li, Ornella Palamara, and Tingjun Yang for their insights on portions of the presented analysis.  
We also acknowledge the ArgoNeuT Collaboration for the use of its LArTPC event display and LArSoft implementations, which were primarily used for this analysis.  
\end{acknowledgments}

\bibliographystyle{apsrev4-1}
\bibliography{refs}{}

\begin{thebibliography}{80}%
\makeatletter
\providecommand \@ifxundefined [1]{%
 \@ifx{#1\undefined}
}%
\providecommand \@ifnum [1]{%
 \ifnum #1\expandafter \@firstoftwo
 \else \expandafter \@secondoftwo
 \fi
}%
\providecommand \@ifx [1]{%
 \ifx #1\expandafter \@firstoftwo
 \else \expandafter \@secondoftwo
 \fi
}%
\providecommand \natexlab [1]{#1}%
\providecommand \enquote  [1]{``#1''}%
\providecommand \bibnamefont  [1]{#1}%
\providecommand \bibfnamefont [1]{#1}%
\providecommand \citenamefont [1]{#1}%
\providecommand \href@noop [0]{\@secondoftwo}%
\providecommand \href [0]{\begingroup \@sanitize@url \@href}%
\providecommand \@href[1]{\@@startlink{#1}\@@href}%
\providecommand \@@href[1]{\endgroup#1\@@endlink}%
\providecommand \@sanitize@url [0]{\catcode `\\12\catcode `\$12\catcode
  `\&12\catcode `\#12\catcode `\^12\catcode `\_12\catcode `\%12\relax}%
\providecommand \@@startlink[1]{}%
\providecommand \@@endlink[0]{}%
\providecommand \url  [0]{\begingroup\@sanitize@url \@url }%
\providecommand \@url [1]{\endgroup\@href {#1}{\urlprefix }}%
\providecommand \urlprefix  [0]{URL }%
\providecommand \Eprint [0]{\href }%
\providecommand \doibase [0]{http://dx.doi.org/}%
\providecommand \selectlanguage [0]{\@gobble}%
\providecommand \bibinfo  [0]{\@secondoftwo}%
\providecommand \bibfield  [0]{\@secondoftwo}%
\providecommand \translation [1]{[#1]}%
\providecommand \BibitemOpen [0]{}%
\providecommand \bibitemStop [0]{}%
\providecommand \bibitemNoStop [0]{.\EOS\space}%
\providecommand \EOS [0]{\spacefactor3000\relax}%
\providecommand \BibitemShut  [1]{\csname bibitem#1\endcsname}%
\let\auto@bib@innerbib\@empty
\bibitem [{\citenamefont {{SBND Collaboration}}(2020)}]{sbnd_web}%
  \BibitemOpen
  \bibfield  {author} {\bibinfo {author} {\bibnamefont {{SBND
  Collaboration}}},\ }\href@noop {} {} (\bibinfo {year} {2020}),\ \bibinfo
  {note}
  {\href{https://sbn-nd.fnal.gov/}{https://sbn-nd.fnal.gov/}}\BibitemShut
  {NoStop}%
\bibitem [{\citenamefont {Acciarri}\ \emph
  {et~al.}(2017{\natexlab{a}})\citenamefont {Acciarri} \emph
  {et~al.}}]{ub_det}%
  \BibitemOpen
  \bibfield  {author} {\bibinfo {author} {\bibfnamefont {R.}~\bibnamefont
  {Acciarri}} \emph {et~al.} (\bibinfo {collaboration} {MicroBooNE}),\ }\href
  {\doibase 10.1088/1748-0221/12/02/P02017} {\bibfield  {journal} {\bibinfo
  {journal} {JINST}\ }\textbf {\bibinfo {volume} {12}},\ \bibinfo {pages}
  {P02017} (\bibinfo {year} {2017}{\natexlab{a}})}\BibitemShut {NoStop}%
\bibitem [{\citenamefont {{ICARUS Collaboration}}(2020)}]{icarus_web}%
  \BibitemOpen
  \bibfield  {author} {\bibinfo {author} {\bibnamefont {{ICARUS
  Collaboration}}},\ }\href@noop {} {} (\bibinfo {year} {2020}),\ \bibinfo
  {note} {\href{http://icarus.fnal.gov/}{http://icarus.fnal.gov/}}\BibitemShut
  {NoStop}%
\bibitem [{\citenamefont {Antonello}\ \emph {et~al.}(2015)\citenamefont
  {Antonello} \emph {et~al.}}]{sbn}%
  \BibitemOpen
  \bibfield  {author} {\bibinfo {author} {\bibfnamefont {M.}~\bibnamefont
  {Antonello}} \emph {et~al.} (\bibinfo {collaboration} {MicroBooNE, SBND,
  ICARUS}),\ }\href@noop {} {\  (\bibinfo {year} {2015})},\ \Eprint
  {http://arxiv.org/abs/1503.01520} {arXiv:1503.01520 [physics.ins-det]}
  \BibitemShut {NoStop}%
\bibitem [{\citenamefont {Machado}\ \emph {et~al.}(2019)\citenamefont
  {Machado}, \citenamefont {Palamara},\ and\ \citenamefont
  {Schmitz}}]{sbnd_phys}%
  \BibitemOpen
  \bibfield  {author} {\bibinfo {author} {\bibfnamefont {P.~A.}\ \bibnamefont
  {Machado}}, \bibinfo {author} {\bibfnamefont {O.}~\bibnamefont {Palamara}}, \
  and\ \bibinfo {author} {\bibfnamefont {D.~W.}\ \bibnamefont {Schmitz}},\
  }\href {\doibase 10.1146/annurev-nucl-101917-020949} {\bibfield  {journal}
  {\bibinfo  {journal} {Ann. Rev. Nucl. Part. Sci.}\ }\textbf {\bibinfo
  {volume} {69}},\ \bibinfo {pages} {363} (\bibinfo {year} {2019})}\BibitemShut
  {NoStop}%
\bibitem [{\citenamefont {Abi}\ \emph {et~al.}(2020{\natexlab{a}})\citenamefont
  {Abi} \emph {et~al.}}]{DUNE_tdr1}%
  \BibitemOpen
  \bibfield  {author} {\bibinfo {author} {\bibfnamefont {B.}~\bibnamefont
  {Abi}} \emph {et~al.} (\bibinfo {collaboration} {DUNE}),\ }\href@noop {} {\
  (\bibinfo {year} {2020}{\natexlab{a}})},\ \Eprint
  {http://arxiv.org/abs/2002.02967} {arXiv:2002.02967 [physics.ins-det]}
  \BibitemShut {NoStop}%
\bibitem [{\citenamefont {{Proton Improvement Plan II}}(2020)}]{pip_web}%
  \BibitemOpen
  \bibfield  {author} {\bibinfo {author} {\bibnamefont {{Proton Improvement
  Plan II}}},\ }\href@noop {} {} (\bibinfo {year} {2020}),\ \bibinfo {note}
  {\href{https://pip2.fnal.gov/}{https://pip2.fnal.gov/}}\BibitemShut {NoStop}%
\bibitem [{\citenamefont {Abi}\ \emph {et~al.}(2020{\natexlab{b}})\citenamefont
  {Abi} \emph {et~al.}}]{dune_tdr2}%
  \BibitemOpen
  \bibfield  {author} {\bibinfo {author} {\bibfnamefont {B.}~\bibnamefont
  {Abi}} \emph {et~al.} (\bibinfo {collaboration} {DUNE}),\ }\href@noop {} {\
  (\bibinfo {year} {2020}{\natexlab{b}})},\ \Eprint
  {http://arxiv.org/abs/2002.03005} {arXiv:2002.03005 [hep-ex]} \BibitemShut
  {NoStop}%
\bibitem [{\citenamefont {Acciarri}\ \emph
  {et~al.}(2017{\natexlab{b}})\citenamefont {Acciarri} \emph
  {et~al.}}]{argo_nue}%
  \BibitemOpen
  \bibfield  {author} {\bibinfo {author} {\bibfnamefont {R.}~\bibnamefont
  {Acciarri}} \emph {et~al.} (\bibinfo {collaboration} {ArgoNeuT}),\ }\href
  {\doibase 10.1103/PhysRevD.95.072005} {\bibfield  {journal} {\bibinfo
  {journal} {Phys. Rev. D}\ }\textbf {\bibinfo {volume} {95}},\ \bibinfo
  {pages} {072005} (\bibinfo {year} {2017}{\natexlab{b}})}\BibitemShut
  {NoStop}%
\bibitem [{\citenamefont {Acciarri}\ \emph
  {et~al.}(2020{\natexlab{a}})\citenamefont {Acciarri} \emph
  {et~al.}}]{argo_nue_xsec}%
  \BibitemOpen
  \bibfield  {author} {\bibinfo {author} {\bibfnamefont {R.}~\bibnamefont
  {Acciarri}} \emph {et~al.} (\bibinfo {collaboration} {ArgoNeuT}),\
  }\href@noop {} {\  (\bibinfo {year} {2020}{\natexlab{a}})},\ \Eprint
  {http://arxiv.org/abs/2004.01956} {arXiv:2004.01956 [hep-ex]} \BibitemShut
  {NoStop}%
\bibitem [{\citenamefont {{MicroBooNE
  Collaboration}}(2018{\natexlab{a}})}]{uB_numi}%
  \BibitemOpen
  \bibfield  {author} {\bibinfo {author} {\bibnamefont {{MicroBooNE
  Collaboration}}},\ }\href@noop {} {} (\bibinfo {year} {2018}{\natexlab{a}}),\
  \bibinfo {note}
  {\href{https://microboone.fnal.gov/wp-content/uploads/MICROBOONE-NOTE-1054-PUB.pdf}{MICROBOONE-NOTE-1054-PUB}}\BibitemShut
  {NoStop}%
\bibitem [{\citenamefont {Abratenko}\ \emph {et~al.}(2020)\citenamefont
  {Abratenko} \emph {et~al.}}]{ub_hnl}%
  \BibitemOpen
  \bibfield  {author} {\bibinfo {author} {\bibfnamefont {P.}~\bibnamefont
  {Abratenko}} \emph {et~al.} (\bibinfo {collaboration} {MicroBooNE}),\ }\href
  {\doibase 10.1103/PhysRevD.101.052001} {\bibfield  {journal} {\bibinfo
  {journal} {Phys. Rev. D}\ }\textbf {\bibinfo {volume} {101}},\ \bibinfo
  {pages} {052001} (\bibinfo {year} {2020})}\BibitemShut {NoStop}%
\bibitem [{\citenamefont {Acciarri}\ \emph {et~al.}(2014)\citenamefont
  {Acciarri} \emph {et~al.}}]{argo_hammer}%
  \BibitemOpen
  \bibfield  {author} {\bibinfo {author} {\bibfnamefont {R.}~\bibnamefont
  {Acciarri}} \emph {et~al.} (\bibinfo {collaboration} {ArgoNeuT}),\ }\href
  {\doibase 10.1103/PhysRevD.90.012008} {\bibfield  {journal} {\bibinfo
  {journal} {Phys. Rev. D}\ }\textbf {\bibinfo {volume} {90}},\ \bibinfo
  {pages} {012008} (\bibinfo {year} {2014})}\BibitemShut {NoStop}%
\bibitem [{\citenamefont {Adams}\ \emph
  {et~al.}(2019{\natexlab{a}})\citenamefont {Adams} \emph {et~al.}}]{ub_mult}%
  \BibitemOpen
  \bibfield  {author} {\bibinfo {author} {\bibfnamefont {C.}~\bibnamefont
  {Adams}} \emph {et~al.} (\bibinfo {collaboration} {MicroBooNE}),\ }\href
  {\doibase 10.1140/epjc/s10052-019-6742-3} {\bibfield  {journal} {\bibinfo
  {journal} {Eur. Phys. J. C}\ }\textbf {\bibinfo {volume} {79}},\ \bibinfo
  {pages} {248} (\bibinfo {year} {2019}{\natexlab{a}})}\BibitemShut {NoStop}%
\bibitem [{\citenamefont {Abratenko}\ \emph {et~al.}(2017)\citenamefont
  {Abratenko} \emph {et~al.}}]{ub_mcs}%
  \BibitemOpen
  \bibfield  {author} {\bibinfo {author} {\bibfnamefont {P.}~\bibnamefont
  {Abratenko}} \emph {et~al.} (\bibinfo {collaboration} {MicroBooNE}),\ }\href
  {\doibase 10.1088/1748-0221/12/10/P10010} {\bibfield  {journal} {\bibinfo
  {journal} {JINST}\ }\textbf {\bibinfo {volume} {12}},\ \bibinfo {pages}
  {P10010} (\bibinfo {year} {2017})}\BibitemShut {NoStop}%
\bibitem [{\citenamefont {Abratenko}\ \emph {et~al.}(2019)\citenamefont
  {Abratenko} \emph {et~al.}}]{ub_cc}%
  \BibitemOpen
  \bibfield  {author} {\bibinfo {author} {\bibfnamefont {P.}~\bibnamefont
  {Abratenko}} \emph {et~al.} (\bibinfo {collaboration} {MicroBooNE}),\ }\href
  {\doibase 10.1103/PhysRevLett.123.131801} {\bibfield  {journal} {\bibinfo
  {journal} {Phys. Rev. Lett.}\ }\textbf {\bibinfo {volume} {123}},\ \bibinfo
  {pages} {131801} (\bibinfo {year} {2019})}\BibitemShut {NoStop}%
\bibitem [{\citenamefont {Acciarri}\ \emph
  {et~al.}(2018{\natexlab{a}})\citenamefont {Acciarri} \emph
  {et~al.}}]{ub_pandora}%
  \BibitemOpen
  \bibfield  {author} {\bibinfo {author} {\bibfnamefont {R.}~\bibnamefont
  {Acciarri}} \emph {et~al.} (\bibinfo {collaboration} {MicroBooNE}),\ }\href
  {\doibase 10.1140/epjc/s10052-017-5481-6} {\bibfield  {journal} {\bibinfo
  {journal} {Eur. Phys. J. C}\ }\textbf {\bibinfo {volume} {78}},\ \bibinfo
  {pages} {82} (\bibinfo {year} {2018}{\natexlab{a}})}\BibitemShut {NoStop}%
\bibitem [{\citenamefont {Acciarri}\ \emph
  {et~al.}(2017{\natexlab{c}})\citenamefont {Acciarri} \emph
  {et~al.}}]{ub_dl1}%
  \BibitemOpen
  \bibfield  {author} {\bibinfo {author} {\bibfnamefont {R.}~\bibnamefont
  {Acciarri}} \emph {et~al.} (\bibinfo {collaboration} {MicroBooNE}),\ }\href
  {\doibase 10.1088/1748-0221/12/03/P03011} {\bibfield  {journal} {\bibinfo
  {journal} {JINST}\ }\textbf {\bibinfo {volume} {12}},\ \bibinfo {pages}
  {P03011} (\bibinfo {year} {2017}{\natexlab{c}})}\BibitemShut {NoStop}%
\bibitem [{\citenamefont {Adams}\ \emph
  {et~al.}(2019{\natexlab{b}})\citenamefont {Adams} \emph {et~al.}}]{ub_dl2}%
  \BibitemOpen
  \bibfield  {author} {\bibinfo {author} {\bibfnamefont {C.}~\bibnamefont
  {Adams}} \emph {et~al.} (\bibinfo {collaboration} {MicroBooNE}),\ }\href
  {\doibase 10.1103/PhysRevD.99.092001} {\bibfield  {journal} {\bibinfo
  {journal} {Phys. Rev. D}\ }\textbf {\bibinfo {volume} {99}},\ \bibinfo
  {pages} {092001} (\bibinfo {year} {2019}{\natexlab{b}})}\BibitemShut
  {NoStop}%
\bibitem [{\citenamefont {Acciarri}\ \emph
  {et~al.}(2017{\natexlab{d}})\citenamefont {Acciarri} \emph
  {et~al.}}]{ub_michel}%
  \BibitemOpen
  \bibfield  {author} {\bibinfo {author} {\bibfnamefont {R.}~\bibnamefont
  {Acciarri}} \emph {et~al.} (\bibinfo {collaboration} {MicroBooNE}),\ }\href
  {\doibase 10.1088/1748-0221/12/09/P09014} {\bibfield  {journal} {\bibinfo
  {journal} {JINST}\ }\textbf {\bibinfo {volume} {12}},\ \bibinfo {pages}
  {P09014} (\bibinfo {year} {2017}{\natexlab{d}})}\BibitemShut {NoStop}%
\bibitem [{\citenamefont {Foreman}\ \emph {et~al.}(2020)\citenamefont {Foreman}
  \emph {et~al.}}]{lariat_light}%
  \BibitemOpen
  \bibfield  {author} {\bibinfo {author} {\bibfnamefont {W.}~\bibnamefont
  {Foreman}} \emph {et~al.} (\bibinfo {collaboration} {LArIAT}),\ }\href
  {\doibase 10.1103/PhysRevD.101.012010} {\bibfield  {journal} {\bibinfo
  {journal} {Phys. Rev. D}\ }\textbf {\bibinfo {volume} {101}},\ \bibinfo
  {pages} {012010} (\bibinfo {year} {2020})}\BibitemShut {NoStop}%
\bibitem [{\citenamefont {Amoruso}\ \emph
  {et~al.}(2004{\natexlab{a}})\citenamefont {Amoruso} \emph
  {et~al.}}]{icarus_michel}%
  \BibitemOpen
  \bibfield  {author} {\bibinfo {author} {\bibfnamefont {S.}~\bibnamefont
  {Amoruso}} \emph {et~al.} (\bibinfo {collaboration} {ICARUS}),\ }\href
  {\doibase 10.1140/epjc/s2004-01597-7} {\bibfield  {journal} {\bibinfo
  {journal} {Eur. Phys. J. C}\ }\textbf {\bibinfo {volume} {33}},\ \bibinfo
  {pages} {233} (\bibinfo {year} {2004}{\natexlab{a}})}\BibitemShut {NoStop}%
\bibitem [{\citenamefont {Acciarri}\ \emph {et~al.}(2019)\citenamefont
  {Acciarri} \emph {et~al.}}]{argo_mev}%
  \BibitemOpen
  \bibfield  {author} {\bibinfo {author} {\bibfnamefont {R.}~\bibnamefont
  {Acciarri}} \emph {et~al.} (\bibinfo {collaboration} {ArgoNeuT}),\ }\href
  {\doibase 10.1103/PhysRevD.99.012002} {\bibfield  {journal} {\bibinfo
  {journal} {Phys. Rev. D}\ }\textbf {\bibinfo {volume} {99}},\ \bibinfo
  {pages} {012002} (\bibinfo {year} {2019})}\BibitemShut {NoStop}%
\bibitem [{\citenamefont {{MicroBooNE
  Collaboration}}(2018{\natexlab{b}})}]{uB_ar39}%
  \BibitemOpen
  \bibfield  {author} {\bibinfo {author} {\bibnamefont {{MicroBooNE
  Collaboration}}},\ }\href@noop {} {} (\bibinfo {year} {2018}{\natexlab{b}}),\
  \bibinfo {note}
  {\href{https://microboone.fnal.gov/wp-content/uploads/MICROBOONE-NOTE-1050-PUB.pdf}{MICROBOONE-NOTE-1050-PUB}}\BibitemShut
  {NoStop}%
\bibitem [{\citenamefont {Anderson}\ \emph {et~al.}(2012)\citenamefont
  {Anderson} \emph {et~al.}}]{argo_det}%
  \BibitemOpen
  \bibfield  {author} {\bibinfo {author} {\bibfnamefont {C.}~\bibnamefont
  {Anderson}} \emph {et~al.},\ }\href {\doibase 10.1088/1748-0221/7/10/P10019}
  {\bibfield  {journal} {\bibinfo  {journal} {JINST}\ }\textbf {\bibinfo
  {volume} {7}},\ \bibinfo {pages} {P10019} (\bibinfo {year}
  {2012})}\BibitemShut {NoStop}%
\bibitem [{\citenamefont {{ESTAR: Stopping Power and Range Tables for
  Electrons}}(2020)}]{csda}%
  \BibitemOpen
  \bibfield  {author} {\bibinfo {author} {\bibnamefont {{ESTAR: Stopping Power
  and Range Tables for Electrons}}},\ }\href@noop {} {} (\bibinfo {year}
  {2020}),\ \bibinfo {note}
  {\href{https://physics.nist.gov/PhysRefData/Star/Text/ESTAR.html}{https://physics.nist.gov/}}\BibitemShut
  {NoStop}%
\bibitem [{\citenamefont {Lepetic}(2020)}]{lepetic_phd}%
  \BibitemOpen
  \bibfield  {author} {\bibinfo {author} {\bibfnamefont {I.}~\bibnamefont
  {Lepetic}},\ }\href@noop {} {Ph.D. thesis},\ \bibinfo  {school} {IIT,
  Chicago} (\bibinfo {year} {2020})\BibitemShut {NoStop}%
\bibitem [{\citenamefont {Dwyer}\ \emph {et~al.}(2018)\citenamefont {Dwyer},
  \citenamefont {Garcia-Sciveres}, \citenamefont {Gnani}, \citenamefont
  {Grace}, \citenamefont {Kohn}, \citenamefont {Kramer}, \citenamefont
  {Krieger}, \citenamefont {Lin}, \citenamefont {Luk}, \citenamefont {Madigan},
  \citenamefont {Marshall}, \citenamefont {Steiner},\ and\ \citenamefont
  {Stezelberger}}]{Dwyer_2018}%
  \BibitemOpen
  \bibfield  {author} {\bibinfo {author} {\bibfnamefont {D.}~\bibnamefont
  {Dwyer}}, \bibinfo {author} {\bibfnamefont {M.}~\bibnamefont
  {Garcia-Sciveres}}, \bibinfo {author} {\bibfnamefont {D.}~\bibnamefont
  {Gnani}}, \bibinfo {author} {\bibfnamefont {C.}~\bibnamefont {Grace}},
  \bibinfo {author} {\bibfnamefont {S.}~\bibnamefont {Kohn}}, \bibinfo {author}
  {\bibfnamefont {M.}~\bibnamefont {Kramer}}, \bibinfo {author} {\bibfnamefont
  {A.}~\bibnamefont {Krieger}}, \bibinfo {author} {\bibfnamefont
  {C.}~\bibnamefont {Lin}}, \bibinfo {author} {\bibfnamefont {K.}~\bibnamefont
  {Luk}}, \bibinfo {author} {\bibfnamefont {P.}~\bibnamefont {Madigan}},
  \bibinfo {author} {\bibfnamefont {C.}~\bibnamefont {Marshall}}, \bibinfo
  {author} {\bibfnamefont {H.}~\bibnamefont {Steiner}}, \ and\ \bibinfo
  {author} {\bibfnamefont {T.}~\bibnamefont {Stezelberger}},\ }\href {\doibase
  10.1088/1748-0221/13/10/p10007} {\bibfield  {journal} {\bibinfo  {journal}
  {Journal of Instrumentation}\ }\textbf {\bibinfo {volume} {13}},\ \bibinfo
  {pages} {P10007} (\bibinfo {year} {2018})}\BibitemShut {NoStop}%
\bibitem [{\citenamefont {{DUNE Collaboration}}(2018)}]{dune_idr3}%
  \BibitemOpen
  \bibfield  {author} {\bibinfo {author} {\bibnamefont {{DUNE
  Collaboration}}},\ }\href@noop {} {\  (\bibinfo {year} {2018})},\ \Eprint
  {http://arxiv.org/abs/1807.10340v1} {arXiv:1807.10340v1 [physics.ins-det]}
  \BibitemShut {NoStop}%
\bibitem [{\citenamefont {Caratelli}(2018)}]{caratelli_phd}%
  \BibitemOpen
  \bibfield  {author} {\bibinfo {author} {\bibfnamefont {D.}~\bibnamefont
  {Caratelli}},\ }\href@noop {} {Ph.D. thesis},\ \bibinfo  {school} {Columbia
  University} (\bibinfo {year} {2018})\BibitemShut {NoStop}%
\bibitem [{\citenamefont {Adams}\ \emph
  {et~al.}(2020{\natexlab{a}})\citenamefont {Adams} \emph {et~al.}}]{ub_pi0}%
  \BibitemOpen
  \bibfield  {author} {\bibinfo {author} {\bibfnamefont {C.}~\bibnamefont
  {Adams}} \emph {et~al.} (\bibinfo {collaboration} {MicroBooNE}),\ }\href
  {\doibase 10.1088/1748-0221/15/02/P02007} {\bibfield  {journal} {\bibinfo
  {journal} {JINST}\ }\textbf {\bibinfo {volume} {15}},\ \bibinfo {pages}
  {P02007} (\bibinfo {year} {2020}{\natexlab{a}})}\BibitemShut {NoStop}%
\bibitem [{\citenamefont {{S.~Agostinelli, \textit{et~al.}}}(2003)}]{g4}%
  \BibitemOpen
  \bibfield  {author} {\bibinfo {author} {\bibnamefont {{S.~Agostinelli,
  \textit{et~al.}}}} (\bibinfo {collaboration} {GEANT4 Collaboration}),\ }\href
  {\doibase 10.1016/S0168-9002(03)01368-8} {\bibfield  {journal} {\bibinfo
  {journal} {Nucl. Instrum. Meth.}\ }\textbf {\bibinfo {volume} {A506}},\
  \bibinfo {pages} {250} (\bibinfo {year} {2003})}\BibitemShut {NoStop}%
\bibitem [{\citenamefont {Snider}\ and\ \citenamefont
  {Petrillo}(2017)}]{larsoft}%
  \BibitemOpen
  \bibfield  {author} {\bibinfo {author} {\bibfnamefont {E.}~\bibnamefont
  {Snider}}\ and\ \bibinfo {author} {\bibfnamefont {G.}~\bibnamefont
  {Petrillo}},\ }\href {\doibase 10.1088/1742-6596/898/4/042057} {\bibfield
  {journal} {\bibinfo  {journal} {Journal of Physics: Conference Series}\
  }\textbf {\bibinfo {volume} {898}},\ \bibinfo {pages} {042057} (\bibinfo
  {year} {2017})}\BibitemShut {NoStop}%
\bibitem [{\citenamefont {Gardiner}(2018)}]{marley}%
  \BibitemOpen
  \bibfield  {author} {\bibinfo {author} {\bibfnamefont {S.}~\bibnamefont
  {Gardiner}},\ }\emph {\bibinfo {title} {Nuclear Effects in Neutrino
  Detection}},\ \href {https://search.proquest.com/docview/2194284425} {Ph.D.
  thesis},\ \bibinfo  {school} {University of California, Davis} (\bibinfo
  {year} {2018})\BibitemShut {NoStop}%
\bibitem [{\citenamefont {Acciarri}\ \emph
  {et~al.}(2017{\natexlab{e}})\citenamefont {Acciarri} \emph
  {et~al.}}]{ub_noise}%
  \BibitemOpen
  \bibfield  {author} {\bibinfo {author} {\bibfnamefont {R.}~\bibnamefont
  {Acciarri}} \emph {et~al.} (\bibinfo {collaboration} {MicroBooNE}),\ }\href
  {\doibase 10.1088/1748-0221/12/08/P08003} {\bibfield  {journal} {\bibinfo
  {journal} {JINST}\ }\textbf {\bibinfo {volume} {12}},\ \bibinfo {pages}
  {P08003} (\bibinfo {year} {2017}{\natexlab{e}})}\BibitemShut {NoStop}%
\bibitem [{\citenamefont {Capozzi}\ \emph {et~al.}(2019)\citenamefont
  {Capozzi}, \citenamefont {Li}, \citenamefont {Zhu},\ and\ \citenamefont
  {Beacom}}]{dune_solar}%
  \BibitemOpen
  \bibfield  {author} {\bibinfo {author} {\bibfnamefont {F.}~\bibnamefont
  {Capozzi}}, \bibinfo {author} {\bibfnamefont {S.~W.}\ \bibnamefont {Li}},
  \bibinfo {author} {\bibfnamefont {G.}~\bibnamefont {Zhu}}, \ and\ \bibinfo
  {author} {\bibfnamefont {J.~F.}\ \bibnamefont {Beacom}},\ }\href {\doibase
  10.1103/PhysRevLett.123.131803} {\bibfield  {journal} {\bibinfo  {journal}
  {Phys. Rev. Lett.}\ }\textbf {\bibinfo {volume} {123}},\ \bibinfo {pages}
  {131803} (\bibinfo {year} {2019})}\BibitemShut {NoStop}%
\bibitem [{\citenamefont {Friedland}\ and\ \citenamefont {Li}(2019)}]{lar_res}%
  \BibitemOpen
  \bibfield  {author} {\bibinfo {author} {\bibfnamefont {A.}~\bibnamefont
  {Friedland}}\ and\ \bibinfo {author} {\bibfnamefont {S.~W.}\ \bibnamefont
  {Li}},\ }\href {\doibase 10.1103/PhysRevD.99.036009} {\bibfield  {journal}
  {\bibinfo  {journal} {Phys. Rev. D}\ }\textbf {\bibinfo {volume} {99}},\
  \bibinfo {pages} {036009} (\bibinfo {year} {2019})}\BibitemShut {NoStop}%
\bibitem [{\citenamefont {Ahmad}\ \emph {et~al.}(2002)\citenamefont {Ahmad}
  \emph {et~al.}}]{sno_flavor}%
  \BibitemOpen
  \bibfield  {author} {\bibinfo {author} {\bibfnamefont {Q.~R.}\ \bibnamefont
  {Ahmad}} \emph {et~al.},\ }\href {\doibase 10.1103/PhysRevLett.89.011301}
  {\bibfield  {journal} {\bibinfo  {journal} {Phys. Rev. Lett.}\ }\textbf
  {\bibinfo {volume} {89}},\ \bibinfo {pages} {011301} (\bibinfo {year}
  {2002})}\BibitemShut {NoStop}%
\bibitem [{\citenamefont {Elkins}\ \emph {et~al.}(2019)\citenamefont {Elkins}
  \emph {et~al.}}]{minerva_n}%
  \BibitemOpen
  \bibfield  {author} {\bibinfo {author} {\bibfnamefont {M.}~\bibnamefont
  {Elkins}} \emph {et~al.} (\bibinfo {collaboration} {MINERvA}),\ }\href
  {\doibase 10.1103/PhysRevD.100.052002} {\bibfield  {journal} {\bibinfo
  {journal} {Phys. Rev. D}\ }\textbf {\bibinfo {volume} {100}},\ \bibinfo
  {pages} {052002} (\bibinfo {year} {2019})}\BibitemShut {NoStop}%
\bibitem [{\citenamefont {Acero}\ \emph {et~al.}(2019)\citenamefont {Acero}
  \emph {et~al.}}]{nova_osc}%
  \BibitemOpen
  \bibfield  {author} {\bibinfo {author} {\bibfnamefont {M.}~\bibnamefont
  {Acero}} \emph {et~al.} (\bibinfo {collaboration} {NOvA}),\ }\href {\doibase
  10.1103/PhysRevLett.123.151803} {\bibfield  {journal} {\bibinfo  {journal}
  {Phys. Rev. Lett.}\ }\textbf {\bibinfo {volume} {123}},\ \bibinfo {pages}
  {151803} (\bibinfo {year} {2019})}\BibitemShut {NoStop}%
\bibitem [{\citenamefont {MacMullin}\ \emph {et~al.}(2012)\citenamefont
  {MacMullin} \emph {et~al.}}]{lar_ng}%
  \BibitemOpen
  \bibfield  {author} {\bibinfo {author} {\bibfnamefont {S.}~\bibnamefont
  {MacMullin}} \emph {et~al.},\ }\href {\doibase 10.1103/PhysRevC.85.064614}
  {\bibfield  {journal} {\bibinfo  {journal} {Phys. Rev. C}\ }\textbf {\bibinfo
  {volume} {85}},\ \bibinfo {pages} {064614} (\bibinfo {year}
  {2012})}\BibitemShut {NoStop}%
\bibitem [{\citenamefont {Bhatia}\ \emph {et~al.}(2012)\citenamefont {Bhatia},
  \citenamefont {Finch}, \citenamefont {Gooden},\ and\ \citenamefont
  {Tornow}}]{lar_np}%
  \BibitemOpen
  \bibfield  {author} {\bibinfo {author} {\bibfnamefont {C.}~\bibnamefont
  {Bhatia}}, \bibinfo {author} {\bibfnamefont {S.~W.}\ \bibnamefont {Finch}},
  \bibinfo {author} {\bibfnamefont {M.~E.}\ \bibnamefont {Gooden}}, \ and\
  \bibinfo {author} {\bibfnamefont {W.}~\bibnamefont {Tornow}},\ }\href
  {\doibase 10.1103/PhysRevC.86.041602} {\bibfield  {journal} {\bibinfo
  {journal} {Phys. Rev. C}\ }\textbf {\bibinfo {volume} {86}},\ \bibinfo
  {pages} {041602} (\bibinfo {year} {2012})}\BibitemShut {NoStop}%
\bibitem [{\citenamefont {Otuka}\ \emph {et~al.}(2014)\citenamefont {Otuka}
  \emph {et~al.}}]{exfor}%
  \BibitemOpen
  \bibfield  {author} {\bibinfo {author} {\bibfnamefont {N.}~\bibnamefont
  {Otuka}} \emph {et~al.},\ }\href {\doibase 10.1016/j.nds.2014.07.065}
  {\bibfield  {journal} {\bibinfo  {journal} {Nucl. Data Sheets}\ }\textbf
  {\bibinfo {volume} {120}},\ \bibinfo {pages} {272} (\bibinfo {year}
  {2014})}\BibitemShut {NoStop}%
\bibitem [{\citenamefont {Brown}\ \emph {et~al.}(2018)\citenamefont {Brown}
  \emph {et~al.}}]{endf}%
  \BibitemOpen
  \bibfield  {author} {\bibinfo {author} {\bibfnamefont {D.}~\bibnamefont
  {Brown}} \emph {et~al.},\ }\href {\doibase
  https://doi.org/10.1016/j.nds.2018.02.001} {\bibfield  {journal} {\bibinfo
  {journal} {Nucl. Data Sheets}\ }\textbf {\bibinfo {volume} {148}},\ \bibinfo
  {pages} {1 } (\bibinfo {year} {2018})}\BibitemShut {NoStop}%
\bibitem [{\citenamefont {Acero}\ \emph {et~al.}(2018)\citenamefont {Acero}
  \emph {et~al.}}]{nova_mult}%
  \BibitemOpen
  \bibfield  {author} {\bibinfo {author} {\bibfnamefont {M.}~\bibnamefont
  {Acero}} \emph {et~al.} (\bibinfo {collaboration} {NOvA}),\ }\href@noop {}
  {\bibfield  {journal} {\bibinfo  {journal} {Phys. Rev. D}\ }\textbf {\bibinfo
  {volume} {98}},\ \bibinfo {pages} {032012} (\bibinfo {year}
  {2018})}\BibitemShut {NoStop}%
\bibitem [{\citenamefont {{NuDat 2.8}}(2020)}]{nndc_chart}%
  \BibitemOpen
  \bibfield  {author} {\bibinfo {author} {\bibnamefont {{NuDat 2.8}}},\
  }\href@noop {} {} (\bibinfo {year} {2020}),\ \bibinfo {note}
  {\href{https://www.nndc.bnl.gov/nudat2}{https://www.nndc.bnl.gov/nudat2}}\BibitemShut
  {NoStop}%
\bibitem [{\citenamefont {Bhandari}\ \emph {et~al.}(2019)\citenamefont
  {Bhandari} \emph {et~al.}}]{lar_captain}%
  \BibitemOpen
  \bibfield  {author} {\bibinfo {author} {\bibfnamefont {B.}~\bibnamefont
  {Bhandari}} \emph {et~al.} (\bibinfo {collaboration} {CAPTAIN}),\ }\href
  {\doibase 10.1103/PhysRevLett.123.042502} {\bibfield  {journal} {\bibinfo
  {journal} {Phys. Rev. Lett.}\ }\textbf {\bibinfo {volume} {123}},\ \bibinfo
  {pages} {042502} (\bibinfo {year} {2019})}\BibitemShut {NoStop}%
\bibitem [{\citenamefont {Friedland}()}]{Friedland_marley}%
  \BibitemOpen
  \bibfield  {author} {\bibinfo {author} {\bibfnamefont {A.}~\bibnamefont
  {Friedland}},\ }\href@noop {} {}\bibinfo {howpublished} {personal
  communication}\BibitemShut {NoStop}%
\bibitem [{\citenamefont {Acciarri}\ \emph {et~al.}(2015)\citenamefont
  {Acciarri} \emph {et~al.}}]{dune_cdr}%
  \BibitemOpen
  \bibfield  {author} {\bibinfo {author} {\bibfnamefont {R.}~\bibnamefont
  {Acciarri}} \emph {et~al.} (\bibinfo {collaboration} {DUNE}),\ }\href@noop {}
  {\  (\bibinfo {year} {2015})},\ \Eprint {http://arxiv.org/abs/1512.06148}
  {arXiv:1512.06148 [physics.ins-det]} \BibitemShut {NoStop}%
\bibitem [{\citenamefont {Acciarri}\ \emph
  {et~al.}(2020{\natexlab{b}})\citenamefont {Acciarri} \emph
  {et~al.}}]{lariat_detpaper}%
  \BibitemOpen
  \bibfield  {author} {\bibinfo {author} {\bibfnamefont {R.}~\bibnamefont
  {Acciarri}} \emph {et~al.} (\bibinfo {collaboration} {LArIAT}),\ }\href
  {\doibase 10.1088/1748-0221/15/04/P04026} {\bibfield  {journal} {\bibinfo
  {journal} {JINST}\ }\textbf {\bibinfo {volume} {15}},\ \bibinfo {pages}
  {P04026} (\bibinfo {year} {2020}{\natexlab{b}})}\BibitemShut {NoStop}%
\bibitem [{\citenamefont {Adams}\ \emph
  {et~al.}(2019{\natexlab{c}})\citenamefont {Adams} \emph
  {et~al.}}]{ub_cosmicreject}%
  \BibitemOpen
  \bibfield  {author} {\bibinfo {author} {\bibfnamefont {C.}~\bibnamefont
  {Adams}} \emph {et~al.} (\bibinfo {collaboration} {MicroBooNE}),\ }\href
  {\doibase 10.1140/epjc/s10052-019-7184-7} {\bibfield  {journal} {\bibinfo
  {journal} {Eur. Phys. J. C}\ }\textbf {\bibinfo {volume} {79}},\ \bibinfo
  {pages} {673} (\bibinfo {year} {2019}{\natexlab{c}})}\BibitemShut {NoStop}%
\bibitem [{\citenamefont {Adams}\ \emph
  {et~al.}(2020{\natexlab{b}})\citenamefont {Adams} \emph {et~al.}}]{ub_cal}%
  \BibitemOpen
  \bibfield  {author} {\bibinfo {author} {\bibfnamefont {C.}~\bibnamefont
  {Adams}} \emph {et~al.} (\bibinfo {collaboration} {MicroBooNE}),\ }\href
  {\doibase 10.1088/1748-0221/15/03/P03022} {\bibfield  {journal} {\bibinfo
  {journal} {JINST}\ }\textbf {\bibinfo {volume} {15}},\ \bibinfo {pages}
  {P03022} (\bibinfo {year} {2020}{\natexlab{b}})}\BibitemShut {NoStop}%
\bibitem [{\citenamefont {Acciarri}\ \emph
  {et~al.}(2018{\natexlab{b}})\citenamefont {Acciarri} \emph
  {et~al.}}]{argo_pi}%
  \BibitemOpen
  \bibfield  {author} {\bibinfo {author} {\bibfnamefont {R.}~\bibnamefont
  {Acciarri}} \emph {et~al.} (\bibinfo {collaboration} {ArgoNeuT}),\ }\href
  {\doibase 10.1103/PhysRevD.98.052002} {\bibfield  {journal} {\bibinfo
  {journal} {Phys. Rev. D}\ }\textbf {\bibinfo {volume} {98}},\ \bibinfo
  {pages} {052002} (\bibinfo {year} {2018}{\natexlab{b}})}\BibitemShut
  {NoStop}%
\bibitem [{\citenamefont {Gramellini}(2018)}]{elena_phd}%
  \BibitemOpen
  \bibfield  {author} {\bibinfo {author} {\bibfnamefont {E.}~\bibnamefont
  {Gramellini}},\ }\href@noop {} {Ph.D. thesis},\ \bibinfo  {school} {Yale
  University} (\bibinfo {year} {2018})\BibitemShut {NoStop}%
\bibitem [{\citenamefont {McGivern}\ \emph {et~al.}(2016)\citenamefont
  {McGivern} \emph {et~al.}}]{minerva_pi}%
  \BibitemOpen
  \bibfield  {author} {\bibinfo {author} {\bibfnamefont {C.}~\bibnamefont
  {McGivern}} \emph {et~al.} (\bibinfo {collaboration} {MINERvA}),\ }\href
  {\doibase 10.1103/PhysRevD.94.052005} {\bibfield  {journal} {\bibinfo
  {journal} {Phys. Rev. D}\ }\textbf {\bibinfo {volume} {94}},\ \bibinfo
  {pages} {052005} (\bibinfo {year} {2016})}\BibitemShut {NoStop}%
\bibitem [{\citenamefont {Abe}\ \emph {et~al.}(2017)\citenamefont {Abe} \emph
  {et~al.}}]{t2k_pi}%
  \BibitemOpen
  \bibfield  {author} {\bibinfo {author} {\bibfnamefont {K.}~\bibnamefont
  {Abe}} \emph {et~al.} (\bibinfo {collaboration} {T2K}),\ }\href {\doibase
  10.1103/PhysRevD.95.012010} {\bibfield  {journal} {\bibinfo  {journal} {Phys.
  Rev. D}\ }\textbf {\bibinfo {volume} {95}},\ \bibinfo {pages} {012010}
  (\bibinfo {year} {2017})}\BibitemShut {NoStop}%
\bibitem [{\citenamefont {Measday}(2001)}]{cap_measday}%
  \BibitemOpen
  \bibfield  {author} {\bibinfo {author} {\bibfnamefont {D.}~\bibnamefont
  {Measday}},\ }\href {\doibase https://doi.org/10.1016/S0370-1573(01)00012-6}
  {\bibfield  {journal} {\bibinfo  {journal} {Physics Reports}\ }\textbf
  {\bibinfo {volume} {354}},\ \bibinfo {pages} {243 } (\bibinfo {year}
  {2001})}\BibitemShut {NoStop}%
\bibitem [{\citenamefont {Weyer}(1990)}]{cap_meyer}%
  \BibitemOpen
  \bibfield  {author} {\bibinfo {author} {\bibfnamefont {H.~J.}\ \bibnamefont
  {Weyer}},\ }\href {\doibase https://doi.org/10.1016/0370-1573(90)90076-E}
  {\bibfield  {journal} {\bibinfo  {journal} {Physics Reports}\ }\textbf
  {\bibinfo {volume} {195}},\ \bibinfo {pages} {295 } (\bibinfo {year}
  {1990})}\BibitemShut {NoStop}%
\bibitem [{\citenamefont {{MicroBooNE
  Collaboration}}(2018{\natexlab{c}})}]{uB_dedx}%
  \BibitemOpen
  \bibfield  {author} {\bibinfo {author} {\bibnamefont {{MicroBooNE
  Collaboration}}},\ }\href@noop {} {} (\bibinfo {year} {2018}{\natexlab{c}}),\
  \bibinfo {note}
  {\href{https://microboone.fnal.gov/wp-content/uploads/MICROBOONE-NOTE-1048-PUB.pdf}{MICROBOONE-NOTE-1048-PUB}}\BibitemShut
  {NoStop}%
\bibitem [{\citenamefont {Argüelles}\ \emph {et~al.}(2019)\citenamefont
  {Argüelles} \emph {et~al.}}]{bsm_white}%
  \BibitemOpen
  \bibfield  {author} {\bibinfo {author} {\bibfnamefont {C.}~\bibnamefont
  {Argüelles}} \emph {et~al.},\ }\href@noop {} {\  (\bibinfo {year} {2019})},\
  \Eprint {http://arxiv.org/abs/1907.08311} {arXiv:1907.08311 [hep-ph]}
  \BibitemShut {NoStop}%
\bibitem [{\citenamefont {Altmannshofer}\ \emph {et~al.}(2019)\citenamefont
  {Altmannshofer}, \citenamefont {Gori}, \citenamefont {Martín-Albo},
  \citenamefont {Sousa},\ and\ \citenamefont {Wallbank}}]{dune_trident}%
  \BibitemOpen
  \bibfield  {author} {\bibinfo {author} {\bibfnamefont {W.}~\bibnamefont
  {Altmannshofer}}, \bibinfo {author} {\bibfnamefont {S.}~\bibnamefont {Gori}},
  \bibinfo {author} {\bibfnamefont {J.}~\bibnamefont {Martín-Albo}}, \bibinfo
  {author} {\bibfnamefont {A.}~\bibnamefont {Sousa}}, \ and\ \bibinfo {author}
  {\bibfnamefont {M.}~\bibnamefont {Wallbank}},\ }\href {\doibase
  10.1103/PhysRevD.100.115029} {\bibfield  {journal} {\bibinfo  {journal}
  {Phys. Rev. D}\ }\textbf {\bibinfo {volume} {100}},\ \bibinfo {pages}
  {115029} (\bibinfo {year} {2019})}\BibitemShut {NoStop}%
\bibitem [{\citenamefont {Ballett}\ \emph {et~al.}(2019)\citenamefont
  {Ballett}, \citenamefont {Hostert}, \citenamefont {Pascoli}, \citenamefont
  {Perez-Gonzalez}, \citenamefont {Tabrizi},\ and\ \citenamefont
  {Zukanovich~Funchal}}]{dune_trident2}%
  \BibitemOpen
  \bibfield  {author} {\bibinfo {author} {\bibfnamefont {P.}~\bibnamefont
  {Ballett}}, \bibinfo {author} {\bibfnamefont {M.}~\bibnamefont {Hostert}},
  \bibinfo {author} {\bibfnamefont {S.}~\bibnamefont {Pascoli}}, \bibinfo
  {author} {\bibfnamefont {Y.~F.}\ \bibnamefont {Perez-Gonzalez}}, \bibinfo
  {author} {\bibfnamefont {Z.}~\bibnamefont {Tabrizi}}, \ and\ \bibinfo
  {author} {\bibfnamefont {R.}~\bibnamefont {Zukanovich~Funchal}},\ }\href
  {\doibase 10.1007/JHEP01(2019)119} {\bibfield  {journal} {\bibinfo  {journal}
  {JHEP}\ }\textbf {\bibinfo {volume} {01}},\ \bibinfo {pages} {119} (\bibinfo
  {year} {2019})}\BibitemShut {NoStop}%
\bibitem [{\citenamefont {Ballett}\ \emph {et~al.}(2017)\citenamefont
  {Ballett}, \citenamefont {Pascoli},\ and\ \citenamefont
  {Ross-Lonergan}}]{bsm_heavy}%
  \BibitemOpen
  \bibfield  {author} {\bibinfo {author} {\bibfnamefont {P.}~\bibnamefont
  {Ballett}}, \bibinfo {author} {\bibfnamefont {S.}~\bibnamefont {Pascoli}}, \
  and\ \bibinfo {author} {\bibfnamefont {M.}~\bibnamefont {Ross-Lonergan}},\
  }\href {\doibase 10.1007/JHEP04(2017)102} {\bibfield  {journal} {\bibinfo
  {journal} {JHEP}\ }\textbf {\bibinfo {volume} {04}},\ \bibinfo {pages} {102}
  (\bibinfo {year} {2017})}\BibitemShut {NoStop}%
\bibitem [{\citenamefont {Batell}\ \emph {et~al.}(2019)\citenamefont {Batell},
  \citenamefont {Berger},\ and\ \citenamefont {Ismail}}]{bsm_higgs}%
  \BibitemOpen
  \bibfield  {author} {\bibinfo {author} {\bibfnamefont {B.}~\bibnamefont
  {Batell}}, \bibinfo {author} {\bibfnamefont {J.}~\bibnamefont {Berger}}, \
  and\ \bibinfo {author} {\bibfnamefont {A.}~\bibnamefont {Ismail}},\ }\href
  {\doibase 10.1103/PhysRevD.100.115039} {\bibfield  {journal} {\bibinfo
  {journal} {Phys. Rev. D}\ }\textbf {\bibinfo {volume} {100}},\ \bibinfo
  {pages} {115039} (\bibinfo {year} {2019})}\BibitemShut {NoStop}%
\bibitem [{\citenamefont {Berryman}\ \emph {et~al.}(2020)\citenamefont
  {Berryman}, \citenamefont {de~Gouvea}, \citenamefont {Fox}, \citenamefont
  {Kayser}, \citenamefont {Kelly},\ and\ \citenamefont {Raaf}}]{bsm_mpd}%
  \BibitemOpen
  \bibfield  {author} {\bibinfo {author} {\bibfnamefont {J.~M.}\ \bibnamefont
  {Berryman}}, \bibinfo {author} {\bibfnamefont {A.}~\bibnamefont {de~Gouvea}},
  \bibinfo {author} {\bibfnamefont {P.~J.}\ \bibnamefont {Fox}}, \bibinfo
  {author} {\bibfnamefont {B.~J.}\ \bibnamefont {Kayser}}, \bibinfo {author}
  {\bibfnamefont {K.~J.}\ \bibnamefont {Kelly}}, \ and\ \bibinfo {author}
  {\bibfnamefont {J.~L.}\ \bibnamefont {Raaf}},\ }\href {\doibase
  10.1007/JHEP02(2020)174} {\bibfield  {journal} {\bibinfo  {journal} {JHEP}\
  }\textbf {\bibinfo {volume} {02}},\ \bibinfo {pages} {174} (\bibinfo {year}
  {2020})}\BibitemShut {NoStop}%
\bibitem [{\citenamefont {Enrico~Bertuzzo}\ \emph {et~al.}(2019)\citenamefont
  {Enrico~Bertuzzo}, \citenamefont {Jana}, \citenamefont {Machado},\ and\
  \citenamefont {Funchal}}]{bsm_dark}%
  \BibitemOpen
  \bibfield  {author} {\bibinfo {author} {\bibfnamefont {E.}~\bibnamefont
  {Enrico~Bertuzzo}}, \bibinfo {author} {\bibfnamefont {S.}~\bibnamefont
  {Jana}}, \bibinfo {author} {\bibfnamefont {P.~A.}\ \bibnamefont {Machado}}, \
  and\ \bibinfo {author} {\bibfnamefont {R.~Z.}\ \bibnamefont {Funchal}},\
  }\href {\doibase https://doi.org/10.1016/j.physletb.2019.02.023} {\bibfield
  {journal} {\bibinfo  {journal} {Physics Letters B}\ }\textbf {\bibinfo
  {volume} {791}},\ \bibinfo {pages} {210 } (\bibinfo {year}
  {2019})}\BibitemShut {NoStop}%
\bibitem [{\citenamefont {Acciarri}\ \emph
  {et~al.}(2020{\natexlab{c}})\citenamefont {Acciarri} \emph
  {et~al.}}]{argo_mcp}%
  \BibitemOpen
  \bibfield  {author} {\bibinfo {author} {\bibfnamefont {R.}~\bibnamefont
  {Acciarri}} \emph {et~al.} (\bibinfo {collaboration} {ArgoNeuT}),\ }\href
  {\doibase 10.1103/PhysRevLett.124.131801} {\bibfield  {journal} {\bibinfo
  {journal} {Phys. Rev. Lett.}\ }\textbf {\bibinfo {volume} {124}},\ \bibinfo
  {pages} {131801} (\bibinfo {year} {2020}{\natexlab{c}})}\BibitemShut
  {NoStop}%
\bibitem [{\citenamefont {Harnik}\ \emph {et~al.}(2019)\citenamefont {Harnik},
  \citenamefont {Liu},\ and\ \citenamefont {Palamara}}]{harnik_mcp}%
  \BibitemOpen
  \bibfield  {author} {\bibinfo {author} {\bibfnamefont {R.}~\bibnamefont
  {Harnik}}, \bibinfo {author} {\bibfnamefont {Z.}~\bibnamefont {Liu}}, \ and\
  \bibinfo {author} {\bibfnamefont {O.}~\bibnamefont {Palamara}},\ }\href
  {\doibase 10.1007/JHEP07(2019)170} {\bibfield  {journal} {\bibinfo  {journal}
  {JHEP}\ }\textbf {\bibinfo {volume} {07}},\ \bibinfo {pages} {170} (\bibinfo
  {year} {2019})}\BibitemShut {NoStop}%
\bibitem [{\citenamefont {Abe}\ \emph {et~al.}(2019)\citenamefont {Abe} \emph
  {et~al.}}]{t2k_spec}%
  \BibitemOpen
  \bibfield  {author} {\bibinfo {author} {\bibfnamefont {K.}~\bibnamefont
  {Abe}} \emph {et~al.} (\bibinfo {collaboration} {T2K}),\ }\href {\doibase
  10.1103/PhysRevD.100.112009} {\bibfield  {journal} {\bibinfo  {journal}
  {Phys. Rev. D}\ }\textbf {\bibinfo {volume} {100}},\ \bibinfo {pages}
  {112009} (\bibinfo {year} {2019})}\BibitemShut {NoStop}%
\bibitem [{\citenamefont {Wang}(2018)}]{lar_neutron}%
  \BibitemOpen
  \bibfield  {author} {\bibinfo {author} {\bibfnamefont {J.}~\bibnamefont
  {Wang}},\ }\href {https://indico.fnal.gov/event/18523/contributions/47916/}
  {\enquote {\bibinfo {title} {{Pulsed Neutron Source for Liquid Argon TPC
  Calibration}},}\ }\bibinfo {howpublished} {{Workshop on Calibration and
  Reconstruction for LArTPC Detectors}} (\bibinfo {year} {2018})\BibitemShut
  {NoStop}%
\bibitem [{\citenamefont {Fischer}\ \emph {et~al.}(2019)\citenamefont {Fischer}
  \emph {et~al.}}]{lar_aced}%
  \BibitemOpen
  \bibfield  {author} {\bibinfo {author} {\bibfnamefont {V.}~\bibnamefont
  {Fischer}} \emph {et~al.} (\bibinfo {collaboration} {ACED Collaboration}),\
  }\href {\doibase 10.1103/PhysRevD.99.103021} {\bibfield  {journal} {\bibinfo
  {journal} {Phys. Rev. D}\ }\textbf {\bibinfo {volume} {99}},\ \bibinfo
  {pages} {103021} (\bibinfo {year} {2019})}\BibitemShut {NoStop}%
\bibitem [{\citenamefont {Benetti}\ \emph {et~al.}(2007)\citenamefont {Benetti}
  \emph {et~al.}}]{lar_ar39}%
  \BibitemOpen
  \bibfield  {author} {\bibinfo {author} {\bibfnamefont {P.}~\bibnamefont
  {Benetti}} \emph {et~al.} (\bibinfo {collaboration} {WARP}),\ }\href
  {\doibase 10.1016/j.nima.2007.01.106} {\bibfield  {journal} {\bibinfo
  {journal} {Nucl. Instrum. Meth. A}\ }\textbf {\bibinfo {volume} {574}},\
  \bibinfo {pages} {83} (\bibinfo {year} {2007})}\BibitemShut {NoStop}%
\bibitem [{\citenamefont {Crespo-Anadón}(2019)}]{ub_trig}%
  \BibitemOpen
  \bibfield  {author} {\bibinfo {author} {\bibfnamefont {J.}~\bibnamefont
  {Crespo-Anadón}} (\bibinfo {collaboration} {MicroBooNE}),\ }\href {\doibase
  10.1088/1742-6596/1312/1/012006} {\bibfield  {journal} {\bibinfo  {journal}
  {J. Phys. Conf. Ser.}\ }\textbf {\bibinfo {volume} {1312}},\ \bibinfo {pages}
  {012006} (\bibinfo {year} {2019})}\BibitemShut {NoStop}%
\bibitem [{\citenamefont {{MicroBooNE Collaboration}}(2019)}]{uB_sn2}%
  \BibitemOpen
  \bibfield  {author} {\bibinfo {author} {\bibnamefont {{MicroBooNE
  Collaboration}}},\ }\href@noop {} {} (\bibinfo {year} {2019}),\ \bibinfo
  {note}
  {\href{https://microboone.fnal.gov/wp-content/uploads/MICROBOONE-NOTE-1030-PUB.pdf}{MICROBOONE-NOTE-1030-PUB}}\BibitemShut
  {NoStop}%
\bibitem [{\citenamefont {Walkowiak}(2000)}]{lar_diff1}%
  \BibitemOpen
  \bibfield  {author} {\bibinfo {author} {\bibfnamefont {W.}~\bibnamefont
  {Walkowiak}},\ }\href {\doibase 10.1016/S0168-9002(99)01301-7} {\bibfield
  {journal} {\bibinfo  {journal} {Nucl. Instrum. Meth. A}\ }\textbf {\bibinfo
  {volume} {449}},\ \bibinfo {pages} {288} (\bibinfo {year}
  {2000})}\BibitemShut {NoStop}%
\bibitem [{\citenamefont {Amoruso}\ \emph
  {et~al.}(2004{\natexlab{b}})\citenamefont {Amoruso} \emph
  {et~al.}}]{lar_diff2}%
  \BibitemOpen
  \bibfield  {author} {\bibinfo {author} {\bibfnamefont {S.}~\bibnamefont
  {Amoruso}} \emph {et~al.},\ }\href {\doibase
  https://doi.org/10.1016/j.nima.2003.11.423} {\bibfield  {journal} {\bibinfo
  {journal} {Nucl. Instrum. Meth. A}\ }\textbf {\bibinfo {volume} {523}},\
  \bibinfo {pages} {275 } (\bibinfo {year} {2004}{\natexlab{b}})}\BibitemShut
  {NoStop}%
\bibitem [{\citenamefont {Cabrera}\ \emph {et~al.}(2019)\citenamefont {Cabrera}
  \emph {et~al.}}]{liquido_phys}%
  \BibitemOpen
  \bibfield  {author} {\bibinfo {author} {\bibfnamefont {A.}~\bibnamefont
  {Cabrera}} \emph {et~al.},\ }\href@noop {} {\  (\bibinfo {year} {2019})},\
  \Eprint {http://arxiv.org/abs/1908.02859} {arXiv:1908.02859
  [physics.ins-det]} \BibitemShut {NoStop}%
\bibitem [{\citenamefont {Oberla}\ and\ \citenamefont {Frisch}(2016)}]{o_tpc}%
  \BibitemOpen
  \bibfield  {author} {\bibinfo {author} {\bibfnamefont {E.}~\bibnamefont
  {Oberla}}\ and\ \bibinfo {author} {\bibfnamefont {H.}~\bibnamefont
  {Frisch}},\ }\href {\doibase 10.1016/j.nima.2016.01.030} {\bibfield
  {journal} {\bibinfo  {journal} {Nucl. Instrum. Meth. A}\ }\textbf {\bibinfo
  {volume} {814}},\ \bibinfo {pages} {19} (\bibinfo {year} {2016})}\BibitemShut
  {NoStop}%
\bibitem [{\citenamefont {Alvarez-Ruso}\ \emph {et~al.}(2018)\citenamefont
  {Alvarez-Ruso} \emph {et~al.}}]{nustec}%
  \BibitemOpen
  \bibfield  {author} {\bibinfo {author} {\bibfnamefont {L.}~\bibnamefont
  {Alvarez-Ruso}} \emph {et~al.} (\bibinfo {collaboration} {NuSTEC}),\ }\href
  {\doibase 10.1016/j.ppnp.2018.01.006} {\bibfield  {journal} {\bibinfo
  {journal} {Prog. Part. Nucl. Phys.}\ }\textbf {\bibinfo {volume} {100}},\
  \bibinfo {pages} {1} (\bibinfo {year} {2018})}\BibitemShut {NoStop}%
\bibitem [{\citenamefont {Mahn}\ \emph {et~al.}(2018)\citenamefont {Mahn},
  \citenamefont {Marshall},\ and\ \citenamefont {Wilkinson}}]{mahn}%
  \BibitemOpen
  \bibfield  {author} {\bibinfo {author} {\bibfnamefont {K.}~\bibnamefont
  {Mahn}}, \bibinfo {author} {\bibfnamefont {C.}~\bibnamefont {Marshall}}, \
  and\ \bibinfo {author} {\bibfnamefont {C.}~\bibnamefont {Wilkinson}},\ }\href
  {\doibase 10.1146/annurev-nucl-101917-020930} {\bibfield  {journal} {\bibinfo
   {journal} {Ann. Rev. Nucl. Part. Sci.}\ }\textbf {\bibinfo {volume} {68}},\
  \bibinfo {pages} {105} (\bibinfo {year} {2018})}\BibitemShut {NoStop}%
\end{thebibliography}%
\end{document}